\pdfoutput=1
\RequirePackage{ifpdf}
\ifpdf % We are running pdfTeX in pdf mode
\documentclass[pdftex]{sigma}
\else
\documentclass{sigma}
\fi

\def\tr{\mathop{\rm tr}\nolimits}

\def\re{\mathop{\rm Re}\nolimits}
\def\im{\mathop{\rm Im}\nolimits}

\def\diag{\mathop{\rm diag}\nolimits}
\def\ad{\mathop{\rm ad}\nolimits}

\def\Res{\mathop{\rm Res}\limits}

\def\openone{\leavevmode\hbox{\small1\kern-3.3pt\normalsize1}}

\def\wedgecomma{\mathop{\wedge}\limits_{'}}

\def\biglb{\big[\hspace*{-.7mm}\big[}
\def\bigrb{\big]\hspace*{-.7mm}\big]}

\def\newpic#1{%
   \def\emline##1##2##3##4##5##6{%
      \put(##1,##2){\special{em:point #1##3}}%
      \put(##4,##5){\special{em:point #1##6}}%
      \special{em:line #1##3,#1##6}}}

\newpic{}

\def\bbbr{{\mathbb R}}

\def\bbbc{{\mathbb C}}

\def\bbbz{{\mathbb Z}}

\def\biglb{\big[\hspace*{-.7mm}\big[}
\def\bigrb{\big]\hspace*{-.7mm}\big]}

\def\openone{\leavevmode\hbox{\small1\kern-3.3pt\normalsize1}}

\def\openone{\leavevmode\hbox{\small1\kern-3.3pt\normalsize1}}

\def\0{{\boldsymbol 0}}

\def\wedgecomma{\mathop{\wedge}\limits_{'}}

\def\bbbc{{\mathbb C}}
\def\bbbr{{\mathbb R}}

\numberwithin{equation}{section}

\begin{document}

\allowdisplaybreaks

\renewcommand{\thefootnote}{$\star$}

\renewcommand{\PaperNumber}{096}

\FirstPageHeading

\ShortArticleName{Polynomial Bundles and GFT for Integrable Equations
on {\bf A.III}-type Symmetric Spaces}

\ArticleName{Polynomial Bundles and Generalised Fourier\\ Transforms for Integrable Equations\\
on {\bf A.III}-type  Symmetric Spaces\footnote{This
paper is a contribution to the Proceedings of the Conference ``Symmetries and Integrability of Dif\/ference Equations (SIDE-9)'' (June 14--18, 2010, Varna, Bulgaria). The full collection is available at \href{http://www.emis.de/journals/SIGMA/SIDE-9.html}{http://www.emis.de/journals/SIGMA/SIDE-9.html}}}

\Author{Vladimir S.~GERDJIKOV~$^\dag$, Georgi G.~GRAHOVSKI~$^{\dag,\ddag}$, Alexander V.~MIKHAILOV~$^\S$\\ and
Tihomir I.~VALCHEV~$^\dag$}

\AuthorNameForHeading{V.S.~Gerdjikov, G.G.~Grahovski, A.V.~Mikhailov and T.I.~Valchev}

\Address{$^\dag$~Institute for Nuclear Research and Nuclear Energy,  Bulgarian Academy of Sciences,\\
\hphantom{$^\dag$}~72 Tsarigradsko chausee, Sofia 1784, Bulgaria}
\EmailD{\href{mailto:gerjikov@inrne.bas.bg}{gerjikov@inrne.bas.bg}, \href{mailto:grah@inrne.bas.bg}{grah@inrne.bas.bg}, \href{mailto:valtchev@inrne.bas.bg}{valtchev@inrne.bas.bg}}

\Address{$^\ddag$~School of Mathematical Sciences, Dublin Institute of Technology, \\
\hphantom{$^\ddag$}~Kevin Street, Dublin 8, Ireland}

\Address{$^\S$~Applied Math. Department, University of Leeds, Woodhouse Lane, Leeds, LS2 9JT, UK}
\EmailD{\href{mailto:a.v.mikhailov@leeds.ac.uk}{a.v.mikhailov@leeds.ac.uk}}

\ArticleDates{Received May 26, 2011, in f\/inal form October 04, 2011;  Published online October 20, 2011}

\Abstract{A special class of integrable nonlinear dif\/ferential equations related to {\bf A.III}-type symmetric spaces and having
additional reductions are analyzed via the inverse scattering method (ISM). Using the dressing method we construct
two classes of soliton solutions associated with the Lax operator. Next, by using the Wronskian relations,  the mapping
between the potential and the minimal sets of scattering data is constructed.
Furthermore, completeness relations for the `squared solutions'  (generalized
 exponentials) are derived. Next, expansions of the potential and its
variation are obtained. This demonstrates that the interpretation
of the inverse scattering method as a generalized Fourier transform holds true. Finally,  the
Hamiltonian structures of these generalized multi-component Heisenberg ferromagnetic (MHF) type integrable models on
{\bf A.III}-type symmetric spaces are brief\/ly analyzed.}


\Classification{37K20; 35Q51; 74J30; 78A60}

{\small \tableofcontents}

\renewcommand{\thefootnote}{\arabic{footnote}}
\setcounter{footnote}{0}

\section{Introduction}

It is well known \cite{ForKu*83}, that two important classes of integrable nonlinear evolution equations
(NLEE) are related to symmetric spaces: the multi-component nonlinear Schr\"odinger equations and their gauge
equivalent Heisenberg ferromagnet (HF) equations:
\begin{gather*}
 i  S_t=\frac{1}{2}[S,S_{xx}],\qquad S^2(x,t) = \openone.
%\label{HFb}
\end{gather*}
Here $S$ takes values in some simple Lie algebra. In the simplest case of the algebra $\mathfrak{su}(2)$
the function $S$ describes the spin of one-dimensional ferromagnet~\cite{BorKis}.

The HF's equation is integrable in the sense of inverse
scattering transform~\cite{ForKu*83, blue-bible}. Its Lax representation has the form:
\begin{gather*}
L(\lambda) \equiv  i \partial_x-\lambda S,\qquad
A(\lambda) \equiv  i \partial_t + \frac{ i \lambda}{2}[S,S_x] + 2\lambda^2 S.
\end{gather*}
Since the time  the complete integrability of HF equations was discovered, many
attempts for constructing its generalization have been made~\cite{golsok2}. The main goal of this
paper is to derive and analyze a special class of  NLEE which is obtained from the
HF type equations by applying  additional algebraic symmetries.

A well known method for constructing new integrable NLEE is based on the so-called
reduction group, introduced in~\cite{mik_toda,miktetr,mik,mik_ll} and further developed in
\cite{GGK*01,lomsan,miklom1,miklom}. It led to the discovery of the 2-dimensional Toda f\/ield theories \cite{mik,mop1}.
Its latest developments
of the method led to the discovery of new automorphic Lie algebras and their classif\/ication \cite{lomsan,miklom1,miklom}.

The  hierarchies of integrable NLEE are generated by the so-called recursion (generating) operator.
Such operators have been constructed and analyzed for a wide class of Lax opera\-tors~$L$ and appeared to generate not only the
Lax representations, but also the hierarchy of NLEE's related to a given Lax operator $L$, their conservation laws and their hierarchy of Hamiltonian
structures, see \cite{DrSok*85,brown-bible,blue-bible,GVY*08,golsok1} and the numerous references therein.
Such operators can be viewed also as  Lax $L$ operators, taken
in the adjoint representation of the underlying Lie algebra~$\mathfrak{g}$.

The (generating) recursion operator appeared f\/irst in the
AKNS-approach \cite{AKNS} as a tool to generate the class of
all $A$-operators as well as the NLEE related to the given Lax
operator~$L(\lambda)$. Next, I.M.~Gel'fand and L.A.~Dickey~\cite{wildgelf} discovered that the class of these
$A$-operators is contained in the diagonal of the resolvent of~$L$. The kernel of the resolvent of~$L$ can be explicitly constructed
in terms of the so-called fundamental analytic solutions (FAS), see~\cite{IP2,GeKu,TMF94}.

\looseness=-1
The construction of the recursion operator for Lax operators, whose explicit dependence on the spectral parameter~$\lambda$ is
comparatively simple  (say, linear, or quadratic) was done a long time ago \cite{AKNS,KN,GeKh,bjp}.
Furthermore, the completeness property for the set of eigenfunctions of the recursion operator (the `squared solutions' of $L$) is of paramount importance. The completeness of the `squared solutions' plays a  fundamental role
in proving that the inverse scattering method is, in fact, a nonlinear analogue of the Fourier transform, which allows one to linearize the NLEE.
Using these relations one is able to derive all fundamental properties of the NLEE on a common basis.

In the symmetry approach to integrable equations the so-called `formal recursion operator' plays a crucial role. It has the general property to map a symmetry into a symmetry of the given integrable NLEE~\cite{olver}. If the considered NLEE possesses an inf\/inite hierarchy of symmetries~\cite{IbrSh*80}, or conservation laws \cite{SvS*82a} of arbitrary (high) order, or can be linearized by a dif\/ferential substitution \cite{SvS*82b}, then it has a formal recursion operator \cite{MShY*87,MSSh*91,SSh*84}. For an extensive and up-to-date surveys on symmetry approach to integrability we refer to \cite{AShY*2000,MS*09} and
the references therein. We just note that in the Symmetry approach the formal recursion operator does not depend on the boundary conditions, imposed on the potential of the associated Lax operator. An alternative approach for construction of recursion operator from a given Lax operator is presented by M.~G\"urses, A.~Karasu and V.~Sokolov in~\cite{GKS}. This method is pure algebraic, it is not related yet with the spectral theory of the Lax operator~$L$ and it is not sensitive to the choice of the class of admissible potentials of~$L$.

The problem of deriving  recursion operators becomes more dif\/f\/icult when we impose additional reductions on $L$. If this additional
reduction is compatible with $L$, being linear or quadratic in $\lambda$, the construction of the recursion operators is not a dif\/f\/icult task (see
\cite{GVY*08}; an alternative construction of $\Lambda$ is given in \cite{GKS,golsok}). The  ef\/fect of the $\bbbz_n$-reduction is as follows: i)~the  relevant `squared solutions' have analyticity properties
in sectors of the complex $\lambda$-plane closing angles $\pi/n$; ii)~the grading of the Lie algebra $\mathfrak{g} \equiv \oplus_{k=0}^{n-1}\mathfrak{g}^{(k)}$ is more involved and as a~consequence the recursion operator is factorized into a product of~$n$ factors $\Lambda =\prod\limits_{k=0}^{n-1}\Lambda_k$, and each of the factors~$\Lambda_k$ maps $\Lambda_k \colon \mathfrak{g}^{(k-1)} \to
\mathfrak{g}^{(k)}$.

The applications of the dif\/ferential geometric and Lie algebraic
methods to soliton type equations led to the discovery of a close
relationship between the multi-component (matrix) integrable equations (of nonlinear Schr\"odinger type)
and the symmetric and homogeneous spaces~\cite{ForKu*83}.  Later on, this approach was extended to other types
of multi-component integrable models, like the derivative NLS,
Korteweg--de~Vries and modif\/ied Korteweg--de~Vries, $N$-wave,
Davey--Stewartson, Kadomtsev--Petviashvili equations
\cite{AthFor,For}.

The purpose of the present paper is to derive nonlinear evolution
equations to generalize Heisenberg's model and study some
of the properties of their Lax operators. We are going to focus
our attention on equations whose Lax representation is related
to $\mathfrak{su}(3)$.

This paper is a natural continuation of our previous papers \cite{ours,ours2,ours3}.
In Section~\ref{section2} below we give some of the necessary preliminaries.
We also study the $\bbbz_2$-reductions of the generalized
Heisenberg ferromagnets ($\bbbz_2$-HF) related to symmetric spaces and
outline the spectral pro\-per\-ties of the relevant Lax operator and
the construction of its fundamental analytic solutions.
 Section~\ref{section3} is dedicated to the soliton solutions of  the ($\bbbz_2$-HF) models.
In their derivation we make use of the dressing method \cite{brown-bible,zakharovshabat,zm1,zm2}.
Due to the additional reductions the Lax opera\-tor~$L$ may possess two types of discrete eigenvalues:
generic ones, coming in quadruplets~$\pm\lambda_k$,~$\pm\lambda_k^*$ and purely imaginary ones
coming as doublets $\pm i\kappa_j$. Therefore we will have two types of soliton solutions which
we will call quadruplet and doublet solitons respectively. We outline the purely algebraic
construction for deriving the $N$-soliton solutions for both types of solitons and provide
the explicit expressions for the one-soliton solutions.
Section~\ref{section4} is devoted to the derivation of the recursion operators~$\Lambda$. Here we f\/irst derive $\Lambda$ using the G\"urses--Karasu--Sokolov
(GKS) method~\cite{GKS}.  Next we analyze  the Wronskian relations as a basic tool in the inverse scattering
method \cite{CaDe*76_1,CaDe*76_2}. From them, there naturally arise
the `squared solutions', which play a fundamental role also in the
analysis of the mapping between the set of  admissible
potentials and the minimal sets of scattering data. The `squared solutions' can also be viewed as
eigenfunctions of the recursion operators~$\Lambda_\pm$, which can  be used as an alternative def\/inition
of the recursion operators. One can check that both approaches lead to equivalent expressions
for~$\Lambda_\pm$.
Section~\ref{section5} is dedicated to the spectral properties of the recursion operators. There we start by
proving the completeness relation for the `squared solutions'. Next we use this relation to derive
the expansions of $\ad_{L_1}^{-1}L_{1,x}$, $\ad_{L_1}^{-1}L_{2,x}$ and $\ad_{L_1}^{-1}\delta L_{1}$
over the `squared solutions'.
In the last Section~\ref{section6} these expansions are treated as generalized Fourier transforms, allowing one to linearize the
NLEE. We also demonstrate that all fundamental properties of the class of NLEE can be formulated through the
recursion operators. These include not only  the description of the class of NLEE related to $L$,
but also their integrals of motion and hierarchy of Hamiltonian structures.
We end by discussion and conclusions. In Appendix~\ref{appendixA} we have collected some useful
formulae for the Gauss factors of the scattering matrix and explicitly derive the operator~$\ad_{L_1}^{-1}$.

\section{Preliminaries}\label{section2}

\subsection{Basic notations and general form of equations}\label{ssec:2.1}

The main object in our paper is the following 2-component system of NLEEs:
\begin{gather}
i u_t + u_{xx}+(uu^*_x+vv^*_x)u_x+(uu^*_x+vv^*_x)_x u=0,\nonumber\\
i v_t + v_{xx}+(uu^*_x+vv^*_x)v_x+(uu^*_x+vv^*_x)_x v=0,
\label{nee}
\end{gather}
where the functions $u:\bbbr\times\bbbr\to\bbbc$ and $v:\bbbr\times\bbbr\to\bbbc$
are assumed to be inf\/initely smooth and satisfy the following boundary conditions
\begin{gather}
\lim_{x\to \pm\infty} u(x,t) =0,\qquad
\lim_{x\to \pm\infty} v(x,t) =e^{i \phi_\pm},\qquad \phi_{\pm}\in\bbbr.
\label{const_bc}
\end{gather}
Moreover, $u$ and $v$ are not functionally independent, but obey the constraint
$|u|^2 + |v|^2 = 1$.

The system (\ref{nee}) represents a reduction of generalized Heisenberg ferromagnet
equations related to the symmetric space $SU(3)/S(U(1)\times U(2))$, see
\cite{ours,ours2}. It possesses a zero curvature representation with Lax operators
in the form
\begin{gather}
L(\lambda)\equiv i \partial_x + \lambda L_1(x,t), \label{lax_1}\\
A(\lambda)\equiv i \partial_t + \lambda A_1(x,t)+ \lambda^2 A_2(x,t),
\label{lax_2}
\end{gather}
where $\lambda$ is a spectral parameter and the matrix coef\/f\/icients (potentials) read
\begin{gather}
L_1 = \left(\begin{array}{ccc}
0 & u & v \\ u^* & 0 & 0 \\
v^* & 0 & 0
\end{array}\right),\qquad
A_2 = \frac{2}{3}\openone-L^2_1 =-
\left(\begin{array}{ccc}
1/3 & 0 & 0 \\ 0 & |u|^2-2/3 & u^*v \\
0 & v^*u & |v|^2-2/3
\end{array}\right),\label{matr_coef1}\\
A_1 = \left(\begin{array}{ccc}
0 & a & b \\ a^* & 0 & 0 \\
b^* & 0 & 0
\end{array}\right),\qquad
a= i u_{x}+i (uu^*_x+vv^*_x)u,\qquad
b= i v_{x}+i (uu^*_x+vv^*_x)v.
\label{matr_coef2}
\end{gather}
The specif\/ic structure of the matrices above is a result of the simultaneous action
of two $\bbbz_2$ reductions on generic Lax opera\-tors~$L$ and~$A$, namely
\begin{gather}
L^{\dag}(\lambda^*) =  -\breve{L}(\lambda)  , \qquad   A^{\dag}(\lambda^*) =
-\breve{A}(\lambda),\label{red1}\\
\mathbf{C} L(-\lambda) \mathbf{C}  =  L(\lambda),\qquad
\mathbf{C} A(-\lambda)\mathbf{C} = A(\lambda),
\qquad \mathbf{C}=\diag(1,-1,-1),\label{red2}
\end{gather}
where the operation $\breve{}$\, is def\/ined as follows
\[\breve{L}(\lambda)\psi(x,t,\lambda) \equiv i \partial_x\psi(x,t,\lambda)
 - \lambda \psi(x,t,\lambda) L_1(x,t,\lambda). \]
The latter reduction represents Cartan's involutive automorphism
\cite{Hel,loos} involved in the de\-f\/i\-ni\-tion of the symmetric space
$SU(3)/S(U(1)\times U(2))$, that is, it induces a $\bbbz_2$-grading in the
Lie algebra $\mathfrak{sl}(3,\bbbc)$
\begin{gather*}%\label{eq:16}
\mathfrak{sl}(3) = \mathfrak{sl}^{0}(3)\oplus\mathfrak{sl}^{1}(3),
\qquad\mathfrak{sl}^{\sigma}(3) = \{ X\in \mathfrak{g} \, | \,
 \mathbf{C} X \mathbf{C} = (-1)^{\sigma} X\}.
\end{gather*}
It is evident that the grading condition
\begin{gather}
\left[\mathfrak{sl}^{\sigma}(3),\mathfrak{sl}^{\varsigma}(3)\right]
\subset \mathfrak{sl}^{\sigma+ \varsigma (\mathrm{mod}\, 2)}(3)
\label{grad_cond}\end{gather}
is fulf\/illed. This grading will be used widely in our further considerations.
 Due to the existence of the grading (\ref{grad_cond}), any function $X(x,t,\lambda)$ with values in
$\mathfrak{sl}(3)$ can be  represented in the form:
\begin{gather}
X(x,t,\lambda) = X^{0}(x,t,\lambda) + X^{1}(x,t,\lambda),
\qquad X^{0,1}(x,t,\lambda)\in\mathfrak{sl}^{0,1}(3).
\label{X_split}
\end{gather}
Each component $X^{0,1}$, in its turn, splits into a term commuting with
$L_1$ and its orthogonal complement, namely
\begin{gather}
X^{0}  =   X^{0,\bot} + \kappa_0 L_2,\qquad L_2 = -A_2 = L^2_1 - \frac{2}{3}\openone,
\qquad \langle X^{0,\bot}, L_2\rangle = 0,\label{X0_split}\\
X^{1}  =    X^{1,\bot} + \kappa_1 L_1,\qquad \langle X^{1,\bot}, L_1\rangle = 0.
\label{X1_split}
\end{gather}
The brackets above stand for Killing form, def\/ined as:
\[\langle X, Y\rangle = \tr (XY).\]
It is evident that the functions $\kappa_0$ and $\kappa_1$ can be expressed as
follows
\[
\kappa_0 = \frac{\langle X^{0}, L_2\rangle}{\langle L_2, L_2\rangle},
\qquad \kappa_1 = \frac{\langle X^{1}, L_1\rangle}{\langle L_1, L_1\rangle}.
\]
One can easily check that the norms of $L_1$ and $L_2$ are
\begin{gather*}
\langle L_1, L_1\rangle = 2,\qquad \langle L_2, L_2\rangle = \frac{2}{3}
\end{gather*}
and hence $\kappa_0$ and $\kappa_1$ are given by
\begin{gather*}
\kappa_0 = \frac{3}{2}\langle X^{0}, L_2\rangle,
\qquad \kappa_1 = \frac{1}{2}\langle X^{1}, L_1\rangle.
%\label{coef01}
\end{gather*}
As we shall convince ourselves in some cases it proves to be more
convenient to deal with Lax operators
\begin{gather}
\tilde{L}(\lambda)  =  i \partial_x + U_0(x,t) + \lambda J,
\qquad J=\diag(1,0,-1), \label{lax_1_g}\\
\tilde{A}(\lambda) = i \partial_t + \tilde{A}_0(x,t)
+ \lambda \tilde{A}_1(x,t) - \lambda^2 I,\qquad
I=\diag(1/3,-2/3,1/3)
\label{lax_2_g}
\end{gather}
gauge equivalent to (\ref{lax_1}), (\ref{lax_2}). The gauge
transformation, which puts $L_1$ and $A_2$ into their diagonal
form, is given by the unitary matrix
\begin{gather*}
g = \frac{\sqrt{2}}{2}\left(\begin{array}{ccc}
1 & 0 & -1 \\ u^* & \sqrt{2} v & u^*\\
v^* & -\sqrt{2} u  & v^*
\end{array}\right).
%\label{g_trans}
\end{gather*}
Further on we are also going to use the explicit form
of $U_0$ given by next formula
\begin{gather}
U_0 = i g^{-1}g_x =
\frac{i}{2}\left(\begin{array}{ccc}
uu^*_x +vv^*_x  & \sqrt{2}(uv_x-vu_x) & uu^*_x +vv^*_x \\
-\sqrt{2}(u^*v^*_x-v^*u^*_x) & 2(u^*u_x +v^*v_x) & -\sqrt{2}(u^*v^*_x-v^*u^*_x)\\
uu^*_x +vv^*_x & \sqrt{2}(uv_x-vu_x) & uu^*_x +vv^*_x
\end{array}\right).
\label{u_0}
\end{gather}

\subsection{Direct scattering problem for the Lax operator (\ref{lax_1})}\label{ssec:2.2}

Here we shall provide a short summary of some basic results and notions concerning
the spectral theory of the Lax operator $L$, the direct scattering
problem and introduce the so-called fundamental analytic solutions. The spectral
properties of $L$ depend on the choice of  the class of admissible potentials, i.e.\
the potentials  are a subject to certain boundary condition. In the case of
boundary conditions of type (\ref{const_bc}), the continuous spectrum of $L$
coincides with the real axis in the complex $\lambda$-plane, see \cite{ours2}.

In order to formulate the direct scattering problem for $L$, one needs to
introduce fundamental sets of solutions\footnote{Further we shall
simply call them fundamental solutions for short.} $\psi$ to the
auxiliary (spectral) linear system
\begin{gather}
L(\lambda)\psi(x,t,\lambda) = i \partial_x \psi(x,t,\lambda)
+ \lambda L_1(x,t) \psi(x,t,\lambda) = 0.
\label{lax_pol}
\end{gather}
Since $L(\lambda)$ and $A(\lambda)$ commute,  fundamental solutions
$\psi$ also satisfy  the equation
\begin{gather}
A(\lambda)\psi(x,t,\lambda) = i \partial_t \psi(x,t,\lambda)
+ (\lambda A_1(x,t) + \lambda^2 A_2(x,t))\psi(x,t,\lambda)
= \psi(x,t,\lambda) f(\lambda).
\label{auxsys_2}\end{gather}
The matrix-valued function
\begin{gather}\label{f_lambda}
f(\lambda)=\lim_{x\to\pm \infty} g_{\pm}^{-1}
(\lambda A_1(x,t) + \lambda^2 A_2(x,t)) g_{\pm}
= -\lambda^2 I
\end{gather}
is called dispersion law of the nonlinear equation.

A special type of fundamental solutions are the so-called Jost
solutions $\psi_{\pm}$ which are normalized as follows
\begin{gather*}
\lim_{x\to\pm\infty}\psi_{\pm}(x,t,\lambda)
e^{-i \lambda J x}g_{\pm}^{-1} =\openone,
%\label{josts}
\end{gather*}
where
\[g_{\pm} = \lim_{x\to\pm\infty}g(x,t) =
\frac{1}{\sqrt{2}}\left(\begin{array}{ccc}
1 & 0 & -1 \\ 0 & \sqrt{2}\,e^{i \phi_\pm} & 0 \\
e^{-i \phi_\pm} & 0 & e^{-i \phi_\pm}
\end{array}\right)
\]
diagonalizes the asymptotics $L_{1,\pm}=\lim\limits_{x\to\pm\infty}L_1(x,t)$.
Due to (\ref{f_lambda}) one can show that the asymptotic behavior of
$\psi_{\pm}$ do not depend on $t$, i.e.\ the def\/inition is correct.
The transition matrix
\begin{gather}
T(t,\lambda) = [\psi_{+}(x,t,\lambda)]^{-1}\psi_{-}(x,t,\lambda).
\label{tmatrix}
\end{gather}
is called scattering matrix. It can be easily deduced from relation
(\ref{auxsys_2}) that the scattering matrix evolves with time according
to the linear dif\/ferential equation
\begin{gather*}
i \partial_t T+[f(\lambda),T]=0,
\end{gather*}
which is integrated straight away to give
\begin{gather*}
T(t,\lambda)=e^{i f(\lambda)t}T(0,\lambda)e^{-i f(\lambda)t}.
\end{gather*}
Since the time parameter $t$ does not play a signif\/icant role in
our further considerations, we shall omit it (it will be f\/ixed).

The action of $\bbbz_2$-reductions (\ref{red1}), (\ref{red2}) imposes
the following restrictions
\begin{gather}
\big[\psi^\dag _{\pm}(x,\lambda^*)\big]^{-1}
 =  \psi_{\pm}(x,\lambda),\qquad
\big[T^\dag (\lambda^*)\big]^{-1} = T(\lambda), \label{jostred1}\\
\mathbf{C}\psi_{\pm}(x,-\lambda)\mathbf{C}  =  \psi_{\pm}(x,\lambda),\qquad
\mathbf{C}T(-\lambda)\mathbf{C} = T(\lambda)\label{jostred2}
\end{gather}
on the Jost solutions and the scattering matrix.

From now on we will assume that $\phi_+=\phi_-=0$. Hence
the asymptotic values of $L_1$ and $g$ become equal to each other:
\[L_{1,\pm} = L_{1,\rm{as}},\qquad
g_{+} = g_{-}=g_{\rm{as}}.\]
The complex $\lambda$-plane is separated by the continuous spectrum of $L$ (real axis)
into two regions of analyticity: the upper half plane $\bbbc_{+}$ and the lower
one $\bbbc_{-}$. We are going to sketch two ways of constructing fundamental solutions
(FAS) $\chi^{+}(x,\lambda)$ and $\chi^{-}(x,\lambda)$ which are analytic functions in
$\bbbc_{+}$ and $\bbbc_{-}$ respectively, see~\cite{ours2} for more detailed explanations.
The f\/irst way is based on introducing some auxiliary functions
\begin{gather*}%\label{eq:xi-pm}
\eta_{\pm}(x,\lambda)=g_{\rm{as}}^{-1} \psi_{\pm}(x,\lambda)
e^{-i \lambda J x},
\end{gather*}
which satisfy the auxiliary system:
\begin{gather*}%\label{eq:eta}
i\partial_x\eta_{\pm} + U(x,\lambda) \eta_{\pm}(x,\lambda)
- \lambda \eta _{\pm}(x,\lambda) J=0  , \qquad
U(x,\lambda)= \lambda g_{\rm{as}}^{-1}  L_1(x)  g_{\rm{as}}
\end{gather*}
with the boundary conditions $\lim\limits_{x\to\pm\infty} \eta_{\pm}(x,\lambda) =\openone$.
Equivalently, $\eta_\pm(x,\lambda)$ can be regarded as solutions to the following Volterra-type integral equations:
\begin{gather*}
\eta_{\pm}(x,\lambda)=\openone+i \int^{x}_{\pm\infty} d y
e^{i \lambda J(x-y)}[U (y,\lambda)-\lambda J]\eta_{\pm}(y,\lambda)
e^{-i  \lambda J(x-y)}.
\end{gather*}
Then we def\/ine $\xi^+(x,\lambda)$ as a solution to the following set
of integral equations:
\begin{gather*}
\xi_{kl}^+(x,\lambda) = \delta_{kl}+i \int^{x}_{-\infty}dy
e^{i \lambda(J_{kk}  -J_{ ll})(x-y)}\left[(U(y,\lambda)
- \lambda J )\xi^+(y,\lambda)\right]_{kl},\qquad k\leq l,\nonumber\\
\xi_{kl}^+(x,\lambda) = i \int^{x}_{\infty} d y
e^{i \lambda(J_{kk}(\lambda) - J_{ll})(x-y)}
\left[(U(y,\lambda) - \lambda J )\xi^+(y,\lambda)\right]_{kl},
\qquad k>l,%\label{eq:xip}
\end{gather*}
and $\xi^{-}(x,\lambda)$ being a solution to
\begin{gather*}
\xi_{kl}^-(x,\lambda)  = \delta_{kl}+i \int^{x}_{\infty}dy
e^{i \lambda (J_{kk} -J_{ll} )(x-y)}\left((U(y,\lambda)
-\lambda J )\xi^-(y,\lambda)\right)_{kl},\qquad
k\leq l,\nonumber\\
\xi_{kl}^-(x,\lambda)  = i \int^{x}_{-\infty} d y
e^{i \lambda (J_{kk} -J_{ll} )(x-y)}\left((U(y,\lambda)
- \lambda J )\xi^-(y,\lambda)\right)_{kl},\qquad k>l.
%\label{eq:xim1}
\end{gather*}
It is easy to check that $\xi^{+}$ and $\xi^{-}$ have the analytic
properties in $\bbbc_{+}$ and $\bbbc_{-}$ respectively due to the
appropriate choice of the lower integration limits. The initial fundamental
analytic solutions~$\chi^{\pm}(x,\lambda)$ are obtained
from $\xi^{\pm}(x,\lambda)$ by applying the transformation:
\begin{gather}\label{eq:chi'}
\chi^{\pm}(x,\lambda) = g_{\rm{as}} \xi^{\pm}(x,\lambda)e^{i \lambda J x}.
\end{gather}
Another way of how one can construct FAS is by multiplying the Jost solutions
by appropriately chosen matrices. It turns out that these matrix factors are
involved in the Gauss decomposition
\begin{gather}\label{eq:T}
T(\lambda)=T^{\mp}(\lambda)D^{\pm}(\lambda)(S^{\pm}(\lambda))^{-1}
\end{gather}
of the scattering matrix $T(\lambda)$. Here is a list of all Gauss factors
in a parametrization we are going to use further in our exposition:
\begin{alignat}{3}
& S^\pm (\lambda) =\exp s^\pm(\lambda) , \qquad && T^\pm (\lambda)  =\exp r^\pm(\lambda) ,& \nonumber\\
& s^+ (\lambda) = \left(\begin{array}{ccc} 0 & s^+_{\alpha_1} &  s^+_{\alpha_3} \\ 0 & 0 &  s^+_{\alpha_2} \\
0 & 0 & 0  \end{array}\right), \qquad&& r^+ (\lambda)  = \left(\begin{array}{ccc} 0 & r^+_{\alpha_1} &  r^+_{\alpha_3} \\ 0 & 0 &
r^+_{\alpha_2} \\ 0 & 0 & 0  \end{array}\right),& \nonumber\\
& s^- (\lambda) = \left(\begin{array}{ccc} 0 & 0 & 0 \\ s^-_{\alpha_1} & 0 &  0 \\ s^-_{\alpha_3} & s^-_{\alpha_2} & 0  \end{array}\right),
\qquad && r^- (\lambda)  = \left(\begin{array}{ccc} 0 & 0 &  0 \\ r^-_{\alpha_1} & 0 &  \\ r^-_{\alpha_3} & r^-_{\alpha_2} & 0
 \end{array}\right), &\nonumber\\
& D^+(\lambda) = \diag (m_{1}^+, m_{2}^+/m_{1}^+, 1/m_{2}^+),  \qquad && D^-(\lambda) = \diag (1/m_{2}^-, m_{2}^-/m_{1}^-, m_{1}^-),& \label{gauss_par}
\end{alignat}
where $m_k^+(\lambda)$ (resp.\ $m_k^-(\lambda)$) are the principal minors of $T(\lambda)$ of order $k=1,2$.
Then~$\chi^{+}$ and~$\chi^{-}$ are expressed as follows
\begin{gather*}%\label{eq:rhp'}
\chi^{\pm}(x,\lambda)=\psi_{-}(x,\lambda)S^{\pm}
=\psi_{+}(x,\lambda)T^{\mp}(\lambda)D^{\pm}(\lambda).
\end{gather*}
The relation above can be rewritten in the following manner, using (\ref{eq:chi'}):
\begin{gather}\label{riemman}
 \xi^{+}(x,\lambda)=\xi^{-}(x,\lambda)G(x,\lambda),\qquad   G(x,\lambda) =
e^{i \lambda J x} (S^-)^{-1}S^+(\lambda)e^{-i \lambda J x}.
\end{gather}
Thus FAS can be regarded as solutions to a local Riemann--Hilbert problem.
The established interrelation between the inverse scattering method and
local Riemann--Hilbert problem proves to be useful in constructing solutions
to NLEEs.

It can be shown that the reduction conditions (\ref{jostred1}), (\ref{jostred2})
and equation (\ref{eq:T}) lead to the following demands on the Gauss factors
\begin{alignat*}{4}
& (S^+(\lambda^*))^\dag = (S^-(\lambda))^{-1},   \qquad && \tilde{\mathbf{C}}S^\pm (-\lambda) \tilde{\mathbf{C}} = S^\mp (\lambda), &&& \nonumber\\
& (T^+(\lambda^*))^\dag = (T^-(\lambda))^{-1},   \qquad && \tilde{\mathbf{C}}T^\pm (-\lambda)\tilde{\mathbf{C}} = T^\mp (\lambda), \qquad &&
\tilde{\mathbf{C}} = g_{\rm as} \mathbf{C} g_{\rm as}^{-1} =
\left(\begin{array}{ccc} 0 & 0 & 1 \\ 0 & 1 & 0 \\ 1 & 0 & 0  \end{array}\right),&\nonumber\\
& (D^+(\lambda^*))^\dag = (D^-(\lambda))^{-1},   \qquad &&  \tilde{\mathbf{C}}D^\pm(-\lambda)\tilde{\mathbf{C}} = D^\pm(\lambda).&&&%\label{eq:TGred}
\end{alignat*}
According to (\ref{gauss_par}) these relations can be also written as:
\begin{alignat*}{3}
& (s^+(\lambda^*))^\dag = -s^-(\lambda) , \qquad && (r^+(\lambda^*))^\dag = -r^-(\lambda) , &\nonumber\\
& \tilde{\mathbf{C}}s^+(-\lambda)\tilde{\mathbf{C}} = s^-(\lambda) , \qquad && \tilde{\mathbf{C}}r^+(-\lambda)\tilde{\mathbf{C}} = r^-(\lambda). & %\label{par_red}
\end{alignat*}
Finally, combining all this information we see that the FAS obey the symmetry
conditions
\begin{gather*}%\label{eq:FAS-red}
(\chi^{+})^\dag (x,\lambda^*) = [\chi^{-}(x,\lambda)]^{-1}, \qquad
\mathbf{C}\chi^{+}(x,-\lambda)\mathbf{C} = \chi^{-}(x,\lambda).
\end{gather*}

\begin{remark}\label{rem:2}
The Riemann--Hilbert problem allows singular solutions as well. The simplest
types of singularities are simple poles and zeroes of the FAS and generically
correspond to discrete eigenvalues of the Lax operator~$L$.  Due to the reduction
symmetries the discrete eigenvalues must form orbits of the reduction group.
Generic orbits contain quadruplets, so if~$\mu$ is an eigenvalue, then~$-\mu$ and~$\pm \mu^*$, are eigenvalues too. However, we can have
degenerate orbits too. If the eigenvalue lies on the imaginary
axis we will have doublets of eigenvalues.
\end{remark}

\section{Dressing method and soliton solutions}\label{section3}

In the present section we are going to derive and analyze
the 1-soliton solution to the 2-component system (\ref{nee}).
For this to be done, we are going to apply the dressing method
proposed in~\cite{zakharovshabat} and developed in
\cite{zm1,zm2,mik,mik_ll}.  In the previous section (see Remark~\ref{rem:2})
we established that the operator $L$ may possess two types of discrete eigenvalues:
one, coming in quadruplets as well as purely imaginary ones
coming as doublets $\pm i\kappa_j$. Therefore we will have two types of soliton solutions:
generic (or quadruplet) ones and doublet ones (kinks).

\subsection{Rational dressing}

The idea of the dressing method is the following. Firstly,
one starts from known solutions $u_0(x,t)$, $v_0(x,t)$ of
(\ref{nee}) and a solution $\psi_0(x,t,\lambda)$ of the auxiliary
linear problems
\begin{gather}
L^{(0)}(\lambda)\psi_0  = i\partial_x\psi_0 + \lambda L^{(0)}_1\psi_0 = 0,\nonumber\\
A^{(0)}(\lambda)\psi_0  = i\partial_t\psi_0 +\big(\lambda A^{(0)}_1 + \lambda^2 A^{(0)}_2\big)\psi_0 = 0,
\label{bare_lax}
\end{gather}
where $L^{(0)}_1$, $A^{(0)}_1$ and $A^{(0)}_2$ have the same form as given in
(\ref{matr_coef1}), (\ref{matr_coef2}) but $u$ and $v$ are
replaced by $u_0$ and $v_0$ respectively. Then one constructs
another function $\psi_1(x,t,\lambda)= \Phi(x,t,\lambda)\psi_0(x,t,\lambda)$
which is assumed to be a fundamental solution to a similar system of equations
\begin{gather}
L^{(1)}(\lambda)\psi_1  = i\partial_x\psi_1 + \lambda L^{(1)}_1\psi_1 = 0,\nonumber\\
A^{(1)}(\lambda)\psi_1  = i\partial_t\psi_1 +\big(\lambda A^{(1)}_1
+ \lambda^2 A^{(1)}_2\big)\psi_1 = 0.
\label{dressed_lax}
\end{gather}
The matrices $L^{(1)}_1$, $A^{(1)}_1$ and $A^{(1)}_2$ have
the same structure as $L^{(0)}_1$, $A^{(0)}_1$ and $A^{(0)}_2$
but instead of $u_0(x,t)$ and $v_0(x,t)$ one plugs some new solutions
$u_1(x,t)$ and $v_1(x,t)$ to (\ref{nee}) which are to be found.

The dressing factor $\Phi(x,t,\lambda)$ is assumed to be regular at
$|\lambda| \to 0,\infty$. Due to the reduction conditions (\ref{red1}),
(\ref{red2}) the dressing factor has the symmetries:
\begin{gather}\label{eq:fas-invol}
{\bf C}\Phi (x,t,-\lambda){\bf C}=\Phi(x,t,\lambda),\\
\Phi (x,t,\lambda)\Phi^\dag (x,t,\lambda^*)=\openone .
\label{eq:her-invol}
\end{gather}
From  (\ref{bare_lax}) and (\ref{dressed_lax}) it follows that
$\Phi (x,t,\lambda)$ is a solution to the following equations:
\begin{gather}\label{eq:dress-eq}
 i \partial_x \Phi + \lambda L_1^{(1)}\Phi
- \lambda\Phi L_1^{(0)} = 0,\\
\label{eq:dress-eqA}
  i\partial_t \Phi +\big(\lambda A_1^{(1)} + \lambda^2A_2^{(1)}\big)
\Phi - \Phi \big(\lambda A_1^{(0)}-\lambda^2 A_2^{(0)}\big) = 0.
\end{gather}
Since $\Phi(x,t,\lambda)$ is regular at $|\lambda|\to \infty$
from~(\ref{eq:dress-eq}) one can derive the following relation
between~$L_1^{(1)}$ and~$L_1^{(0)}$:
\begin{gather}\label{eq:dress-tr}
L_1^{(1)}
(x,t)=\Phi(x,t,\infty)L_1^{(0)}(x,t)\Phi^\dag(x,t,\infty).
\end{gather}
This equation will play a fundamental role in our further considerations
since it allows one to generate a new solution to (\ref{nee})
from a given one.

If we consider a $\lambda$-independent dressing factor then from
(\ref{eq:dress-eq}) and (\ref{eq:dress-eqA}) we can deduce that~$\Phi$
do not depend on~$x$ and~$t$ either. Thus we have a simple rotation,
related to the~$U(2)$ symmetry of the model. In order to get something
more nontrivial we shall assume that $\Phi(x,t,\lambda)$ is a~rational
function of~$\lambda$ with a~minimal number of simple poles.  From
the reduction condition~(\ref{eq:fas-invol}) it follows that if~$\mu$ is a pole of~$\Phi$ then~$-\mu$ is a pole too. It is natural to assume that
these poles do not overlap with the continuous spectrum of $L$, i.e.\
$\im \mu \neq 0$.  On the other hand (\ref{eq:her-invol}) leads to the conclusion that~$\Phi^{-1}$
has poles at~$\mu^*$ and~$-\mu^*$.

We f\/irst consider the generic case when $\mu \neq -\mu^*$, so
the poles of $\Phi$ do not coincide with the poles of $\Phi^{-1}$ and
choose the following anzatz for the dressing factor and its inverse:
\begin{gather}\label{eq:dress-anz1}
\Phi(x,t,\lambda) = \openone + \frac{\lambda M(x,t)}{\lambda - \mu} + \frac{\lambda {\bf
C}M(x,t){\bf C}}{\lambda + \mu},\\
%\label{eq:dress-anz2}
\Phi^{-1}(x,t,\lambda) =  \openone + \frac{\lambda M^\dag(x,t)}{\lambda - \mu^*}
+\frac{\lambda {\bf C}M^\dag(x,t){\bf C}}{\lambda + \mu^*}.\nonumber
\end{gather}
From (\ref{eq:dress-anz1}) it follows that the matrix $M$ satisf\/ies the following
equation:
\begin{gather}\label{eq:dress-m}
\left(\openone+\frac{\mu^* M(x,t)}{\mu^*- \mu} +\frac{\mu^* {\bf
C}M(x,t){\bf C}}{\mu^* + \mu}\right)M^\dag(x,t)=0.
\end{gather}
If we assume that the residue $M$ is an invertible matrix (${\rm rank}\, M = 3$),
then it is proportional to~$\openone$ and the dressing is trivial.
In order to obtain a non-trivial dressing $M$ ought to be singular.
For our purposes it suf\/f\/ices to consider the case ${\rm rank}\, M = 1$. Then
the residue can be represented as a product $M=|{\bf n}\rangle\langle {\bf m}|$
of a vector $|{\bf n}\rangle=\left(n_1,n_2,n_3\right)^T$ and a co-vector
$\langle {\bf m}|=(m^*_1,m^*_2,m^*_3)$.  Substitution of this representation
into~(\ref{eq:dress-m}) leads to a linear equation for the vector $|{\bf n}\rangle$:
\begin{gather}\label{eq:n}
|{\bf m}\rangle+\frac{\mu^* |{\bf n}\rangle\langle {\bf m|m}\rangle}{\mu^*-
\mu} +\frac{\mu^* |{\bf C} {\bf n}\rangle\langle {\bf m}|{\bf C m}\rangle}
{\mu^* + \mu}=0.
\end{gather}
Introduce the functions
\[
f_0(x,t)={1\over \omega} \langle {\bf m(x,t)} |{\bf m(x,t)}\rangle, \qquad f_1(x,t)={1\over \kappa} \langle {\bf m(x,t)}|C |{\bf m(x,t)}\rangle, \qquad \mu = \omega + i\kappa.
\]
Then, the solution of (\ref{eq:n}) reads:
\begin{gather}\label{sol:n}
|{\bf n}(x,t)\rangle= {2( if_1(x,t)|{\bf m(x,t)}\rangle-f_0(x,t)|C |{\bf m(x,t)}\rangle)\over \mu^* (f_0^2+f_1^2)}.
\end{gather}
The vector $|{\bf m}\rangle$ is an element of the projective
space ${\mathbb CP}^2$. Indeed, it is easily seen from above that
a rescaling $|{\bf m}\rangle \to f|{\bf m}\rangle$ with any
nonzero complex $f$ does not af\/fect the matrix~$M$.

Taking into account the anzatz (\ref{eq:dress-anz1}) one can
rewrite (\ref{eq:dress-tr}) as:
\begin{gather}\label{L_1}
L_1^{(1)}=(\openone+M+{\bf C}M{\bf C})L_1^{(0)}
(\openone+M+{\bf C}M{\bf C})^\dag.
\end{gather}
 Notice that the dressing preserves the matrix structure of the Lax operator $L$, since the dressing factor $\openone+M+{\bf C}M{\bf C}$ is a block-diagonal matrix. From this equality and from (\ref{L_1}) it follows that the potentials $u_1$, $v_1$ can be expressed by $u_0$, $v_0$ in the following way:
\begin{gather}\label{u1}
 \begin{pmatrix}
u_1\\v_1\end{pmatrix}
={\cal M}\begin{pmatrix}
u_0\\v_0\end{pmatrix},\qquad {\cal
M}=\frac{G}{\mu^* (G^*)^2}\begin{pmatrix}
H&-F\\ F^* & H^*\end{pmatrix},
\end{gather}
where
\begin{gather*}
G = \omega \langle{\bf m}|{\bf m}\rangle+ i\kappa \langle{\bf m}|C|{\bf m}\rangle,\ \quad
H = |\mu|^2|m_1|^2+(\mu^*)^2|m_2|^2+\mu^2|m_3|^2,\ \quad
F = 4 i\omega \kappa m_3 m^*_2  .
\end{gather*}
It is easy to verify that the matrix ${\cal M}$ is unitary for a generic choice of
$\mu$ and the nonzero vector~$|{\bf m}\rangle$.

Thus we expressed all quantities needed in terms of $|\mathbf{m}\rangle$. It remains
to f\/ind $|\mathbf{m}\rangle$ itself. For that purpose we rewrite equations
(\ref{eq:dress-eq}), (\ref{eq:dress-eqA}) in the form:
\begin{gather}
\Phi(x,t,\lambda)\big(i\partial_x +\lambda
L_1^{(0)}\big)\Phi^{-1}(x,t,\lambda)  =\lambda L_1^{(1)},\nonumber\\
\Phi(x,t,\lambda)\big(i\partial_t +\lambda
A_1^{(0)} + \lambda^2 A^{(0)}_2\big)\Phi^{-1}(x,t,\lambda)
 =\lambda A_1^{(1)} + \lambda^2 A^{(1)}_2.\label{eq:dress-eqL1}
\end{gather}
It is obviously satisf\/ied at $\lambda=0$. From (\ref{eq:dress-eqL1}), it
follows that the residues at $\lambda=\pm \mu^*$, $\lambda=\pm {\mu}$ should
vanish. It is suf\/f\/icient to vanish the residue at  $\lambda= \mu^*$
(vanishing of the other residues follows from the symmetry conditions
(\ref{eq:fas-invol})). Computing the residue at  $\lambda= \mu^*$ and
taking into account equation (\ref{eq:dress-m}), we get
\begin{gather}\label{eq_m}
\left(\openone+\frac{\mu^* M}{\mu^*- \mu} + \frac{\mu^* {\bf
C}M{\bf C}}{\mu^* + \mu}\right)\big(i\partial_x
+\mu^* L_1^{(0)}\big)|{\bf m}\rangle=0,\\
\left(\openone + \frac{\mu^* M}{\mu^*- \mu} + \frac{\mu^* {\bf
C}M{\bf C}}{\mu^* + \mu}\right)\big(i\partial_t +\mu^*
A_1^{(0)}+(\mu^*)^2 A_2^{(0)}\big)|{\bf m}\rangle=0,
\label{eq_mA}\end{gather}
i.e.\ $|{\bf m}(x,t)\rangle$ is an eigenfunction of the dressed Lax operator. A general solution of equations~(\ref{eq_m}),~(\ref{eq_mA}) is
\begin{gather*}
 %\label{m(x,t)}
|{\bf m}(x,t)\rangle=f(x,t) \psi_0(x,t;\mu^*)|{\bf m}_0\rangle,
\end{gather*}
where~$f(x,t)$ is an arbitrary non-vanishing complex function and $|{\bf
m}_0\rangle\in\bbbc^3$ is a non-zero, but otherwise arbitrary complex vector.
Without any loss of generality we can set $f(x,t)=1$ (see the discussion above
about the projective nature of the vector $|{\bf m}\rangle$.

Thus we proved the following proposition:
\begin{proposition}\label{proposition1}
Let $u_0(x,t)$, $v_0(x,t)$ form a solution of the system \eqref{nee} and
$\psi_0(x,t,\lambda)$ be a~simultaneous fundamental solution of~\eqref{bare_lax}.
 Let also $\re\mu\neq 0 $, $\im \mu\neq 0$ and $|{\bf m}_0\rangle\in \bbbc^3$ be a~complex vector. Then $u_1(x,t),v_1(x,t)$ given by \eqref{u1} for $m_k$ being components
of the vector  $|{\bf m}\rangle= \psi_0(x,t,\mu^*)|{\bf m}_0\rangle$ satisfy \eqref{nee}
as well. The corresponding solution $\psi_1(x,t,\lambda)$ of the associa\-ted linear system \eqref{dressed_lax} is given by $\psi_1=\Phi(x,t,\lambda)\psi_0(x,t,\lambda)$ where $\Phi(x,t,\lambda)$ is defined by~\eqref{eq:dress-anz1}, \eqref{sol:n}, \eqref{eq_m} and \eqref{eq_mA}.
\end{proposition}

The new solution $u_1(x,t)$, $v_1(x,t)$ of equations (\ref{nee}) and the
fundamental solution $\psi_1(x,t;\lambda)$ of the corresponding linear system
are parameterized by a complex number $\mu$ and a complex vector $|{\bf
m}_0\rangle\in \bbbc^3$.

Let us now consider the special case when $\mu = i\kappa$. Then the sequence of steps
necessary to determine $\Phi$ slightly changes. Indeed, suppose we have a dressing factor in the form:
\begin{gather}
\Phi(x,t,\lambda)=\openone + \lambda\left(\frac{P(x,t)}{\lambda-i\kappa}+\frac{\mathbf{C}P(x,t)\mathbf{C}}{\lambda+i\kappa} \right).
\label{eq:dress-anz1a}
\end{gather}
Due to the reduction (\ref{eq:her-invol}) its inverse has the same poles as $\Phi$ itself  and therefore the equation $\Phi\Phi^{-1}=\openone$ already contains second order poles. Vanishing of the poles of second and f\/irst order respectively leads to the following algebraic conditions:
\begin{gather}
P\mathbf{C}P^{\dag} = 0,\nonumber%\label{algsys_m2}
\\
\left(\openone + P + \frac{\mathbf{C}P\mathbf{C}}{2}\right) \mathbf{C}P^{\dag}\mathbf{C} + P\left(\openone + \mathbf{C}P^{\dag}\mathbf{C} + \frac{P^{\dag}}{2}\right) = 0.
\label{algsys_m1}
\end{gather}
$P$ is a degenerate matrix which means that it can be presented as $P=|\mathbf{q}\rangle\langle \mathbf{p}|$. Then the former algebraic constraint transforms into a quadratic relation for the vector $|\mathbf{p}\rangle$
\begin{gather}
\langle \mathbf{p}|\mathbf{C}|\mathbf{p}\rangle = 0.
\label{algsys_m2a}
\end{gather}
Equation (\ref{algsys_m1}) is reduced to a linear system for $|\mathbf{q}\rangle$
\begin{gather*}
\left(\openone + \frac{\mathbf{C}|\mathbf{q}\rangle\langle \mathbf{p}|\mathbf{C}}{2}
\right)\mathbf{C}|\mathbf{p}\rangle = i\sigma |\mathbf{q}\rangle,
%\label{algsys_m1a}
\end{gather*}
where  $\sigma$ is some auxiliary real function to be found.
The above linear system allows one to express $|\mathbf{q}\rangle$ through $|\mathbf{p}\rangle$ and $\sigma$
\begin{gather}
|\mathbf{q}\rangle = \left(i\sigma - \frac{\langle\mathbf{p}|\mathbf{p}\rangle}{2}\mathbf{C}\right)^{-1}\mathbf{C}|\mathbf{p}\rangle.
\label{n_malpha}\end{gather}
In order to f\/ind $|\mathbf{p}\rangle$ and $\sigma$ we consider again the partial dif\/ferential equations (\ref{eq:dress-eqL1}). The requirement that the poles of second
and f\/irst order in (\ref{eq:dress-eqL1}) vanish identically yields to dif\/ferential
relations for $|\mathbf{p}\rangle$ and $\sigma$:
\begin{alignat*}{3}
 & i\partial_x \langle\mathbf{p}|
- i\nu \langle\mathbf{p}|L_{1}^{(0)} =0,
%\label{pdex_m2}
\qquad &&
 i\partial_t \langle\mathbf{p}| - \langle\mathbf{p}|\big(i\nu A^{(0)}_1
 + (i\nu)^2 A^{(0)}_2\big) =0, &
 %\label{pdet_m2}
 \\
&  i\partial_x\sigma = \kappa \langle \mathbf{p}|L^{(0)}_1\mathbf{C}|\mathbf{p}\rangle, \qquad &&
%\label{pdex_m1}\\
  i\partial_t\sigma = \kappa \langle \mathbf{p}|\big(A^{(0)}_1 + 2i\nu A^{(0)}_2\big)
\mathbf{C}|\mathbf{p}\rangle. &
%\label{pdet_m1}
\end{alignat*}
It is not hard to verify that $|\mathbf{p}(x,t)\rangle$ and $\sigma(x,t)$ are expressed
through the initial solution $\psi_0(x,t,\lambda)$ as follows:
\begin{gather}
|\mathbf{p}(x,t)\rangle  =  \psi_0(x,t,-i\kappa)|\mathbf{p}_{0}\rangle, \label{m_d}\\
\sigma(x,t)  =  -\kappa\langle \mathbf{p}_0|\psi^{-1}_0(x,t,i\kappa)\dot{\psi}_0(x,t, i\kappa)\mathbf{C}|\mathbf{p}_{0}\rangle + \sigma_0,
\label{sigma_d}\end{gather}
where $|\mathbf{p}_{0}\rangle$ and $\sigma_0$ are constants of integration and
$\dot{\psi}_0:= \partial_{\lambda}\psi_0$.

After substituting (\ref{m_d}) into (\ref{algsys_m2a}) and taking into account the
f\/irst $\bbbz_2$ reduction we see that the components of the polarization vector $|\mathbf{p}_0\rangle$ are no longer independent but satisfy the constraint:
\begin{gather}
\langle \mathbf{p}_0|\mathbf{C}|\mathbf{p}_0\rangle = 0\quad\Leftrightarrow\quad
|p_{0,1}|^2 = |p_{0,2}|^2 + |p_{0,3}|^2.
\label{algsys_m2b}\end{gather}

Thus to calculate the soliton solution itself one just substitutes the result
for $|\mathbf{p}\rangle$ and $\langle \mathbf{p}|$ into~$M$ and uses formula (\ref{L_1}).
Now we are able to formulate a result quite similar to Proposition~\ref{proposition1}:
\begin{proposition}\label{proposition2}
Let it be given functions $u_0(x,t)$, $v_0(x,t)$ to satisfy \eqref{nee} and
$\psi_0(x,t,\lambda)$ to be a~fundamental solution of \eqref{bare_lax}. Let also
$\kappa\in \bbbr\backslash \{0\}$, $\sigma_0\in\bbbr$ and $|{\bf p}_0\rangle$ be
a non-vanishing complex vector satisfying \eqref{algsys_m2b}. Then $u_1(x,t)$, $v_1(x,t)$ given by
\begin{gather*}
\left(\begin{array}{l}
u_1\\v_1\end{array}\right)
=\frac{|p_1|^2 + i\sigma}{(|p_1|^2 - i\sigma)^2}\left(\begin{array}{cc}
i\sigma + |p_2|^2 - |p_3|^2 & 2p^*_2p_3\\
2p_2p^*_3 & i\sigma - |p_2|^2 + |p_3|^2\end{array}\right)\left(\begin{array}{l}
u_0\\v_0\end{array}\right),
\end{gather*}
where $p_k$ are the components of the vector
$|{\bf p}\rangle= \psi_0(x,t,-i\kappa)|{\bf p}_0\rangle$ is a solution of the system~\eqref{nee} too. The solution $\psi_1(x,t,\lambda)$ of the linear system \eqref{dressed_lax} is given by $\psi_1=\Phi\psi_0$ where $\Phi$ is defined by \eqref{eq:dress-anz1a},
\eqref{n_malpha}, \eqref{m_d} and \eqref{sigma_d}.
\end{proposition}
As it is seen the new solution is parametrized by the polarization vector $|\mathbf{m}_0\rangle$, the real number~$\sigma_0$ and the pole $i\kappa$.

One can apply the dressing procedure repeatedly to build a sequence of exact
solutions to the system which is parametrized by a set of complex constants $\mu_k$ and constant vectors
$|{\bf m}_k\rangle \in {\mathbb CP}^2$ (resp.\ a set of real constants $\kappa_j$, $\sigma_{0,j}$ and complex
three-vectors $|{\bf p}_k\rangle$ in the case of purely imaginary poles).

\subsection{One-soliton solutions}

Let us apply the dressing procedure to a trivial solution $u_0=0$, $v_0=1$
of equation (\ref{nee}). In this case
\begin{gather}
 \label{psi0}
\psi_0(x,t,\lambda)=\left(\begin{array}{ccc}
\cos (\lambda x)\exp\left(-\frac{i\lambda^2
t}{3}\right) & 0 & i\sin (\lambda x)\exp\left(-\frac{i\lambda^2 t}{3}\right)\\
0 & \exp\left(\frac{2i\lambda^2 t}{3}\right) & 0 \\
i\sin (\lambda x)\exp\left(-\frac{i\lambda^2 t}{3}\right) &
0 & \cos (\lambda x)\exp\left(-\frac{i\lambda^2 t}{3}\right)
\end{array}\right).
\end{gather}
We are going to consider the generic case  f\/irst, i.e.\ we have 4 distinct poles
of $\Phi$ and $\Phi^{-1}$ to form a `quadruplet' $\{\mu,-\mu,\mu^*,-\mu^*\}$.
It is convenient to decompose a constant complex vector~$|{\bf m}_0\rangle $ according
to the eigen-spaces of the endomorphism $\psi_0$ (\ref{psi0}):
\begin{gather}
\label{m0}|{\bf m}_0\rangle=\alpha \begin{pmatrix}
1\\0\\1 \end{pmatrix} +\beta \begin{pmatrix}
1\\0\\-1 \end{pmatrix}+\gamma\begin{pmatrix}
0\\1\\0 \end{pmatrix},
\end{gather}
where $\alpha$, $\beta$, $\gamma$ are arbitrary complex constants.

If the vector $|{\bf m}_0\rangle$ is proportional to one of the eigenvectors of the
endomorphism $\psi_0$, then the corresponding matrix $M$ does not depend on the
variables $x$ and $t$ (due to the projective nature of the vector $|{\bf m}\rangle$)
and the corresponding solution (\ref{u1}) is a simple unitary rotation of the constant
solution $u_0=0$, $v_0=1$.

Elementary solitons correspond to vectors  $|{\bf m}_0\rangle$, belonging to an
essentially  two-dimensional invariant subspaces of the endomorphism~$\psi_0$.
In other words, elementary soliton solutions correspond to the choices of
vector $|{\bf m}_0\rangle$ with only one zero coef\/f\/icient in the expansion~(\ref{m0}). Let us consider each of these three cases in more detail.

\subsubsection*{Case (i):  $\alpha\ne 0$, $\beta\ne 0$, $\gamma=0$}
In this case, the solution does not depend on the variable $t$. It follows from
(\ref{u1}) that the solution can be written in the following form
\begin{gather}
 \label{solugamma0}
u_1 = 0,\\ v_1=\left(\frac{\omega(|\alpha|^2e^{2\kappa x}+|\beta|^2e^{-2\kappa
x})+\kappa(\alpha^*\beta - \beta^*\alpha)\sin(2\omega
x)+i\kappa(\alpha^*\beta + \beta^*\alpha)\cos(2\omega x)}{
\omega(|\alpha|^2e^{2\kappa x}+|\beta|^2e^{-2\kappa
x})-\kappa(\alpha^*\beta - \beta^*\alpha)\sin(2\omega
x)-i\kappa(\alpha^*\beta + \beta^*\alpha)\cos(2\omega x)}\right)^2.\nonumber
\end{gather}

\begin{figure}[t]
\centering
\includegraphics[width=0.37\textwidth]{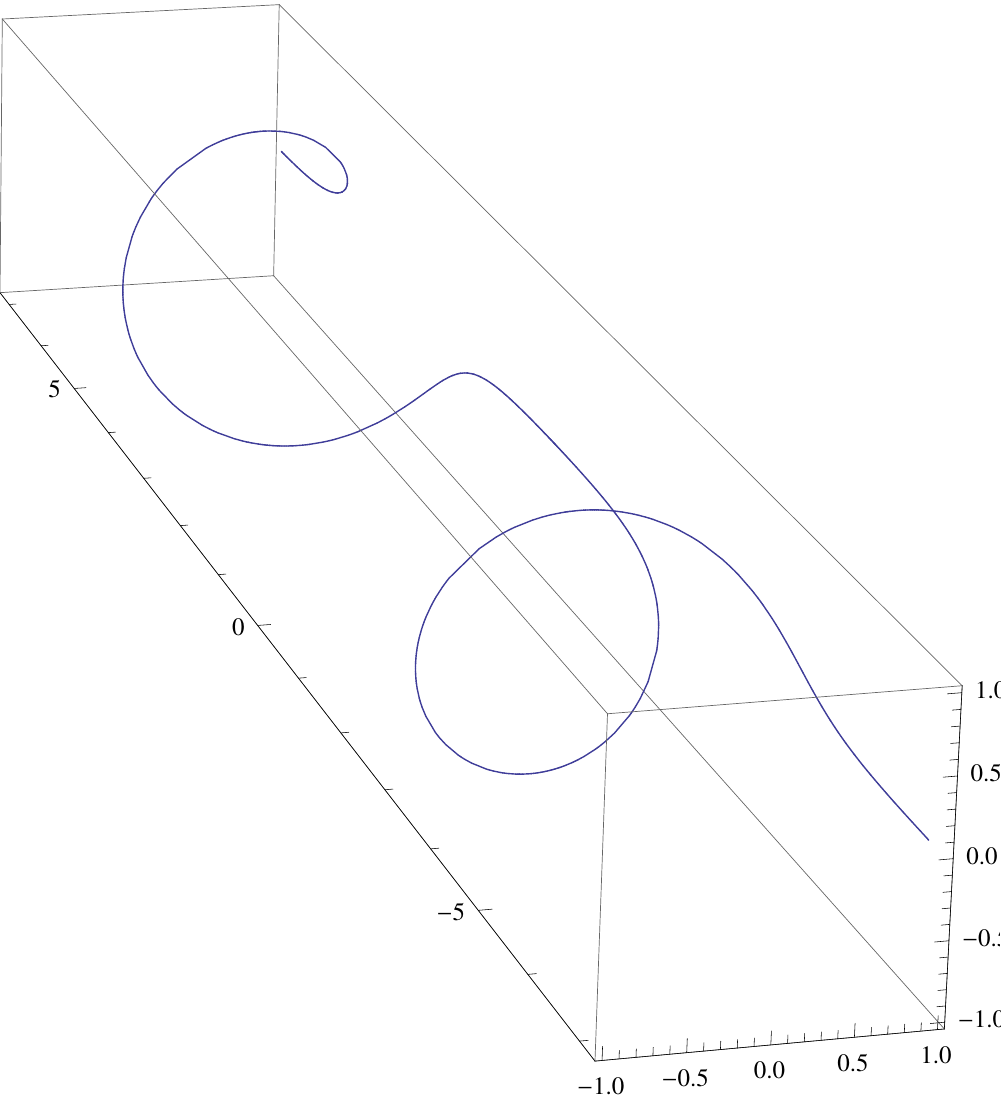}
\caption{Real and imaginary parts of $v_1$ (cf.\ formula~(\ref{solugamma0})) as a function of $x$. Here
$\kappa=1$, $\omega=10^{-3}$, $\alpha =1$, $\beta=1+i$, $\gamma=0$.}
\label{fig:gamma03d}
\end{figure}

\begin{figure}[t]
\centering
\includegraphics[width=0.4\textwidth]{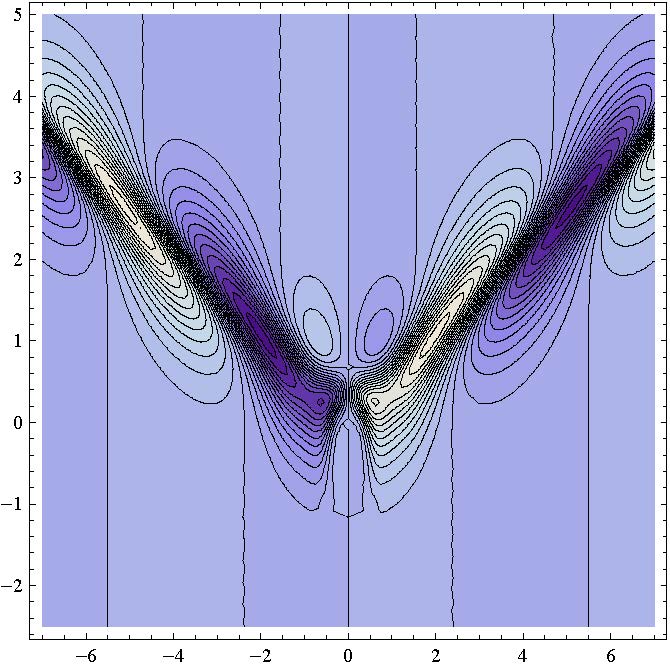} \qquad \includegraphics[width=0.4\textwidth]{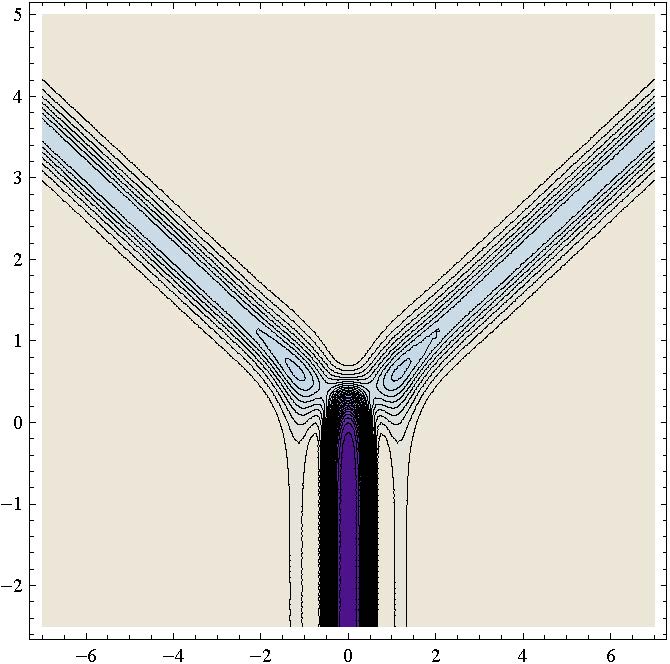}
\caption{Contour plot of ${\rm Re}\, v_1(x,t)$ (left panel) and ${\rm Re}\,
u_1(x,t)$ (right panel)for a generic soliton solution (see formula~\ref{solbeta0}) as a function of~$x$ and $t$ where $\alpha=\beta=\gamma=\kappa=\omega=1$.}
\label{fig:sol-1}
\end{figure}

The function $v_1$ can be written in the form $v_1=\exp(4i\theta(x))$ where
\[  \theta(x)={\rm Arctan} \left(\frac{\kappa\cos(2 \omega
x+\phi_\alpha-\phi_\beta)}{\omega\cosh(2\kappa x+\ln|\alpha/\beta|)}\right),
\qquad\phi_\alpha={\rm arg}\,\alpha, \qquad \phi_\beta={\rm arg}\,\beta.\]
It is easy to check that $u=0$, $v=\exp(if(x))$ is an exact solution of
(\ref{nee}) for any dif\/ferentiable function~$f(x)$. It resembles the case of
the three-wave equation~\cite{zman-3w} where one wave of an arbitrary shape is
an exact solution of the system and  the two other waves are identically zero.
A plot of the solution~(\ref{solugamma0}) is shown on Fig.~\ref{fig:gamma03d}.
The solution~(\ref{solugamma0}) has a simple spectral characterization and an
explicitly given analytic fundamental solution of the corresponding linear problem.

\subsubsection*{Case (ii): $\alpha\ne 0$, $\beta= 0$, $\gamma\ne 0$}

In this case, from (\ref{u1}) it follows that the solution can be written
in the form
\begin{gather}\label{solbeta0}
u_1=\frac{4i\omega\kappa Q^*\exp\{i\omega x+it(\kappa^2-\omega^2)
+i(\phi_\alpha - \phi_\gamma)\}}{(\omega-i\kappa)Q^2},\qquad
v_1=1-\frac{8\omega\kappa^2}{(\omega-i\kappa)Q^2},
\end{gather}
where
\[ Q=2\omega e^{\kappa(x-2\omega t)+\ln |\alpha/\gamma|}
+(\omega+i\kappa)e^{-\kappa(x-2\omega t)-\ln |\alpha/\gamma|}.
\]

\subsubsection*{Case (iii): $\alpha= 0,\beta\ne 0,\gamma\ne 0$}
The solution now can be obtained from the solution in the case (ii), by
changing $\alpha\to\beta$ and $x\to -x$.

In the cases (ii) the solution (\ref{solbeta0}) is a soliton of width
$1/\kappa$ moving with velocity $2\omega$. The corresponding soliton in the
case (iii) moves with a velocity $-2\omega$.

In the generic case, when all three constants are non-zero, the solution
(\ref{u1}) represents a~nonlinear interaction of the above described solitons.
For $\kappa>0$ it is a decay of unstable time independent soliton from the case
(i)~into two solitons, corresponding to the cases~(ii) and~(iii) (see Fig.~\ref{fig:sol-1}). For $\kappa<0$,
the solution is a fusion of two colliding solitons into a stationary one.

Let us now consider the case of imaginary poles, i.e.\ $\mu = i\kappa$. There exist two
essentially dif\/ferent cases.

\begin{figure}[t]
\centering
\includegraphics[width=0.76\textwidth]{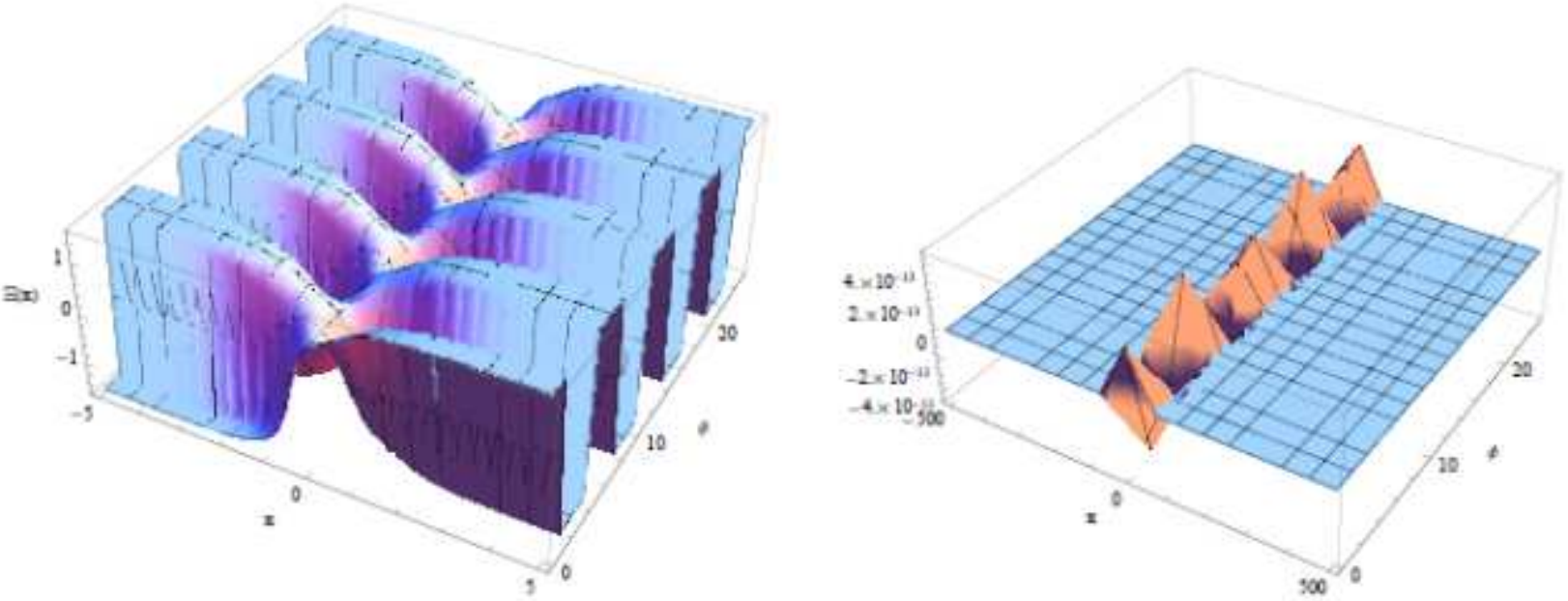}
\caption{Plots of the phase $\Xi(x)$ (left panel) and ${\rm Im}\, v(x)$  (right panel) for a stationary doublet soliton solution (\ref{eq:kink} ) as a function of
$x$ and the phase $\phi$ where $\kappa=\sigma_0=1$.}
\label{fig:sol-3}
\end{figure}

1.\ Suppose $p_{0,2} = \gamma = 0$. Then from (\ref{algsys_m2b}) it follows that $|p_{0,1}| = |p_{0,3}|$. It suf\/f\/ices to pick up $p_{0,1} = 1$
and for the third component we have $p_{0,3} = \exp(i\varphi)$, $\varphi\in\bbbr$.
In this case the 1-soliton solution is stationary:
\begin{gather*}
u(x) = 0,\qquad v(x) = \left(\frac{\cosh 2\kappa x + \sinh 2\kappa x \cos\varphi  + i(\sigma_0 - 2\kappa x \sin\varphi)}{\cosh 2\kappa x + \sinh 2\kappa x \cos\varphi  - i(\sigma_0 - 2\kappa x \sin\varphi)}\right)^2.
\end{gather*}
The function $v$ can be presented as
\begin{gather}
v(x) = \exp(4i\,\Xi(x) ),\qquad  \Xi(x)= \arctan\left(\frac{\sigma_0 - 2\kappa x\sin\varphi}{\cosh 2\kappa x + \sinh 2\kappa x \cos\varphi}\right).  \label{eq:kink}
\end{gather}
A plot of the solution \eqref{eq:kink}
is shown on Fig.~\ref{fig:sol-3}.
In particular, when $\varphi = 0; \pi$, i.e.\ $p_{0,1} =\pm p_{0,3}$ we have for the doublet solution:
\begin{gather*}
u(x)=0,\qquad v(x) = \left(\frac{e^{\pm 2\kappa x} + i\sigma_0}
{e^{\pm 2\kappa x} - i\sigma_0}\right)^2.
\end{gather*}
It is evident that the presence of the constant $\sigma_0$ here is essential since
otherwise the solution coincides with the vacuum.

\begin{figure}[t]
\centering
\includegraphics[width=0.8\textwidth]{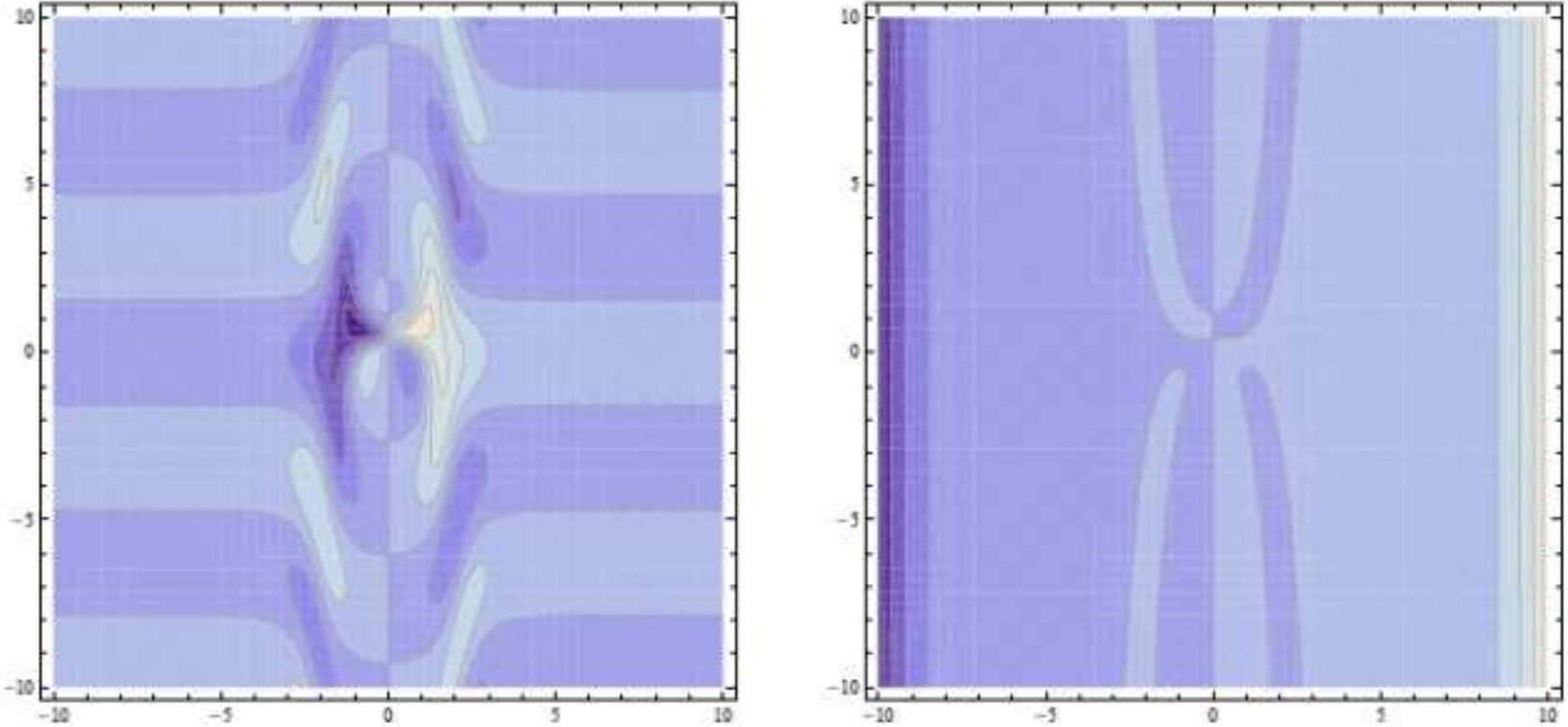}
\caption{Contour plot of ${\rm Re}\, u(x,t)$ (left panel) and ${\rm Re}\,
v(x,t)$ (right panel) for doublet soliton solution of the type~(\ref{eq:kink2}) as a function of~$x$ and~$t$. Here $\kappa=\sigma_0=1$, $\theta_0=0$ and $\tilde{\varphi} = 0$.}
\label{fig:sol-4}
\end{figure}

2.\ Now let us assume $p_{0,2} \neq 0$. For simplicity we f\/ix $p_{0,2} = 1$. Then the norms of~$p_{0,1}$ and~$p_{0,3}$ are interrelated through
\[|p_{0,1}|^2 - |p_{0,3}|^2 = 1. \]
This is why it proves to be convenient to parametrize them as follows:
\begin{gather*}
p_{0,1} = \cosh\theta_0 e^{i(\varphi_0 + \tilde{\varphi})},
\qquad p_{0,3} = |\sinh\theta_0| e^{i(\varphi_0 - \tilde{\varphi})},
\qquad \theta_0, \varphi_0, \tilde{\varphi}\in\bbbr.
\end{gather*}
The 1-soliton solution reads:
\begin{gather*}
u(x,t)  = \frac{2\Delta^*}{\Delta^2}e^{i\left(\kappa^2t + \varphi_0\right)} \left[\sinh(\kappa x + \theta_0)\cos\tilde{\varphi}
+ i\sinh(\kappa x - \theta_0)\sin\tilde{\varphi}\right],\nonumber\\
v(x,t)  = \frac{\Delta^*\left(\Delta^* - 2\right)}{\Delta^2} = 1 + \frac{2(2i\sigma -1)}{\Delta} + \frac{4i\sigma(i\sigma -1)}{\Delta^2},
\end{gather*}
where
\begin{gather*}
\Delta (x,t) =  \cosh^2(\kappa x + \theta_0)\cos^2\tilde{\varphi}
+ \cosh^2(\kappa x - \theta_0)\sin^2\tilde{\varphi} - i(\sigma_0 - 2\kappa^2 t
+ \kappa x\sinh 2 \theta_0\sin2\tilde{\varphi}),\\
\sigma(x,t)  =  \sigma_0 -2\kappa^2 t + \kappa x\sinh 2\theta_0 \sin2\tilde{\varphi}.
\end{gather*}
The above solution can be signif\/icantly simplif\/ied if one assumes that $p_{0,3}/p_{0,1}\in\bbbr$. In these boundary cases the doublet solution reads:
\begin{gather}
u(x,t)  = \frac{2\left(\cosh^2(\kappa x + \theta_0) + i(\sigma_0 - 2\kappa^2 t)\right)}{\left(\cosh^2(\kappa x + \theta_0) - i(\sigma_0 - 2\kappa^2 t)\right)^2}\,e^{i\left(\kappa^2t + \varphi_0\right)}\sinh(\kappa x + \theta_0),\nonumber\\
v(x,t)  = \frac{\left(\cosh^2(\kappa x + \theta_0) + i(\sigma_0 - 2\kappa^2 t)\right)\left(\cosh^2(\kappa x + \theta_0) + i\sigma_0 - 2 - 2i\kappa^2 t\right)}{\left(\cosh^2(\kappa x + \theta_0) - i(\sigma_0 - 2\kappa^2 t)\right)^2}\label{eq:kink2}
\end{gather}
if $p_{0,3}/p_{0,1}>0$ ($\tilde{\varphi} = 0$) and
\begin{gather*}
u(x,t)  = \frac{2\left(\cosh^2(\kappa x - \theta_0) + i(\sigma_0 - 2\kappa^2 t)\right)}{\left(\cosh^2(\kappa x - \theta_0) - i(\sigma_0 - 2\kappa^2 t)\right)^2}\,e^{i\left(\kappa^2t + \varphi_0 + \pi/2\right)}\sinh(\kappa x - \theta_0),\nonumber\\
v(x,t)  = \frac{\left(\cosh^2(\kappa x - \theta_0) + i(\sigma_0 - 2\kappa^2 t)\right)\left(\cosh^2(\kappa x - \theta_0) + i\sigma_0 - 2 - 2i\kappa^2 t\right)}{\left(\cosh^2(\kappa x - \theta_0) - i(\sigma_0 - 2\kappa^2 t)\right)^2},%\label{eq:kink2a}
\end{gather*}
when $p_{0,3}/p_{0,1}<0$ ($\tilde{\varphi} =\pi/2$). A~plot of the solution \eqref{eq:kink2} is depicted on Fig.~\ref{fig:sol-4}.

\subsection{Multisoliton solutions}

As we mentioned above, the dressing procedure can be applied several times consequently.
Thus after dressing the $1$-soliton solution one derives a $2$-soliton
solution, after dressing the $2$-soliton solution one obtains a $3$-soliton
solution and so on.  Of course, in doing this one is allowed to apply
either of dressing factors (\ref{eq:dress-anz1}) and (\ref{eq:dress-anz1a}).
Therefore the multisoliton solution will be a~certain combination of quadruplet
and doublet solitons.

Another way of derivation the multisoliton solution consists in using a dressing
factor with multiple poles. For example, if one wants to generate $N$ quadruplet
solitons one should use a~dressing factor in the form:
\begin{gather*}
\Phi(x,t,\lambda)=\openone+\sum^{N}_{k=1}\lambda\left(\frac{M_k(x,t)}{\lambda-\mu_k}
+\frac{\mathbf{C}M_k(x,t)\mathbf{C}}{\lambda+\mu_k} \right),
%\label{phi_gen}
\end{gather*}
which is evidently compatible with the reduction condition (\ref{eq:fas-invol}).
Due to (\ref{eq:her-invol}) its inverse has the form:
\begin{gather*}
\Phi^{-1}(x,t,\lambda)=\openone + \sum^N_{k=1}\lambda\left(\frac{M^{\dag}_k(x,t)}{\lambda-\mu_k^*}+\frac{\mathbf{C}M^{\dag}_k(x,t)\mathbf{C}}{\lambda+\mu_k^*}\right).
%\label{phi_inv_gen}
\end{gather*}
In order to determine the residues of $\Phi$ one follows the same steps as in the
case of a $2$-poles dressing factor. Firstly, the identity $\Phi\Phi^{-1}=\openone$
implies that the residues of $\Phi$ and $\Phi^{-1}$ satisfy certain algebraic
relations, namely:
\begin{gather}
M_k\left[\openone + \sum^N_{l=1}\mu_k\left(\frac{M^{\dag}_l}{\mu_k-\mu_l^*}+\frac{\mathbf{C}M_l^{\dag}\mathbf{C}}
{\mu_k+\mu_l^*}\right)\right]=0,\qquad k=1,\ldots, N,
\label{algsys_n}
\end{gather}
to ensure vanishing of the residue of $\Phi\Phi^{-1}$ at $\lambda = \mu_k$.
Of course due to the  $\bbbz_2$ reductions we will have an additional set of algebraic
relations which are obtained from (\ref{algsys_n}) by hermitian conjugation.

As discussed before in order to obtain a nontrivial dressing the residues must be
degenerate matrices. Thus one introduces the factorization $M_k=|\mathbf{n}_k\rangle\langle \mathbf{m}_k|$ and then reduces (\ref{algsys_n}) to a~linear system for $|\mathbf{n}_k\rangle$
\begin{gather*}
|\mathbf{m}_k\rangle = \sum^N_{l=1}\mathcal{B}_{lk}|\mathbf{n}_{\,l}\rangle,\qquad
\mathcal{B}_{lk} := \mu^*_k\left(\frac{\langle \mathbf{m}_{\,l}|\mathbf{m}_{k}\rangle}{\mu_l-\mu_k^*}
- \frac{\langle \mathbf{m}_{\,l}|\mathbf{C}|\mathbf{m}_{k}\rangle}{\mu_l+\mu_k^*}\mathbf{C}\right) .
%\label{matr_sys_gen}
\end{gather*}
By solving it one can express the vectors $|\mathbf{n}_{l}\rangle$, $l=1,\ldots,N$ through all $|\mathbf{m}_{k}\rangle$ and that way determine the whole dressing factor in terms of $|\mathbf{m}_{k}\rangle$. This step contains the main technical dif\/f\/iculty in the whole
scheme.

The vectors $|\mathbf{m}_{k}\rangle$ can be found from the natural requirement
of vanishing of the poles in the dif\/ferential equations~(\ref{eq:dress-eqL1}). The
result reads
\begin{gather*}
|\mathbf{m}_{k}(x,t)\rangle = \psi_0(x,t,\mu^*_k)|\mathbf{m}_{k,0}\rangle.
%\label{Fl_res}
\end{gather*}
Thus we have proved that as in the $2$-poles case the dressing factor is determined if
one knows the seed solution $\psi_{0}(x,t,\lambda)$. The quadruplet $N$-soliton solution itself
can be derived through the following formula
\begin{gather*}
L_1^{(1)} = \left(\openone + \sum^N_{k=1} (M_k + \mathbf{C}M_k\mathbf{C})\right) L_{1}^{(0)}
\left(\openone + \sum^N_{k=1} (M_k + \mathbf{C}M_k\mathbf{C})\right)^{\dag}.
\end{gather*}
From all said above it follows that the algorithm for obtaining the $N$-soliton solution
can be presented symbolically as follows
\begin{gather*}
L_{1}^{(0)}\to (|\mathbf{m}_1\rangle,\ldots,|\mathbf{m}_N\rangle) \to (|\mathbf{n}_1\rangle,\ldots,|\mathbf{n}_N\rangle)
\to (M_1,\ldots, M_N) \to L_1^{(1)}.
\end{gather*}

Similarly, one is able to generate a sequence of $N$ doublet
solitons by making use of the following factor
\begin{gather*}
\Phi(x,t,\lambda)=\openone+\sum^{N}_{l=1}\lambda\left(\frac{P_l(x,t)}{\lambda-i\kappa_l}+\frac{\mathbf{C}P_l(x,t)\mathbf{C}}{\lambda + i\kappa_l}   \right).
%\label{phi_gen_d}
\end{gather*}
Following the same steps as in the $1$-soliton case one can convince
himself that the vectors $|\mathbf{q}_{\,l}\rangle$ involved in the
decomposition $P_l = |\mathbf{q}_l\rangle\langle \mathbf{p}_l|$ satisfy
the linear system:
\begin{gather*}
\mathbf{C}|\mathbf{p}_l\rangle = \sum^{N}_{s=1}\mathcal{D}_{sl}|\mathbf{q}_{s}\rangle, %\label{matr_sys_gen2}
\end{gather*}
where
\begin{gather*}
\mathcal{D}_{ss} :=i\sigma_{s} - \frac{\langle \mathbf{p}_{s}|\mathbf{p}_{s}\rangle}{2}\mathbf{C},\qquad
\mathcal{D}_{sl} := \kappa_l\left(\frac{\langle \mathbf{p}_{s}|\mathbf{C}|\mathbf{p}_{l}\rangle}{\kappa_s-\kappa_l}
- \frac{\langle \mathbf{p}_{\,s}|\mathbf{p}_{l}\rangle}{\kappa_s+\kappa_l}\mathbf{C}\right),\qquad s\neq l.
\end{gather*}
On the other hand the vectors $|\mathbf{p}_{l}\rangle$ and the real functions $\sigma_l$
are expressed in terms of the initial solution $\psi_0$ and its derivatives with respect
to $\lambda$ as follows:
\begin{gather*}
|\mathbf{p}_{l}(x,t)\rangle  =  \psi_0(x,t,-i\kappa_l)|\mathbf{p}_{l,0}\rangle,\\ %\label{Gl_res}\\
\sigma_l(x,t)  =  -\kappa_l\langle \mathbf{p}_{l,0}|\psi^{-1}_0(x,t,i\kappa_l)\dot{\psi}_0(x,t,i\kappa_l)\mathbf{C}|\mathbf{p}_{l,0}\rangle + \sigma_{l,0},
\end{gather*}
where $|\mathbf{p}_{l,0}\rangle$ and $\sigma_{l,0}$ for $l=1,\ldots, N$ are constants
of integration to parametrize the solution. The components of $|\mathbf{p}_l\rangle$
are not independent but satisfy
\[\langle \mathbf{p}_l|\mathbf{C}|\mathbf{p}_l\rangle = 0\quad\Rightarrow
\quad \langle \mathbf{p}_{l,0}|\mathbf{C}|\mathbf{p}_{l,0}\rangle = 0.\]
In order to  derive doublet $N$-soliton solution one uses:
\begin{gather*}
L_1^{(1)} = \left(\openone + \sum^{N}_{l=1} (P_l + \mathbf{C}P_l\mathbf{C})\right) L_{1}^{(0)} \left(\openone + \sum^{N}_{l=1} (P_l + \mathbf{C}P_l\mathbf{C})\right)^{\dag}.
\end{gather*}
In this case the diagram which describes the algorithm to obtain the multisoliton
solution reads
\begin{gather*}
L_{1}^{(0)}\to \{|\mathbf{p}_{\,l}\rangle, \sigma_l\}^{N}_{l=1}\to \{|\mathbf{q}_{\,l}\rangle\}^{N}_{l=1} \to \{P_l\}^{N}_{l=1}
\to L_1^{(1)}.
\end{gather*}

It is clear that apart of the `pure' multisolitons discussed above there is
a `mixed' type of multisolitons, i.e.\ multisolitons to contain both species
of $1$-soliton solutions. In order to obtain such solution one needs to
use a dressing factor possessing both types of poles.

\section{Generalized Fourier transform}\label{section4}

In this section we aim to develop the generalized Fourier
transform interpretation of the inverse scattering method
for the Lax operator (\ref{lax_1}). For this to be done we
are going to use some basic notions like Wronskian relations,
`squared solutions', recursion operators etc. The `squared
solutions' also known as adjoint solutions are generalizations
of exponential functions in usual Fourier analysis, namely they
are a complete system. They occur naturally in Wronskian relations
which interrelate the scattering data and the potential $L_1(x)$
(as well as their variations) in a form which resembles a scalar
product and is called skew-scalar product. The Wronskian relations
allows one to express the expansion coef\/f\/icients of $L_1(x)$ and
its variation along these generalized exponents in terms of the
scattering data and its variations as we shall see later on. Within
this framework the role of the operator $id\slash dx$ whose
eigenfunctions are $\exp(i kx)$ is played by the recursion operators.
We shall present two substantially dif\/ferent ways of obtaining the
recursion operators and compare them.

\subsection{Recursion operators and integrable hierarchy}

Recursion operators appear in many situations in the theory
of integrable systems \cite{AKNS,GVY*08,olver}. Their existence is tightly related to
the fact that each integrable NLEE belongs to a whole family of
inf\/inite number of integrable NLEEs associated to the same
Lax operator $L$. One can also assign to this family a whole
inf\/inite set of symmetries and integrals of motion (Hamiltonians).
In this subsection we are to derive the recursion operators by
analyzing the interrelations between members of the hierarchy of NLEEs.
In doing this we are following the ideas in \cite{GKS,wang}.

The hierarchy of integrable NLEEs associated to $L$ is generated through
the zero curvature representation which we rewrite in the original Lax
form
\begin{gather*}
i L_t = [L,V].
%\label{lax_orig}
\end{gather*}
Each choice of $V$ corresponds to an individual member of the hierarchy.
Following the ideas in~\cite{GKS} we interrelate two adjacent f\/lows~$V$ and
$\tilde{V}$ through the equality
\begin{gather}
\tilde{V}(\lambda) = k(\lambda)V(\lambda) + B(\lambda),
\label{vtildev}
\end{gather}
where $k(\lambda)$ is a scalar function and the remainder $B$ has
the structure of the Lax operator $A(\lambda)$
\begin{gather*}
B(\lambda) = \lambda B_1 + \lambda^2 B_2.
%\label{remainder}
\end{gather*}
All quantities above must be invariant with respect to the action of
reductions (\ref{red1}), (\ref{red2}). This is why we pick up
$k(\lambda) = \lambda^2$ while the operator $B$ obey the condition
\begin{gather*}
B^{\dag}(\lambda^*) = B(\lambda),\qquad
\mathbf{C}B(-\lambda)\mathbf{C} = B(\lambda).
\end{gather*}
This means that the coef\/f\/icients of $B$ are hermitian matrices
to fulf\/ill $B_1\in\mathfrak{sl}^{1}(3)$, $B_2\in\mathfrak{sl}^{0}(3)$.

After substituting (\ref{vtildev}) into the Lax representation
one obtains the following relation
\begin{gather}
i L_{\tau}= i \lambda^2 L_t + [L,B],
\label{recurop_rel}\end{gather}
where $\tau$ is the evolution parameter of the f\/low
$\tilde{V}$. Then the recursion operator $\mathcal{R}$
is def\/ined as the mapping
\begin{gather*}
L_{1,\tau} = \mathcal{R} L_{1,t}.
%\label{recurop_def}
\end{gather*}
In order to f\/ind it we need to solve (\ref{recurop_rel}) and
thus calculate the remainder $B$.

Since relation (\ref{recurop_rel}) holds identically with respect to
$\lambda$ it splits into the following set of recurrence relations:
\begin{gather}
i L_{1,t} + [L_1,B_2]  =  0,\label{recurel_1}\\
i B_{2,x} + [L_1,B_1]  =  0,\label{recurel_2}\\
L_{1,\tau} - B_{1,x}   =  0.\label{recurel_3}
\end{gather}
In accordance with our conventions explained in the Preliminaries
(see formulae (\ref{X_split})--(\ref{X1_split})) the coef\/f\/icients~$B_1$
and $B_2$ have splittings
\begin{gather}
B_1  =  B^{\bot}_1 + b_1L_1, \qquad\langle B^{\bot}_1,L_1\rangle = 0,\nonumber
%\label{b1}
\\
B_2  =  B^{\bot}_2 + b_2L_2,\qquad \langle B^{\bot}_2,L_2\rangle
= 0, \label{b2}
\end{gather}
of terms~$B^{\bot}_{1}$ and~$B^{\bot}_{2}$ not commuting with~$L_1$
and their orthogonal complements. After substituting~(\ref{b2}) into~(\ref{recurel_1}) it becomes possible to express~$B^{\bot}_2$ through
$L_{1,t}$ as follows:
\begin{gather}
B^{\bot}_2=-i \,\ad^{-1}_{L_1} L_{1,t},
\label{b2bot}\end{gather}
where $\ad^{-1}_{L_1}$ acts on the orthogonal parts only, i.e.\
the kernel of $\ad_{L_1}$ is factorized. In this case it can be
proven the presentation below holds true
\[
\ad^{-1}_{L_1} = \frac{1}{4}\left(5\ad_{L_1} - \ad^3_{L_1}\right).
\]

The coef\/f\/icient $b_2$ is determined from relation~(\ref{recurel_2}). Indeed,
after plugging~(\ref{b2}) into~(\ref{recurel_2}) we get
\begin{gather}
i \partial_x B^{\bot}_2 + i b_{2,x}L_2 + i b_2 L_{2,x}
+ [L_1, B_1] = 0.\label{b1_0}
\end{gather}
The $L_1$-commuting part of (\ref{b1_0}) is separated by
taking $\langle\,\cdot\,,L_2\rangle$ in both hand sides of the equation
to give
\begin{gather}
\frac{2}{3}b_{2,x} + \langle \partial_x B^{\bot}_2, L_2 \rangle =0
\quad\Rightarrow\quad b_2 = -\frac{3}{2}\partial^{-1}_x
\langle \partial_x B^{\bot}_2, L_2 \rangle,
\label{b2_1}
\end{gather}
where we have denoted by $\partial^{-1}_x$ the integral $\int^x_{\pm\infty}d y$.
On the other hand by projecting (\ref{recurel_2}) with $\pi=\ad^{-1}_{L_1}\ad_{L_1}$
one extracts its orthogonal part
\begin{gather*}
i \pi\partial_x B^{\bot}_2 + i b_{2}L_{2,x}
+ [L_1, B^{\bot}_1] = 0.
\end{gather*}
Taking into account (\ref{b2_1}) one can f\/ind out
the following recurrence relation
\begin{gather}
B^{\bot}_1 = \Lambda_2 B^{\bot}_{2},\qquad
\Lambda_2 = -i \ad^{-1}_{L_1}\left(
\pi\partial_x(\cdot) -\frac{3}{2}L_{2,x}\partial^{-1}_x
\langle \partial_x (\cdot), L_2 \rangle\right).
\label{b1bot}\end{gather}
Formally speaking, one should write $\Lambda^{\pm}_2$ instead of
$\Lambda_2$ for the integro-dif\/ferential operator introduced above
since the lower integration limit in $\partial^{-1}_x$ dif\/fers.
This choice of signs, however, is unessential for the considerations
in the current subsection and this is why prefer a more simplif\/ied
notation.

After a similar treatment of (\ref{recurel_3}) we obtain
\begin{gather}
2b_{1,x}  +  \langle \partial_x B^{\bot}_1, L_1 \rangle =0
\quad\Rightarrow\quad b_1 = -\frac{1}{2}\partial^{-1}_x
\langle \partial_x B^{\bot}_1, L_1 \rangle,\nonumber%\label{b1_1}
\\
L_{1,\tau} = i \ad_{L_1} \Lambda_1 B^{\bot}_1,\qquad
\Lambda_1 = -i \ad^{-1}_{L_1} \left(\pi\partial_x(\cdot)
-L_{1,x}\frac{1}{2}\partial^{-1}_x\langle\partial_x(\cdot),
L_1\rangle\right).\label{l1t_tau}
\end{gather}
Finally combining (\ref{b2bot}), (\ref{b1bot}) and (\ref{l1t_tau})
one derives
\begin{gather}
L_{1,\tau} = \ad_{L_1} \Lambda_1\Lambda_2\ad^{-1}_{L_1}L_{1,t}\quad
\Rightarrow\quad\mathcal{R} = \ad_{L_1} \Lambda_1\Lambda_2\ad^{-1}_{L_1}.
\label{recurs_op_f}
\end{gather}
The recursion operator can be also presented conveniently as a $4\times 4$ matrix
to map the column vector $(u,v,u^*,v^*)^T_t$ to the column vector $(u,v,u^*,v^*)^T_{\tau}$.
Then it can be verif\/ied that $\mathcal{R}$ obeys the splitting:
\[ \mathcal{R} = \pi\mathcal{A} (\mathcal{B}_{\rm{loc}}+\mathcal{B}_{\rm{nonl}})
+\mathcal{K}.
 \]
The factors $\pi$ and $\mathcal{A}$ look as follows
\begin{gather*}
\pi = \left(\begin{array}{cccc} 1-|u|^2/2  & -uv^*/2 & -u^2/2 & -uv/2 \\
-u^*v/2 & 1-|v|^2/2 & -uv/2 &  -v^2/2  \\
-(u^*)^2/2 & -u^*v^*/2 & 1-|u|^2/2 &  -u^*v/2 \\
-u^*v^*/2 & -(v^*)^2/2 & -uv^*/2 &  1-|v|^2/2 \end{array}\right),\\
\mathcal{A}:= i \partial_x
\left(\begin{array}{cccc}
0   & -v  & -u   & 0 \\
-u  & 0   & v    & 0 \\
v^* & 0   & u^*  & 0 \\
0   & u^* & -v^* & 0
\end{array}\right)\partial_x.
\end{gather*}
The matrices
\[\mathcal{B}_{\rm loc} = \frac{i }{4} \left(\begin{array}{cccc}
0  & -4u^* & 2v(3|v|^2-1) & -6u^*v^2 \\
2v^*(1-3|v|^2) & 6u(v^*)^2 & 0 &  4u  \\
-4u^* & 0 & 3u(|v|^2-|u|^2) &  -v(1+6|u|^2) \\
- u^* & - v^* & 0 &  0 \end{array}\right) %\label{local_part}
\]
and
\[\mathcal{B}_{\rm nonl} = - \frac{3i }{4}
\begin{pmatrix}
2u^*v \\ 2uv^* \\ |u|^2-|v|^2 \\ -1/3 \end{pmatrix}
\partial^{-1}_x\left[(u^*_x,v^*_x,-u_x,-v_x) \right]
\]
are a local and a nonlocal part respectively originating from the
orthogonal part $B^{\bot}_{2}$ and the $L_1$-commuting part $b_2L_2$.
Finally
\[
\mathcal{K} = -\frac{1}{2}\begin{pmatrix}
u\\ v\\ u^*\\ v^* \end{pmatrix}_x \partial^{-1}_x
\left[(u^*,v^*,u,v)\mathcal{A} (\mathcal{B}_{\rm{loc}}
+\mathcal{B}_{\rm{nonl}}) \right]
\]
is connected with the term $b_1L_1$ in the splitting of $B_1$.

\subsection{Wronskian relations and `squared solutions'}

Wronskian relations \cite{CaDe*76_1,CaDe*76_2} provide an important
tool for analyzing the relevant class of NLEE and the
mapping $\mathfrak{F}: \mathfrak{M}\to \mathfrak{T} $, where
$\mathfrak{M} $ is the set of allowed potentials of~$L$
(in our case~$L_1$) and~$\mathfrak{T} $ is the minimal
set of scattering data.

In deriving them we will need along with the f\/irst equation
in (\ref{lax_pol}) also two other related equations:
\begin{gather*}%\label{eq:Linv}
i \frac{\partial \hat{\chi} }{\partial x}
- \lambda \hat{\chi} (x,\lambda)  L_1(x)=0, \qquad
\hat{\chi} (x,\lambda) \equiv \chi^{-1} (x,\lambda),\\
i \frac{\partial \delta\chi }{\partial x}
+ \lambda L_1(x)\delta\chi (x,\lambda)
+\lambda\delta L_1(x)\chi (x,\lambda)=0,
%\label{eq:dL1}
\end{gather*}
where the variation of $\chi(x,\lambda)$ is due to the
variation $\delta L_1(x)$.

The f\/irst type of Wronskian relations interrelates the asymptotics of FAS with $L_1$and its powers
as shown in the examples below
\begin{gather}
  i\left(  \hat\chi J \chi(x,\lambda)
- J\right)\big|_{-\infty}^{\infty}
 =  \lambda \int_{-\infty}^{\infty}  d x\,
\hat{\chi} [ L_1 ,J]  \chi(x,\lambda),\label{eq:Wr2}\\
  \hat{\chi} L_1(x) \chi(x,\lambda)
\big|_{-\infty}^{\infty}  =  \int_{-\infty}^{\infty}  d x\,
\hat{\chi} L_{1,x} \chi(x,\lambda), \label{eq:Wr5}\\
  \hat{\chi} L_2(x) \chi(x,\lambda)
\big|_{-\infty}^{\infty}  =  \int_{-\infty}^{\infty}  d x\,
\hat{\chi} L_{2,x} \chi(x,\lambda).\label{eq:Wr5a}
\end{gather}
We remind the reader that $J=\diag(1,0,-1)$ and $I=\diag(1/3,-2/3,1/3)$
are the diagonal forms of (the asymptotics of) $L_1$ and $L_2 = L^2_1
- 2\openone/3$ respectively, that we introduced at the very beginning in
section Preliminaries (see formulae (\ref{lax_1_g}), (\ref{lax_2_g})).

A second class of Wronskian relations connects the variation
$\delta \chi(x,\lambda)$ with the variation of~$L_1$, i.e.
\begin{gather*}%\label{eq:Wrd}
  \hat{\chi} \delta \chi (x,\lambda)
\big|_{-\infty}^{\infty} =
i \lambda \int_{-\infty}^{\infty}  d x\,
\hat{\chi} \delta L_1  \chi(x,\lambda) .
\end{gather*}
In the left hand sides of the Wronskian relations are
involved the scattering data and its variation while the
right hand sides can be viewed as Fourier type integrals.
To make this obvious let us take the Killing form of the
Wronskian relation (\ref{eq:Wr2}) with a Cartan--Weyl generator
$E_\alpha$ and use the invariance of the Killing form
\begin{gather*}%\label{eq:Wr2'}
 i \left\langle  \hat{\chi} J \chi (x,\lambda)
- J, E_\alpha\right \rangle\big|_{-\infty}^{\infty}
=  \lambda \int_{-\infty}^{\infty}  d x\,
\left\langle [L_1,J], e_\alpha (x,\lambda)\right\rangle.
\end{gather*}
The quantities
\begin{gather*}
e_\alpha(x,\lambda)= \chi E_\alpha \hat{\chi}(x,\lambda)
%\label{sq_sol}
\end{gather*}
introduced above are called `squared solutions'. Due to the fact
that we have two FAS $\chi^{+}(x,\lambda)$ and $\chi^{-}(x,\lambda)$
we obtain two types of 'squared solutions' $e^{\pm}_\alpha(x,\lambda)$.
Similarly, taking the Killing form in (\ref{eq:Wr5}) and
(\ref{eq:Wr5a}) we f\/ind
\begin{gather}
  \left\langle \hat{\chi} L_1\chi (x,\lambda),
E_\alpha \right\rangle \big|_{-\infty}^{\infty}
= \int_{-\infty}^{\infty}  d x\, \left\langle
 L_{1,x} , e_\alpha (x,\lambda)  \right\rangle, \label{eq:Wr5'}\\
  \left\langle \hat{\chi} L_2\chi (x,\lambda),
E_\alpha \right\rangle \big|_{-\infty}^{\infty}
= \int_{-\infty}^{\infty}  d x\,
\left\langle L_{2,x}, e_\alpha(x,\lambda)\right\rangle,\label{eq:Wr5a'}\\
  \left\langle \hat{\chi} \delta \chi (x,\lambda),E_\alpha
\right\rangle \big|_{-\infty}^{\infty}
= i \lambda  \int_{-\infty}^{\infty}  d x\,
\left\langle  \delta L_1 , e_\alpha (x,\lambda)\right\rangle.\label{eq:Wrd2}
\end{gather}
We are interested more specif\/ically in variations that are
due to the time evolution of $L_1(x)$, i.e.
\[\delta L_1(x) = L_1(x,t+\delta t) - L_1(x,t)
\simeq \delta t \frac{\partial L_1 }{\partial t}.\]
Therefore up to f\/irst order terms of $\delta t$ we obtain
\begin{gather*}%\label{eq:Wrd2t}
 \left\langle \hat{\chi} \chi_t (x,\lambda),E_\alpha
\right\rangle \big|_{x=-\infty}^{\infty}
= i \lambda  \int_{-\infty}^{\infty} d x\,
\left\langle L_{1,t}, e_\alpha(x,\lambda)\right\rangle.
\end{gather*}
Now we can explain why the Wronskian relations are important
for analyzing the mapping $\mathfrak{F}: \mathfrak{M}_{L_1}\to
\mathfrak{T} $. Indeed, taking $\chi(x,\lambda)$ to be a
fundamental analytic solution of $L$ we can express the
left hand sides of (\ref{eq:Wr5'}) and (\ref{eq:Wr5a'})
(resp.~(\ref{eq:Wrd2})) through the Gauss factors $S^\pm$,
$T^\mp$ and $D^\pm$ (resp.\ through the Gauss factors and
their variations). The right hand side of (\ref{eq:Wr5'})
and (\ref{eq:Wr5a'}) (resp.~(\ref{eq:Wrd2})) can be interpreted
as a Fourier-like transformation of the potential $L_1(x)$
(resp.\ of the variation $\delta L_1(x)$). As a natural
generalization of the usual exponents there appear the
`squared solutions'. The `squared solutions' are analytic
functions of~$\lambda$. This fact underlies the proof of
their completeness in the space of allowed potentials, as
we shall see.

\subsection[The skew-scalar product and the mapping $\mathfrak{M}_{L_1} \to \mathfrak{T}_j$]{The skew-scalar product and the mapping $\boldsymbol{\mathfrak{M}_{L_1} \to \mathfrak{T}_j}$}

It is obvious, that only some of the matrix elements of
the squared solutions $e_\alpha (x,\lambda)$ contribute
to the right-hand sides of the Wronskian relations. To
make this clearer we will use the $\bbbz_2$-grading
of the Lie algebra which hints that we should split
the squared solutions as in (\ref{X_split}), namely:
\begin{gather}\label{e_split}
e_\alpha (x,\lambda) = \mathcal{H}_\alpha (x,\lambda)
+ \mathcal{K}_\alpha (x,\lambda),\qquad\mathcal{H}_\alpha (x,\lambda)
\in \mathfrak{sl}^{0}(3),\qquad\mathcal{K}_\alpha (x,\lambda) \in \mathfrak{sl}^{1}(3).
\end{gather}In addition, according to (\ref{X0_split}), (\ref{X1_split})
each components $\mathcal{H}_\alpha (x,\lambda)$ and
$\mathcal{K}_\alpha (x,\lambda)$ can be split into
\begin{gather}
\mathcal{H}_\alpha (x,\lambda)  = H_\alpha (x,\lambda) +  h_{\alpha} L_2(x),
\qquad h_{\alpha} = \frac{3}{2} \langle L_2(x), \mathcal{H}_\alpha (x,\lambda)\rangle,\nonumber\\
\mathcal{K}_\alpha (x,\lambda)  = K_\alpha (x,\lambda) +  k_{\alpha} L_1(x),\qquad
k_{\alpha} =  \frac{1}{2}\,\langle L_1(x), \mathcal{K}_\alpha (x,\lambda)\rangle,\label{eq:HK1}
\end{gather}
where $H_\alpha (x,\lambda)$ and $K_\alpha (x,\lambda)$
do not commute with $L_1(x)$. It is not dif\/f\/icult to realize that
only $H_\alpha (x,\lambda)$ and $K_\alpha (x,\lambda)$
contribute to the right-hand sides of the Wronskian relations.

In what follows we will make use also of the skew-scalar product
\begin{gather*}%\label{eq:ssp}
\biglb X, Y \bigrb = \int_{-\infty}^{\infty} d y\,
\langle X(y), [L_1(y), Y(y) ] \rangle ,
\end{gather*}
where $X$ and $Y$ are functions with values in $\mathfrak{sl}(3)\slash\ker{\ad_{L_1}}$
and vanishing for $x\to\pm \infty$.

We will denote the linear space of such functions by $\mathfrak{M}_{L_1}$.
Note, that we can express the right hand sides of the Wronskian
relations using the skew-scalar product:
\begin{gather*}
 \left\langle \hat{\chi} L_1 \chi (x,\lambda),
E_\alpha \right\rangle \big|_{x=-\infty}^{\infty}
 = \biglb e_\alpha (x,\lambda),\ad_{L_1}^{-1}L_{1,x}\bigrb,\nonumber\\
  \left\langle \hat{\chi} L_2 \chi (x,\lambda),
E_\alpha \right\rangle \big|_{x=-\infty}^{\infty}
 = \biglb e_\alpha (x,\lambda),\ad_{L_1}^{-1} L_{2,x}\bigrb,\nonumber\\
  \left\langle \hat{\chi} \delta \chi (x,\lambda),
E_\alpha \right\rangle\big|_{x=-\infty}^{\infty}
 = i \lambda\biglb e_\alpha (x,\lambda),
\ad_{L_1}^{-1}\delta L_{1} \bigrb,\nonumber \\
 \left\langle \hat{\chi} \chi_t (x,\lambda),
E_\alpha \right\rangle \big|_{x=-\infty}^{\infty}
 = i \lambda\biglb e_\alpha (x,\lambda),\ad_{L_1}^{-1}L_{1,t}\bigrb .%\label{eq:Wr6}
 \end{gather*}
We should also point out that the skew-scalar product is
non-degenerate on $\mathfrak{M}_{L_1}$.

We f\/inish this subsection by listing the ef\/fective results from the Wronskian relations.
Below $\alpha \in \Delta^+$:
\begin{gather}
\rho_\alpha^{(1),\pm}(\lambda) \equiv \langle \hat{D}^\pm\hat{T}^\mp J T^\mp D^\pm, E_{\pm\alpha} \rangle
=-\biglb \ad_{L_1}^{-1} L_{1,x} ,K_{\pm\alpha} ^\pm (x,\lambda)\bigrb, \nonumber\\
\rho_\alpha^{(2),\pm}(\lambda) \equiv \langle \hat{D}^\pm\hat{T}^\mp I T^\mp D^\pm, E_{\pm\alpha}\rangle
= -\biglb \ad_{L_1}^{-1} L_{2,x} ,H_{\pm\alpha} ^{\pm}(x,\lambda)\bigrb;\label{eq:rho}\\
\tau_\alpha^{(1),\pm}(\lambda) \equiv \langle \hat{S}^\pm J S^\pm , E_{\mp \alpha} \rangle
= \biglb \ad_{L_1}^{-1} L_{1,x} ,K_{\mp\alpha} ^\pm(x,\lambda)\bigrb,\nonumber\\
\tau_\alpha^{(2),\pm}(\lambda) \equiv \langle \hat{S}^\pm I S^\pm , E_{\mp \alpha} \rangle   = \biglb \ad_{L_1}^{-1} L_{2,x} ,H_{\mp\alpha}^{\pm}(x,\lambda)\bigrb;\label{eq:tau}
\\
\delta'\rho_\alpha^{\pm}(\lambda) \equiv \langle \hat{D}^\pm \hat{T}^\mp \delta T^\mp D^\pm , E_{\pm\alpha} \rangle
= -i \lambda \biglb \ad_{L_1} ^{-1} \delta L_1  ,K_{\pm\alpha }^\pm (x,\lambda)\bigrb, \nonumber\\
\delta'\tau_\alpha^{\pm}(\lambda) \equiv \langle \hat{S}^\pm \delta S^\pm , E_{\mp\alpha} \rangle
= i \lambda \biglb \ad_{L_1} ^{-1} \delta L_1  ,K_{\mp\alpha} ^\pm(x,\lambda)\bigrb.\label{eq:ssp-3}
\end{gather}

In deriving the above formulae, besides the splitting (\ref{e_split}) we also used the fact that $\ad_{L_1}^{-1} \delta L_{1}, \ad_{L_1}^{-1} L_{1,x} \in \mathfrak{sl}^{0}(3)$
and $\ad_{L_1}^{-1} L_{2,x} \in \mathfrak{sl}^{1}(3)$.

The quantities $\rho_\alpha^{(k),\pm}$, $\tau_\alpha^{(k),\pm}$, $k=1,2$, can be viewed as ref\/lection coef\/f\/icients.
These formulae provide the basis in the analysis of the mapping $\mathfrak{M}_{L_1} \to \mathfrak{T}_k $, $k=1,2$.
Indeed $L_{1,x}\in \mathfrak{M}_{L_1} $, while the sets
of coef\/f\/icients
\begin{gather*}
\mathfrak{T}_1  \equiv \mathfrak{T}_1^{(1)} \cup \mathfrak{T}_1^{(2)},  \qquad
\mathfrak{T}_1^{(k)}  \equiv \big\{
\tau_\alpha^{(k),\pm} (\lambda), \  \lambda \in \bbbr, \  \alpha >0 \big\},\nonumber \\
\mathfrak{T}_2  \equiv \mathfrak{T}_2^{(1)} \cup \mathfrak{T}_2^{(2)},  \qquad
\mathfrak{T}_2^{(k)}   \equiv \big\{
\rho_\alpha^{(k),\pm} (\lambda), \ \lambda \in \bbbr,
\  \alpha >0 \big\}%\label{eq:frT}
\end{gather*}
completed with additional sets characterizing the discrete spectrum of~$L$
are candidates for the minimal sets of scattering data of the Lax operator.

The above considerations show, that following \cite{AKNS} we can treat the
`squared solutions' as generalized exponentials. Therefore the mapping
$\mathfrak{M}_{L_!}$ can be viewed as a generalized Fourier transform.

Of course, the justif\/ications of the above statements must
be based on a proof of the completeness relation for the
`squared solutions' which will be presented in the next section.

\subsection{Recursion operators -- an alternative approach}

Our goal in this subsection is to present an alternative def\/inition
of the recursion operators. We introduce them as operators $\Lambda^{\pm}$
whose eigenfunctions are the `squared solutions'. More precisely, we
shall see that the following equality holds:
\begin{gather*}%\label{eq:Lam-def}
\Lambda^\pm K_{\alpha}^{\pm} (x,\lambda)
= \lambda^{2} K_{\alpha}^{\pm} (x,\lambda).
\end{gather*}

Our consideration starts from analysis of the equation
\begin{gather*}
i\partial_x e_\alpha + \lambda [L_1(x), e_\alpha (x,\lambda)] =0
\end{gather*}
satisf\/ied by each of the `squared solutions'.  We have
omitted the superscript $\pm$ in the notation of the squared
solutions since it does not play a signif\/icant role in most
of our further considerations. We shall use it again at the
end of subsection when it will become essential.

Due to the grading condition
(\ref{grad_cond}) the components $\mathcal{H}_{\alpha}$ and $\mathcal{K}_{\alpha}$
are interrelated through the system
\begin{gather}
i\partial_x \mathcal{H}_\alpha +
\lambda [L_1(x), \mathcal{K}_\alpha (x,\lambda)]  =0,\qquad
i\partial_x \mathcal{K}_\alpha +
\lambda [L_1(x), \mathcal{H}_\alpha (x,\lambda)] =0.
\label{HK_sys}
\end{gather}
After we extract terms proportional to $L_1$  from the above equations
we obtain
\begin{gather*}
 \langle L_2, \partial_x H_{\alpha} \rangle
 + \partial_x h_{\alpha} =0  \quad\Rightarrow \quad
h_{\alpha} = h_{\alpha,0} - \frac{3}{2}\partial_x^{-1}  \langle L_2, \partial_x H_\alpha \rangle,\nonumber\\
 \langle L_1, \partial_x K_\alpha \rangle   + \partial_xk_{\alpha} =0  \quad\Rightarrow \quad
k_{\alpha} = k_{\alpha,0} - \frac{1}{2}\partial_x^{-1}  \langle L_1,\partial_x K_\alpha \rangle,
\end{gather*}
where $h_{\alpha,0}$ and $k_{\alpha,0}$ are some integration constants. On
the other hand the orthogonal part of~(\ref{HK_sys}) reads:
\begin{gather*}
i\pi\partial_x H_\alpha  +  i h_{\alpha} L_{2,x}
  = -\lambda [L_1(x),  K_\alpha ],\qquad
i\pi\partial_x K_\alpha  + i\, k_{\alpha} L_{1,x}
  =  -\lambda [L_1(x), H_\alpha ].
\end{gather*}
After substituting $h_{\alpha}$ and $k_{\alpha}$ in the equations above
we get
\begin{gather}
\Lambda_1 K_\alpha  = \lambda H_\alpha
- k_{\alpha,0} \Lambda_2 L_{2},\label{KH_perp2}\\
\Lambda_2 H_\alpha  = \lambda K_\alpha
-  h_{\alpha,0}\Lambda_1 L_{1},\label{HK_perp2}
\end{gather}
where the operators $\Lambda_2$ and $\Lambda_1$ are given in (\ref{b1bot})
and (\ref{l1t_tau}) respectively. This is the right place to mention
that remarkably the operators $\Lambda_{1,2}$ are invertible. Indeed,
they admit the factorization
\begin{gather*}
\Lambda_1 X  =  -i \ad_{L_1}^{-1}\mathcal{J}_1\partial_x X,
\qquad \mathcal{J}_1 X = X - \frac{1}{2} L_{1,x} \partial_x^{-1}
\langle L_1, X\rangle,\\
\Lambda_2 Y  =  -i \ad_{L_1}^{-1}\mathcal{J}_2 \partial_x Y,\qquad
\mathcal{J}_2 Y = X - \frac{3}{2} L_{2,x} \partial_x^{-1}
\langle L_2, X\rangle.
\end{gather*}
It is not hard to see that the operators $\mathcal{J}_1 $ and
$\mathcal{J}_2 $ can be inverted as follows
\[
\mathcal{J}_1^{-1} X = X + \frac{1}{2} L_{1,x}\partial_x^{-1}
\langle L_1, X\rangle,\qquad
\mathcal{J}_2^{-1} Y = X + \frac{3}{2} L_{2,x}\partial_x^{-1}
\langle L_2, X\rangle .
\]
Therefore $\Lambda_{1}$ and $\Lambda_{2}$ are invertible as well, namely
\begin{gather*}
\Lambda_1^{-1} X = i \partial_x^{-1}\mathcal{J}_1^{-1}\ad_{L_1} X,
\qquad \Lambda_2^{-1} X = i \partial_x^{-1}\mathcal{J}_2^{-1}\ad_{L_1} Y.
\end{gather*}

Let us apply $\Lambda_2$ to (\ref{KH_perp2}) and $\Lambda_1$ to
(\ref{HK_perp2}). After restoring the index notation where needed
the result reads:
\begin{gather*}
\Lambda^{\pm}_2 \Lambda^{\pm}_1 K_\alpha^{\pm}
 =  \lambda^{2} K_\alpha^{\pm}
-\lambda h_{\alpha,0}\Lambda^{\pm}_1 L_{1}
-k_{\alpha,0} \Lambda^{\pm}_2\Lambda^{\pm}_2 L_2,%\label{KK_eq}
\\
\Lambda^{\pm}_1\Lambda^{\pm}_2 H_\alpha^{\pm}
 =  \lambda^{2} H_\alpha^{\pm}
- \lambda k_{\alpha,0} \Lambda^{\pm}_2 L_2
- h_{\alpha,0} \Lambda^{\pm}_1\Lambda^{\pm}_2 L_2.%\label{HH_eq}
\end{gather*}

This is the right place to remind the reader that the constants $h_{\alpha,0}$
and $k_{\alpha,0} $ are determined by the asymptotic of the relevant `squared
solution' for $x\to \infty$ (or  for $x\to -\infty$), depending on the proper
def\/inition of the recursion operator.  More detailed analysis shows that for
each $\Lambda_1^\pm$ and $\Lambda_2^\pm$ there exist certain
roots for which the constants $k_{\alpha,0}$ and $h_{\alpha,0}$ vanish. More specif\/ically
the following equalities holds:
\begin{gather*}
\Lambda_1^+ K^{\pm} _{\mp\alpha} (x,\lambda)  = \lambda H^{\pm}_{\mp\alpha} (x,\lambda)   ,    \qquad
\Lambda_1^- K^{\pm} _{\pm\alpha} (x,\lambda)  = \lambda H^{\pm}_{\pm\alpha}(x,\lambda) ,  \nonumber\\
\Lambda_2^+ H^{\pm} _{\mp\alpha} (x,\lambda)  = \lambda K^{\pm}_{\mp\alpha} (x,\lambda)   ,    \qquad
\Lambda_2^- K^{\pm} _{\pm\alpha}(x,\lambda)  = \lambda K^{\pm}_{\pm\alpha} (x,\lambda)  %\label{eq:HK6'}
\end{gather*}
for all roots $\alpha>0$ and therefore:
\begin{gather*}
\Lambda_2^+ \Lambda_1^+ K^{\pm}_{\mp\alpha}(x,\lambda)   = \lambda^{2} K^{\pm} _{\mp\alpha}(x,\lambda)   ,    \qquad
\Lambda_2^- \Lambda_1^- K^{\pm} _{\pm\alpha}(x,\lambda)   = \lambda^{2} K^{\pm} _{\pm\alpha} (x,\lambda),
\nonumber \\
\Lambda_1^+ \Lambda_2^+ H^{\pm} _{\mp\alpha}(x,\lambda)  = \lambda^{2} H^{\pm} _{\mp\alpha}(x,\lambda)   ,    \qquad
\Lambda_1^- \Lambda_2^- H^{\pm} _{\pm\alpha}(x,\lambda)  = \lambda^{2} H^{\pm}_{\pm\alpha} (x,\lambda).%\label{eq:Lam'}
\end{gather*}
The operator $\Lambda^{\pm}$ introduced as
\begin{gather}
\Lambda^{\pm}X = \Lambda^{\pm}_1\Lambda^{\pm}_2X, \quad X\in\mathfrak{sl}^{0}(3),\qquad
\tilde{ \Lambda}Y = \Lambda^{\pm}_2\Lambda^{\pm}_1 Y, \quad Y\in\mathfrak{sl}^{1}(3) \label{Lambda_pm}
\end{gather}
is the recursion operator we have been looking for. If we compare
(\ref{Lambda_pm}) and (\ref{recurs_op_f}) we can conclude that
both approaches lead to compatible results which coincide up to
some conjugation by $\ad_{L_1}$.

\section{Spectral theory of the recursion operators}\label{section5}

\subsection{Asymptotics of the fundamental analytic solution}

For performing the contour integration, in order to derive the completeness relations for the squared solutions, one needs the asymptotics of the squared solutions for $|\lambda|\to \infty$. The starting point here is the asymptotic expansion of the FAS for~$\lambda\to\infty$:
\begin{gather*}%\label{eq:chi_exp}
\chi_{\rm as}(x,\lambda)=\openone + \sum_{k=1}^\infty \chi_k(x)\lambda^{-k}.
\end{gather*}

Introduce:
\begin{gather*}%\label{eq:chi-t}
\tilde{\chi}_{\rm as}(x,\lambda) = g^{-1}(x) \chi_{\rm as} (x,\lambda)  e ^{ i  J \lambda x},
\end{gather*}
which satisfy:
\begin{gather}\label{eq:chi_teq}
 i  \frac{d \tilde{\chi}_{\rm as}} {d x} + i  g^{-1}g_{x} \tilde{\chi}_{\rm as}(x,\lambda) +
\lambda [J,\tilde{\chi}_{\rm as}(x,\lambda)] =0.
\end{gather}
Together with properly chosen asymptotic conditions $\tilde{\chi}_{\rm as}(x,\lambda)$ the last equation will provide the
asymptotics of the fundamental analytic solution. Therefore it will allow an asymptotic expansion of the form:
\begin{gather*}%\label{eq:chit-as}
\tilde{\chi}_{\rm as}(x,\lambda) = \sum_{s=0}^{\infty} \lambda^{-s} \tilde{\chi}_{k,\rm as}(x) .
\end{gather*}
This will lead to a recurrent relations for the expansion coef\/f\/icients $\tilde{\chi}_{k,\rm as}(x)$.
Inserting this expansion into equation~(\ref{eq:chi_teq}) for the f\/irst two coef\/f\/icients of $\tilde{\chi}_{\rm as}(x,\lambda)$ we get:
\begin{gather}\label{eq:chit-k}
[J,\tilde{\chi}_{0,\rm as}(x)] =0, \\
 i  \frac{d \tilde{\chi}_{0,\rm as}} {d x} -  i  g^{-1}_{x}g \tilde{\chi}_{0,\rm as}(x) +
[J,\tilde{\chi}_{1,\rm as}(x)] =0 ,\label{eq:chit-1} \\
 i  \frac{d \tilde{\chi}_{1,\rm as}} {d x} -  i  g^{-1}_{x}g \tilde{\chi}_{1,\rm as}(x) +
[J,\tilde{\chi}_{2,\rm as}(x)] =0.\nonumber%\label{eq:chit-2}
\end{gather}
From (\ref{eq:chit-k}) it follows that $\tilde{\chi}_{0,\rm as}(x)$ must be a diagonal matrix and using the diagonal part of~(\ref{eq:chit-1}) one can derive it in an explicit form:
\begin{gather}\label{eq:chi0}
\tilde{\chi}_{0,\rm as}(x) = \diag \big( e ^{\rho(x)}, e ^{-2\rho(x)}, e ^{\rho(x)}\big),
\qquad \rho(x) = \frac{1}{2}\int_{\pm \infty} ^x d y (u^*u_y + v^*v_y),
\end{gather}
and for the of\/f-diagonal part of $\tilde{\chi}^{\rm f}_{1,\rm as}(x)$ we have (see equation~(\ref{u_0})):
\begin{gather*}
 \tilde{\chi}^{\rm f}_{1,\rm as}(x) = \frac{ i }{2}\left(\!\!\begin{array}{c@{\,\,\,}c@{\,\,\,}c}
0 & \sqrt{2}(uv_x-vu_x) e ^{-\rho(x)}  & (u^*u_x+v^*v_x) e ^{\rho(x)}/2 \\
\sqrt{2}(u^*v^*_x-v^*u^*_x) e ^{\rho(x)} & 0 & -\sqrt{2}(u^*v^*_x-v^*u^*_x)e^{2\rho(x)}\\
-(u^*u_x+v^*v_x) e ^{\rho(x)}/2 & \sqrt{2}(vu_x - uv_x)e^{-2\rho(x)} & 0 \end{array}\!\!\right).%\label{eq:chi1}
\end{gather*}
As a result the asymptotic behavior of $\chi(x,\lambda)$ for $\lambda\to\infty$ is given by:
\begin{gather}\label{eq:chi-asm}
\chi(x,\lambda) \mathop{\simeq}\limits_{\lambda\to\infty} g(x) \left( \tilde{\chi}_{0,\rm as}(x)
+ \frac{1}{\lambda} \tilde{\chi}_{1,\rm as}(x) + \cdots \right)  e ^{ -i  J \lambda x}.
\end{gather}
Note, that the fundamental analytic solution $\chi^\pm$ does not allow a canonical normalization for $|\lambda |\to\infty$.
This dif\/f\/iculty can be overcome by applying a suitable gauge transformation.

Using the asymptotic behavior (\ref{eq:chi-asm}) of the FAS, one can derive the asymptotics of the squared solutions $e_\alpha(x,\lambda)=(\chi(x,\lambda)E_\alpha \chi^{-1}(x,\lambda))^{\rm f}$. Skipping the details:
\begin{gather*}
e^\pm _\alpha (x,\lambda)  \mathop{\simeq}\limits_{\lambda\to\infty}    e ^{ i  (J,\alpha) \lambda x
}g(x)\tilde{\chi}_{\rm as}(x)E_\alpha \tilde{\chi}_{\rm as}^{-1}(x)g^{-1}(x) \nonumber\\
\phantom{e^\pm _\alpha (x,\lambda)}{} =   e ^{ i  (J,\alpha) \lambda x +3\rho (x) (I, \alpha) }g(x)
\left(  E_\alpha + \mathcal{O}\big(\lambda^{-1}\big)\right) g^{-1}(x).%\label{eq:e-asm}
\end{gather*}

\subsection{Completeness of squared solutions}

Below we will derive the completeness relation for the `squared solutions' for a class of poten\-tials~$L_1(x)$ of
the Lax operator $L$ which for every f\/ixed value of $t$ satisfy the following conditions:
\begin{description}\itemsep=0pt
  \item[C1)] $L_1(x) - L_{1,\pm}$ is complex valued function of Schwartz type, i.e.\ it is inf\/initely smooth
  function of $x$ falling of\/f for $|x|\to\infty$ faster than any power of $x$;
  \item[C2)] $L_1(x)$ is such that the corresponding functions $m_1^\pm (\lambda)$ and $m_2^\pm (\lambda)$ have at most f\/inite number
  of simple zeroes. For simplicity we assume also that $m_1^+ (\lambda)$ and $m_2^+ (\lambda)$ (resp.~$m_1^- (\lambda)$ and~$m_2^- (\lambda)$)
  have no common zeroes;
  \item[C3)] the corresponding functions $m_1^\pm (\lambda)$ and $m_2^\pm (\lambda)$ have no zeroes on the real axis of the complex $\lambda$-plane.
\end{description}

\begin{remark}\label{rem:3}
It is well known that the zeroes of the functions $m_1^\pm (\lambda)$ and $m_2^\pm (\lambda)$ determine the discrete spectrum
of $L$. The proof of this fact  comes out of the scope of the present paper.
For other classes of Lax operators it is well known, see e.g.~\cite{IP2,GVY*08}.
\end{remark}

\begin{figure}[t]
\centering
\includegraphics[width=0.45\textwidth]{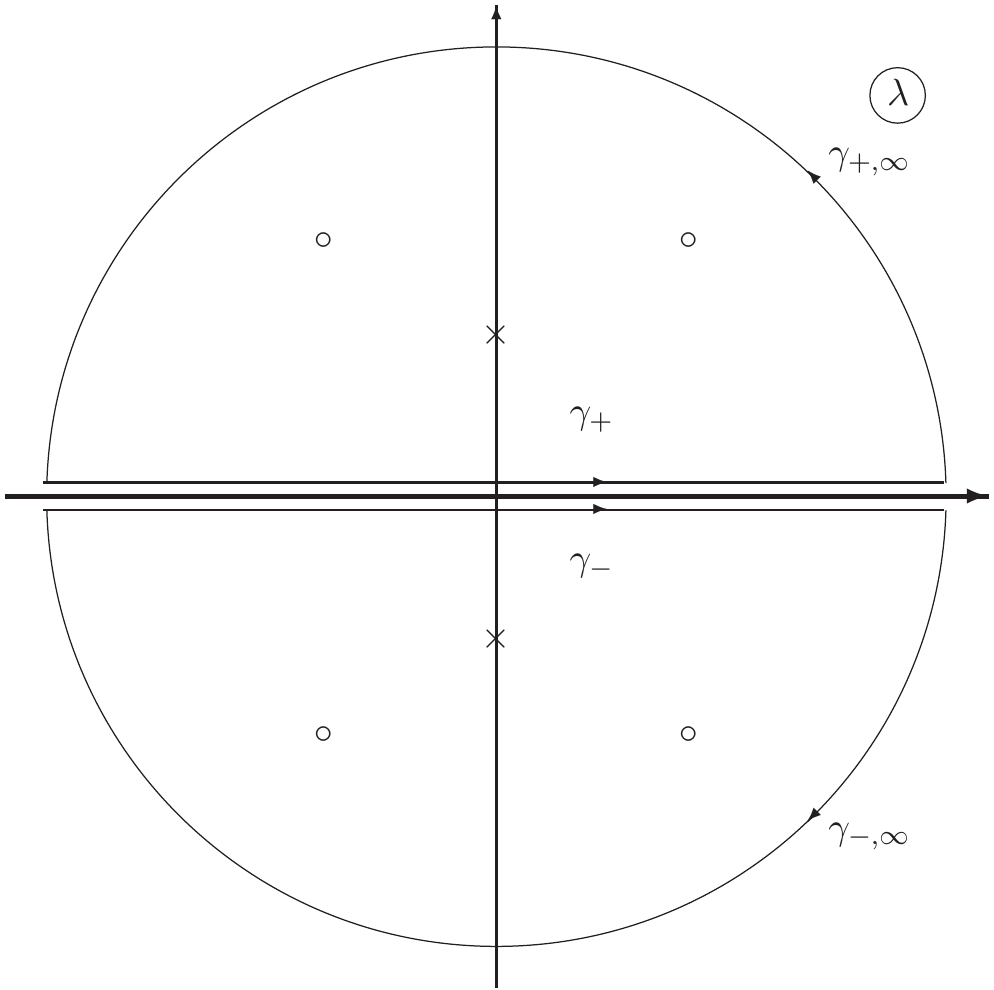}
\caption{The bold line def\/ines the continuous spectrum of $L$, the
regular lines show the integration contours $\gamma_\pm$;
by $\gamma_{\pm,\infty}$  we denote the `inf\/inite'
semi-circles. By $\circ$ we denote the discrete eigenvalues
relevant for one generic soliton which come in quadruplets $\pm \lambda_j^\pm$;
the second type of solitons correspond to purely imaginary
pairs of discrete eigenvalues $\pm i\nu_a$ denoted by $\times.$}
\label{fig:contour}
\end{figure}

Below for def\/initeness we introduce notations for the discrete eigenvalues. As we already mentioned, due to
the two $\bbbz_2$-reductions, they are of two types. Let there be $N_1$ discrete eigenvalues of
generic type which come in quadruplets (see Fig.~\ref{fig:contour}). We will denote by $\lambda_1^+, \lambda_2^+, \dots,$
$\lambda_{N_1}^+$ those of them that are in the f\/irst quadrant, i.e.\ $0 < \arg \lambda_k^+ < \pi/2$, $1\leq k \leq N_1$.
The eigenvalues $\lambda_{N_1+k}^+ = -(\lambda_k^+)^*$ belong also to $\bbbc_+$ but are in  the second quadrant.
Besides we have also $\lambda_j^- = (\lambda_j^+)^*$, $j=1,\dots, 2N_1$ which lie in $\bbbc_-$.
The second type of discrete eigenvalues are purely imaginary, i.e. $\lambda_{2N_1+a}^+ = i\nu_a$, $a=1,\dots ,N_2$;
obviously  $\lambda_{2N_1+a}^- =- i\nu_a$. Therefore the Lax operator $L$ will have $4N_1 + 2N_2$ discrete eigenvalues.

Consider the Green function:
\begin{gather}\label{eq:G}
G^\pm(x,y,\lambda)  =  G_1^\pm(x,y,\lambda) \theta(y-x) - G_2^\pm(x,y,\lambda) \theta(x-y), \\
G_1^\pm(x,y,\lambda)   =  \sum_{\alpha \in \Delta^+} e_{\pm\alpha} ^\pm
(x,\lambda) \otimes e_{\mp \alpha}^\pm (y,\lambda), \\
G_2^\pm (x,y,\lambda)  =  \sum_{\alpha \in \Delta^-} e_{\pm\alpha}^-(x,\lambda)\otimes
e_{\mp\alpha}^-(y,\lambda) + \sum_{j=1}^r h_{j}^\pm (x,\lambda)\otimes h_{j}^\pm(y,\lambda), \\
h_{j}^\pm (x,\lambda)  =  \chi^\pm(x,\lambda) H_j \hat{\chi}^\pm (x,\lambda). \label{eq:5.19}
\end{gather}
In order to derive the completeness relations for the squared solutions, one needs to apply a~contour integration to the integral:
\begin{gather*}%\label{eq:J-int}
\mathcal{J}_G(x,y)=\frac{1}{2\pi i}\oint_{\gamma_+}d \lambda \,G^+(x,y,\lambda) - \frac{1}{2\pi i}\oint_{\gamma_-}d \lambda \,G^-(x,y,\lambda),
\end{gather*}
where the integration contours $\gamma_\pm = \bbbr\cup \gamma_{\pm,\infty } $ are shown on Fig.~\ref{fig:contour}.

Let us recall that the Lax operator $L$ has two types of discrete eigenvalues
whose number is $4N_1 +2N_2$; the f\/irst type correspond to
generic solitons and come in quadruplets $\pm \lambda_j^\pm$;
the second type   correspond to purely imaginary pairs of discrete eigenvalues~$\pm i\nu_a$.
Thus according to Cauchy's residuum theorem
\begin{gather*}%\label{eq:J-res}
\mathcal{J}_G(x,y)  =\sum_{k=1}^{2N_1 +N_2} \left( \Res _{\lambda =\lambda_k^+}\, G^+(x,y,\lambda)
 + \Res_{\lambda =\lambda_k^{-}}\, G^-(x,y,\lambda)\right),
 \end{gather*}
presuming that the radii of the `inf\/inite' semicircles are large enough so that all discrete eigenvalues are
inside the contours $\gamma_\pm$. Condition {\bf C3} above ensures that
$G^\pm(x,y,\lambda)$ has no poles on the continuous spectrum.

Next we integrate along the contours:
\begin{gather*}
\mathcal{J}_G(x,y)  = \frac{1}{2\pi i}\int_{-\infty}^\infty  d\lambda \,( G^+(x,y,\lambda) -G^-(x,y,\lambda) )\nonumber \\
\hphantom{\mathcal{J}_G(x,y)  =}{} +\frac{1}{2\pi i}\oint_{\gamma_{+,\infty}}d \lambda \,G^+_{\rm as} (x,y,\lambda) -
\frac{1}{2\pi i}\oint_{\gamma_{-,\infty}}d \lambda \,G^-_{\rm as} (x,y,\lambda).%\label{eq:J-rres}
 \end{gather*}

In order to calculate the integrals over $\gamma_{\pm,\infty}$ it is enough to use the
asymptotic behavior  of the Green function for $\lambda \to \infty$. From equations~(\ref{eq:chi0})
and~(\ref{eq:chi-asm}) we have:
\begin{gather*}
G^\pm _{1, \rm as}(x,t,\lambda)  \mathop{\simeq}\limits_{\lambda\to\infty}
\sum_{\alpha \in \Delta_+ }   e ^{i   (J,\alpha) \lambda (x-y) + 3(\rho(x)-\rho(y))(I,\alpha)} \nonumber\\
\hphantom{G^\pm _{1, \rm as}(x,t,\lambda)  \mathop{\simeq}\limits_{\lambda\to\infty}}{}
\times \left( g(x)\otimes g(y)\right)
\left( E_{\pm \alpha} \otimes E_{\mp \alpha} + \mathcal{O}(\lambda^{-1})\right) (g^{-1}(x)\otimes g^{-1}(y))%\label{eq:G1-as}
\end{gather*}
and
\begin{gather*}
G^\pm _{2, \rm as}(x,t,\lambda) \mathop{\simeq}\limits_{\lambda\to\infty}
\sum_{j=1}^{r} (g(x)\otimes g(y))  \left( H_j\otimes H_j^\vee + \mathcal{O}(\lambda^{-1})\right) (g^{-1}(x)\otimes g^{-1}(y))\\ %\label{eq:G2-as}\\
+ \sum_{\alpha \in \Delta_0\cup \Delta_1^-}  e ^{i   (J,\alpha) \lambda (x-y) +3(\rho(x)-\rho(y))(I,\alpha)} (g(x)\otimes g(y))
 \left( E_{\pm \alpha} \otimes E_{\mp \alpha} + \mathcal{O}(\lambda^{-1})\right) (\hat{g}(x)\otimes \hat{g}(y)).\nonumber
\end{gather*}
As a result we get:
\begin{gather*}
\frac{1}{2\pi i}\oint_{\gamma_{+,\infty}}d \lambda \,G^+_{\rm as} (x,y,\lambda) -
\frac{1}{2\pi i}\oint_{\gamma_{-,\infty}}d \lambda \,G^-_{\rm as} (x,y,\lambda) \nonumber\\
\qquad{}= \delta(x-y) \left( g(x)\otimes g(y)\right) \sum_{\alpha\in\Delta_+}^{} \left( E_{\alpha} \otimes E_{-\alpha} -
E_{-\alpha} \otimes E_{\alpha} \right) \left( \hat{g}(x)\otimes \hat{g}(y)\right).%\label{eq:Gpminf}
\end{gather*}

The next step is to simplify the integral along the continuous spectrum; namely we will calculate the jump of the Green
function with the result:
\begin{gather*}
G^+(x,y,\lambda) -  G^-(x,y,\lambda)  =  G^+_1(x,y,\lambda) -  G^-_1(x,y,\lambda) \nonumber\\
\phantom{G^+(x,y,\lambda) -  G^-(x,y,\lambda)}{} = \sum_{ \alpha\in \Delta_+}^{} \left( e^+_{\alpha}(x,\lambda)\otimes  e^+_{-\alpha}(x,\lambda)
- e^-_{-\alpha}(x,\lambda)\otimes  e^-_{\alpha}(x,\lambda) \right).%\label{eq:jG}
\end{gather*}

The proof of this fact goes as follows. From the expression for the Green function (equations~(\ref{eq:G})--(\ref{eq:5.19}))
and using the relation $\theta(x-y) + \theta(y-x) =1$ we easily get:
\begin{gather*}
G^+(x,y,\lambda) -  G^-(x,y,\lambda) =  G^+_1(x,y,\lambda) -  G^-_1(x,y,\lambda) \nonumber\\
\qquad{} - \theta(x-y) (
G_1^+(x,y,\lambda) +  G_2^+(x,y,\lambda) -  G^-_1(x,y,\lambda) -  G^-_2(x,y,\lambda)  ),%\label{eq:jG-1}
\end{gather*}
so it will be enough to prove that the coef\/f\/icient in front of $\theta(x-y)$ vanishes. Indeed, we have:
\begin{gather*}
G_1^+(x,y,\lambda) +  G_2^+(x,y,\lambda) -  G^-_1(x,y,\lambda) -  G^-_2(x,y,\lambda)   \nonumber \\
\qquad{}= (\chi^+(x,\lambda)\otimes \chi^+(y,\lambda)) \Pi_2 (\hat{ \chi}^+(x,\lambda)\otimes \hat{ \chi}^+(y,\lambda))\nonumber\\
\qquad\quad{}-
(\chi^-(x,\lambda)\otimes \chi^-(y,\lambda)) \Pi_2 (\hat{ \chi}^-(x,\lambda)\otimes \hat{ \chi}^-(y,\lambda)).%\label{eq:jG-3}
\end{gather*}
Here
\begin{gather*}%\label{eq:Pi2}
\Pi_2= \sum_{\alpha\in \Delta_+}^{}(E_{\alpha} \otimes E_{-\alpha} + E_{-\alpha} \otimes E_{\alpha})
+ \sum_{k=1}^{2} \frac{H_k\otimes H_k }{\langle H_k, H_k\rangle}
\end{gather*}
and $H_1 =J$,  $H_2=J_4$.
Note that $\Pi_2$ is the second Casimir endomorphism for the algebra $sl(3)$.
It remains to use equation~(\ref{riemman}) and the basic property of $\Pi_2$ that states:
\begin{gather*}%\label{eq:pi01}
(G(\lambda)\otimes G(\lambda)) \Pi_2 (\hat{ G}(\lambda)\otimes \hat{ G}(\lambda)) = \Pi_2,
 \end{gather*}
for any element $G(\lambda)$ of the corresponding group $SL(3)$.

We evaluate  the contribution from the discrete spectrum assuming that the squared solutions~$e^\pm _\alpha(x,\lambda)$ have at most simple poles at the points of the discrete spectrum.
Therefore in the vicinity of the eigenvalues $\lambda_k^\pm$ we use the following expansions of the `squared solutions':
\[
e^\pm_\alpha (x,\lambda) = (\lambda -\lambda_k^\pm)^{p_\alpha} \left(
\frac{e^\pm_{\alpha,k}}{\lambda -\lambda_k^\pm} + \dot{e}^\pm_{\alpha,k} \right),
\]
where $p_\alpha$ are integers and $p_\alpha =-p_{-\alpha}$. Thus we obtain:
\begin{gather*}
\Res _{\lambda =\lambda_k^+}\, G^+(x,y,\lambda) = \sum_{\alpha\in\Delta_+ }^{}
\big( e^\pm_{\alpha,k}\otimes \dot{e}^\pm_{-\alpha,k} + \dot{e}^\pm_{\alpha,k}\otimes e^\pm_{-\alpha,k} \big).%\label{eq:ResG}
\end{gather*}

Therefore the completeness relations takes the form:
\begin{gather}
\delta(x-y) \left( g(x)\otimes g(y)\right) \Pi_0 \left( \hat{g}(x)\otimes \hat{g}(y)\right) \nonumber\\
\qquad{} =   \frac{i}{2\pi }\int_{-\infty}^{\infty} d\lambda \; \sum_{\alpha\in \Delta_+}^{} \left( e^+_{\alpha}(x,\lambda)\otimes  e^+_{-\alpha}(x,\lambda)
- e^-_{-\alpha}(x,\lambda)\otimes  e^-_{\alpha}(x,\lambda) \right)\nonumber\\
\qquad\quad {} + \sum_{k=1}^{2N_1+N_2} \sum_{\alpha\in\Delta_+}^{}  \big( e^+_{\alpha,k}(x)\otimes \dot{e}^+_{-\alpha,k}(y) + \dot{e}^+_{\alpha,k}(x)
 \otimes e^+_{-\alpha,k}(y)   \nonumber\\
 \qquad\quad {}+ e^-_{-\alpha,k}(x)\otimes \dot{e}^-_{\alpha,k}(y) + \dot{e}^-_{-\alpha,k}(x)\otimes e^-_{\alpha,k}(y)\big),\label{eq:CR}
\end{gather}
where
\begin{gather*}%\label{eq:Pi0}
\Pi_0= \sum_{\alpha\in \Delta_+}^{} \frac{E_{\alpha} \otimes E_{-\alpha} - E_{-\alpha} \otimes E_{\alpha} }{(\alpha,J)}
\end{gather*}
and
\begin{gather*}%\label{eq:edot}
 e^\pm_{\alpha,k}  =  e^-_{\alpha}(x,\lambda_k^\pm) ,  \qquad
  \dot{e}^\pm_{\alpha,k}  = \left. \frac{\partial  e^\pm_{\alpha}(x,\lambda) }{ \partial \lambda }\right|_{\lambda =\lambda_k^\pm}.
\end{gather*}

\subsection{Expansion over the `squared solutions'}

Using the completeness relations of the squared solutions, one can expand any generic
element~$F(x) $ of the phase space $ \mathcal{ M} $ over the complete sets of squared solutions.
We remind that~$F(x) $ is a generic element of $\mathcal{ M} $ if it  tends fast
enough to zero for $|x|\to\infty $ and is `perpendicular', i.e.
\begin{gather*}%\label{eq:FF}
\langle F(x), L_1\rangle = \langle F(x), L_2\rangle =0.
 \end{gather*}
It can be parametrized  as follows:
\begin{gather*}%\label{eq:F-gen}
F(x) = g(x) \sum_{\alpha \in \Delta_{+}\cup  \Delta_{-}} F_{\alpha}(x) E_{\alpha} g(x).
\end{gather*}
Let us now multiply both sides of the completeness relation (\ref{eq:CR}) by $\openone \otimes [L_1(y),F(y)]$
on the right,  take the Killing form of the second factors in the tensor product and integrate over~$y$. In the left hand side
after some manipulations we get:
\begin{gather*}
\int_{-\infty}^{\infty} dy\; \left\langle (g(x)\otimes g(y)) \Pi_0 (\hat{g}(x) \otimes \hat{g}(y) ),
\openone \otimes [L_1(y),F(y)] \right\rangle_2 \delta(x-y) \nonumber\\
\qquad{} = g(x) \sum_{\alpha \in \Delta_{+}\cup  \Delta_{-} }^{} \frac{ E_\alpha}{(\alpha, J)} \hat{g}(x)
\int_{-\infty}^{\infty} dy\; \langle g(y) E_{-\alpha} \hat{g}(y), [L_1(y), F(y)] \rangle \delta(x-y) \nonumber\\
\qquad{}= g(x) \sum_{\alpha \in \Delta_{+}\cup  \Delta_{-} }^{}   F_\alpha(x) E_\alpha  \hat{g}(x) =F(x).%\label{eq:CR-1}
\end{gather*}
The same operations applied to the right hand sides of (\ref{eq:CR}) replaces each of the terms
$ e^\pm_{\alpha}(x,\lambda) \otimes  e^\pm_{-\alpha}(y,\lambda) $ by $  e^\pm_{\alpha}(x,\lambda)
\biglb  e^\pm_{\alpha}(y,\lambda) ,F(y)\bigrb $. Thus the expansion coef\/f\/icients of~$F(x)$
are provided by its skew-scalar products with the `squared solutions'.

The result is:
\begin{gather}
F(x)  = \frac{i}{2\pi} \int_{-\infty}^\infty d\lambda \sum_{\alpha \in \Delta_{+}} \big(e^+_{\alpha}(x,\lambda)
\gamma^+_{F;\alpha}(\lambda) - e^-_{-\alpha}(x,\lambda) \gamma^-_{F;\alpha}(\lambda) \big)\nonumber \\
\phantom{F(x)  =}{} + \sum_{k=1 }^{2N_1+N_2}
\sum_{\alpha \in \Delta_{1}^{+}} \big( Z_{F;\alpha, k}^+(x) + Z_{F;\alpha, k}^-(x)\big),\label{eq:compl}
\end{gather}
where
\begin{alignat}{3}
& \gamma^\pm _{F;\alpha}(\lambda) = \biglb e^\pm_{\mp \alpha} (y,\lambda), F(y)\bigrb,
\qquad&&  Z_{F,\alpha ;k}^\pm (x)  = e^\pm_{\pm \alpha;k}(x ) \dot{\gamma}^\pm_{F, \alpha ;k} +
 \dot{e}^\pm_{\pm \alpha;k}(x ) \gamma^\pm_{F, \alpha ;k} , &\nonumber \\
& \gamma^\pm _{F,\alpha ;k}  = \biglb e^\pm_{\mp \alpha; k} (y ), F(y)\bigrb, \qquad&&
\dot{\gamma}^\pm _{F,\alpha ;k}  = \biglb \dot{e}^\pm_{\mp \alpha; k} (y ), F(y)\bigrb.&\label{eq:gammas}
\end{alignat}

Similarly, exchanging $x$ and $y$,
multiplying both sides of the completeness relation (\ref{eq:CR}) by $ [L_1(y),F(y)] \otimes\openone $
on the right,  take the Killing form of the f\/irst factors in the tensor product and integrate over $y$ we get:
\begin{gather}
F(x)  = -\frac{i}{2\pi} \int_{-\infty}^\infty d\lambda \sum_{\alpha \in \Delta_{+}} \left(e^+_{\alpha}(x,\lambda)
\theta^+_{F;\alpha}(\lambda) - e^-_{-\alpha}(x,\lambda) \theta^-_{F;\alpha}(\lambda) \right) \nonumber\\
\phantom{F(x)  =}{} - \sum_{k=1 }^{2N_1+N_2}
\sum_{\alpha \in \Delta_{1}^{+}} ( Y_{F;\alpha, k}^+(x) + Y_{F;\alpha, k}^-(x)),\label{eq:compl'}
\end{gather}
where
\begin{alignat}{3}
& \theta^\pm _{F;\alpha}(\lambda) = \biglb e^\pm_{\pm \alpha} (y,\lambda), F(y)\bigrb,
\qquad && Y_{F,\alpha ;k}^\pm (x) = e^\pm_{\mp \alpha;k}(x ) \dot{\theta}^\pm_{F, \alpha ;k} +
 \dot{e}^\pm_{\mp \alpha;k}(x ) \theta^\pm_{F, \alpha ;k} , &\nonumber\\
& \theta^\pm _{F,\alpha ;k}  = \biglb e^\pm_{\pm \alpha; k} (y ), F(y)\bigrb, \qquad&&
 \dot{\theta}^\pm _{F,\alpha ;k}  = \biglb \dot{e}^\pm_{\pm \alpha; k} (y ), F(y)\bigrb.&\label{eq:thetas}
\end{alignat}

The completeness relation (\ref{eq:compl}) allows one to prove the following
\begin{corollary}\label{cor:1}\qquad
\begin{enumerate}\itemsep=0pt
  \item[$i)$] The function $F(x)\equiv 0 $ if and only if all its expansion coefficients
 \eqref{eq:gammas} vanish, i.e.:
\begin{gather}
\gamma^+ _{F;-\alpha}(\lambda)=\gamma^- _{F;\alpha}(\lambda)=0,
\qquad \alpha \in \Delta_{+};\nonumber\\
 Z_{F,\alpha; k}^+(x)=
Z_{F,\alpha;k}^-(x)=0, \qquad k=1,\dots, \tilde{ N},\label{eq:pr'}
\end{gather}
where  $\tilde{N}=2N_1+N_2$.

  \item[$ii)$] The function $F(x)\equiv 0 $ if and only if all its expansion coefficients
  \eqref{eq:thetas} vanish, i.e.:
\begin{gather*}
\theta^+ _{F;-\alpha}(\lambda)=\theta^- _{F;\alpha}(\lambda)=0,
\qquad \alpha \in \Delta_{+};\nonumber\\
 Y_{F,\alpha; k}^+(x)=
Y_{F,\alpha;k}^-(x)=0, \qquad k=1,\dots, \tilde{ N}.%\label{eq:pr''}
\end{gather*}
\end{enumerate}
\end{corollary}

\begin{proof}
Inserting $F(x) \equiv 0 $ into the r.h.s.\ of the inversion formulae~(\ref{eq:gammas}) we obtain the relations~(\ref{eq:pr'}). Next, let us assume that equation~(\ref{eq:pr'}) holds and let us insert it into the r.h.s.\ of equation~(\ref{eq:compl}).
This immediately gives $F(x)=0$. Thus $i)$ is proved. $ii)$ is proved analogously using
equations~(\ref{eq:thetas}) and (\ref{eq:compl'}).
\end{proof}

In other words we established an one-to-one correspondence between the element $F(x) \in
\mathcal{ M} $ and its expansion coef\/f\/icients.

Now, let us take $F(x)= {\rm ad}_{L_1}^{-1}  L_{j,x}(x)$, $j=1,2$.
Then the corresponding expansion coef\/f\/icients are given by:
\begin{alignat*}{4}
& \rho_\alpha^{(j),\pm} (\lambda) = \biglb e^\pm_{\pm \alpha}(x,\lambda), \ad_{L_1}^{-1} L_{j,x} \bigrb, \qquad &&
\rho_{\alpha;k}^{(j),\pm}   = 0, \qquad & & \dot{\rho}_{\alpha; k}^{(j),\pm} = \biglb \dot{e}^\pm_{\pm \alpha;k}(x), \ad_{L_1}^{-1} L_{j,x} \bigrb,&\nonumber \\
& \tau_\alpha^{(j),\pm} (\lambda) = -\biglb e^\pm_{\mp \alpha}(x,\lambda), \ad_{L_1}^{-1} L_{j,x} \bigrb, \qquad &&
\tau_{\alpha;k}^{(j),\pm}   = 0, \qquad &&  \dot{\tau}_{\alpha; k}^{(j),\pm} = -\biglb \dot{e}^\pm_{\mp \alpha;k}(x), \ad_{L_1}^{-1} L_{j,x} \bigrb.&
% \label{eq:excLx}
\end{alignat*}

From the Wronskian relations we have:
\begin{alignat*}{4}
& \rho_\alpha^{(1),\pm} (\lambda) = \langle \hat{D}^\pm \hat{T}^\mp J T^\mp D^\pm,E_{\mp\alpha}\rangle , \qquad&&
\rho_{\alpha;k}^{(1),\pm}  = 0 , \qquad&& \dot{\rho}_{\alpha;k}^{(1),\pm} =
\langle \hat{D}^\pm \hat{T}^\mp J T^\mp D^\pm,E_{\mp\alpha}\rangle, &\\
& \rho_\alpha^{(2),\pm} (\lambda) = 3\langle \hat{D}^\pm \hat{T}^\mp I T^\mp D^\pm,E_{\mp\alpha}\rangle , \qquad &&
\rho_{\alpha;k}^{(2),\pm}  = 0 , \qquad &&\dot{\rho}_{\alpha;k}^{(2),\pm} =
3\langle \hat{D}^\pm \hat{T}^\mp I T^\mp D^\pm,E_{\mp\alpha}\rangle. & %\label{eq:ex-coef1}
\end{alignat*}

Taking into account that $\ad_{L_1}^{-1} L_{1,x}(x) \in \mathfrak{sl}^{(0)}$ and  $\ad_{L_1}^{-1}
L_{2,x}(x) \in \mathfrak{sl}^{(1)}$ we obtain the following expansions of these functions
 over the  `squared solutions' (for details see the appendix):
\begin{gather*}
\ad_{L_1}^{-1}  L_{1,x}(x)  = \frac{i}{2\pi} \int_{-\infty }^{\infty } d \lambda
\sum_{\alpha\in\Delta_+ } (\alpha, J)  \left( s^{+}_{\alpha}(\lambda ) H_{\alpha}^+ (x, \lambda ) +
s_{\alpha}^{ -}(\lambda ) H_{-\alpha}^- (x, \lambda ) \right) \nonumber\\
\phantom{{\rm ad}_{L_1}^{-1}  L_{1,x}(x)  =}{} - \sum_{k=1}^{N}
\sum_{\alpha\in \Delta_+ } (\alpha, J)  \big(  s^{ +}_{\alpha;k} H_{\alpha;k}^+(x) -
s^{ -}_{\alpha;k} H_{-\alpha;k}^-(x)\big),%\label{eq:exp-tau1}
\\
\ad_{L_1}^{-1}  L_{1,x}(x)  = -\frac{i}{2\pi} \int_{-\infty }^{\infty } d \lambda
\sum_{\alpha\in\Delta_+ } (\alpha, J) \left( r^{+}_{\alpha}(\lambda ) H_{-\alpha}^+ (x, \lambda ) +
r_{\alpha}^{ -}(\lambda ) H_{\alpha}^- (x, \lambda ) \right) \nonumber\\
\phantom{{\rm ad}_{L_1}^{-1}  L_{1,x}(x)  =}{}  +\sum_{k=1}^{N}
\sum_{\alpha\in \Delta_+ } (\alpha, J) \big(  r^{ +}_{\alpha;k} H_{-\alpha;k}^+(x) -
r^{ -}_{\alpha;k} H_{\alpha;k}^-(x)\big).%\label{eq:exp-rho1}
\end{gather*}
Similarly:
\begin{gather*}
\ad_{L_1}^{-1}  L_{2,x}(x)  = \frac{i}{2\pi} \int_{-\infty }^{\infty } d \lambda
\sum_{\alpha\in\Delta_+ } \big( \tau^{(2)+}_{\alpha}(\lambda ) K_{\alpha}^+ (x, \lambda ) -
\tau_{\alpha}^{(2),-}(\lambda ) K_{-\alpha}^- (x, \lambda ) \big) \nonumber\\
\phantom{\ad_{L_1}^{-1}  L_{2,x}(x)  =}{} - \sum_{k=1}^{N}
\sum_{\alpha\in \Delta_+ } \big(  \tau^{(2),+}_{\alpha;k} K_{\alpha;k}^+(x) +\tau^{(2),-}_{\alpha;k} K_{-\alpha;k}^-(x)\big),%\label{eq:exp-tau2}
\\
\ad _{L_1}^{-1}  L_{2,x}(x)  = -\frac{i}{2\pi} \int_{-\infty }^{\infty } d \lambda
\sum_{\alpha\in\Delta_+ } \big( \rho^{(2)+}_{\alpha}(\lambda ) K_{-\alpha}^+ (x, \lambda ) -
\rho_{\alpha}^{(2),-}(\lambda ) K_{\alpha}^- (x, \lambda ) \big)\nonumber \\
\phantom{\ad _{L_1}^{-1}  L_{2,x}(x)  =}{} +\sum_{k=1}^{N}
\sum_{\alpha\in \Delta_+ } \left(  \rho^{(2),+}_{\alpha;k} K_{-\alpha;k}^+(x) +\rho^{(2),-}_{\alpha;k} K_{\alpha;k}^-(x)\right).%\label{eq:exp-rho2}
\end{gather*}
Here
 \begin{gather*}
\tau_\alpha^{(2),\pm}(\lambda)   =  \pm (\alpha, I) s^\pm_\alpha(\lambda) \qquad \mbox{for}\quad \alpha \in \delta_1^+, \\
\tau_{\alpha_3}^{(2),\pm}(\lambda)   = s^\pm_{\alpha_1} s^\pm_{\alpha_2} (\lambda),   \qquad \tau_{\alpha; k}^{(2),\pm}
 =\Res_{\lambda =\lambda_k^\pm} \tau_\alpha^{(2),\pm}(\lambda),\\
\rho_\alpha^{(2),\pm}(\lambda)   =  \pm (\alpha, I) r^\pm_\alpha(\lambda) \qquad \mbox{for}\quad  \alpha \in \delta_1^+, \\
\rho_{\alpha_3}^{(2),\pm}(\lambda)   = r ^\pm_{\alpha_1} r^\pm_{\alpha_2} (\lambda),   \qquad \rho_{\alpha; k}^{(2),\pm}
 =\Res_{\lambda =\lambda_k^\pm} \rho_\alpha^{(2),\pm}(\lambda)
\end{gather*}
and $\delta_1^+ =\{ \alpha_1, \alpha_2\}$.

Next, taking $F(x)= \ad_{L_1}^{-1}  \delta L_1(x)$ we derive the corresponding expansions of the
variation of~$L_1(x)$ over the squared solutions.
Thus we obtain the following expansions for $\ad_{L_1}^{-1}  \delta L_1(x)$:
\begin{gather*}
i\ad_{L_1}^{-1}  \delta L_1(x) = -\frac{1}{2\pi} \int_{-\infty }^{\infty } \frac{ d \lambda}{\lambda}
\sum_{\alpha\in\Delta_+ } \big(
H_{\alpha}^+  (x, \lambda )\delta' \tau^{(1)+}_{\alpha}(\lambda ) - H_{-\alpha}^- (x, \lambda )
\delta  \tau_{\alpha}^{(1)-}(\lambda ) \big) \nonumber\\
\phantom{i\ad_{L_1}^{-1}  \delta L_1(x) =}{} - \sum_{k=1}^{N}
\sum_{\alpha\in \Delta_+ } \left\{ \frac{ 1}{\lambda_k^+} \left[ H_{\alpha; k}^+(x) \left( \delta' \tau^{+}_{\alpha;k}
-\frac{ \tau^{+}_{\alpha;k} \delta \lambda_k^+ }{\lambda_k^+} \right) + \dot{H}_{\alpha; k}^+(x) \tau^{+}_{\alpha;k}
\delta \lambda_k^+ \right] \right. \nonumber\\
 \left.
\phantom{i\ad_{L_1}^{-1}  \delta L_1(x) =}{}
 +\frac{ 1}{\lambda_k^-} \left[ H_{-\alpha; k}^-(x) \left( \delta' \tau^{-}_{\alpha;k}
-\frac{ \tau^{-}_{\alpha;k} \delta \lambda_k^- }{\lambda_k^-} \right) + \dot{H}_{-\alpha; k}^-(x) \tau^{-}_{\alpha;k}
\delta \lambda_k^- \right] \right\},%\label{eq:exp-dtau}
\\
i\ad_{L_1}^{-1}  \delta L_1(x) = \frac{1}{2\pi} \int_{-\infty }^{\infty } \frac{ d \lambda}{\lambda}
\sum_{\alpha\in\Delta_+ } \big(
H_{-\alpha}^+  (x, \lambda )\delta' \rho^{(1)+}_{\alpha}(\lambda ) - H_{\alpha}^- (x, \lambda )
\delta  \rho_{\alpha}^{(1)-}(\lambda ) \big) \nonumber\\
\phantom{i\ad_{L_1}^{-1}  \delta L_1(x) =}{} + \sum_{k=1}^{N}
\sum_{\alpha\in \Delta_+ } \left\{ \frac{ 1}{\lambda_k^+} \left[ H_{-\alpha; k}^+(x) \left( \delta' \rho^{+}_{\alpha;k}
-\frac{ \rho^{+}_{\alpha;k} \delta \lambda_k^+ }{\lambda_k^+} \right) + \dot{H}_{-\alpha; k}^+(x) \rho^{+}_{\alpha;k}
\delta \lambda_k^+ \right] \right. \nonumber\\
 \left.
\phantom{i\ad_{L_1}^{-1}  \delta L_1(x) =}{}
+\frac{ 1}{\lambda_k^-} \left[ H_{\alpha; k}^-(x) \left( \delta' \rho^{-}_{\alpha;k}
-\frac{ \rho^{-}_{\alpha;k} \delta \lambda_k^- }{\lambda_k^-} \right) + \dot{H}_{\alpha; k}^-(x) \rho^{-}_{\alpha;k}
\delta \lambda_k^- \right] \right\}.%\label{eq:exp-drho}
\end{gather*}
The corresponding expansion coef\/f\/icients are given by:
\begin{alignat*}{3}
& \delta' \tau_\alpha^{\pm}(\lambda)   =  \delta s^\pm_\alpha(\lambda) \quad \mbox{for} \  \alpha \in \delta_1^+, \qquad &&
\delta' \tau_{\alpha_3}^{\pm}(\lambda)  = \delta s^\pm_{\alpha_3} \pm \frac{1}{2} (\delta s^\pm_{\alpha_1} s^\pm_{\alpha_2} -
s^\pm_{\alpha_1} \delta s^\pm_{\alpha_2}), &\nonumber \\
& \delta' \rho_\alpha^{\pm}(\lambda)  =  \delta r^\pm_\alpha(\lambda) \quad \mbox{for} \  \alpha \in \delta_1^+, \qquad&&
\delta' \rho_{\alpha_3}^{\pm}(\lambda)  = \delta r^\pm_{\alpha_3} \pm \frac{1}{2} (\delta r^\pm_{\alpha_1} r^\pm_{\alpha_2} -
s^\pm_{\alpha_1} \delta s^\pm_{\alpha_2}).&%\label{eq:dtau}
\end{alignat*}
The calculation of the expansion coef\/f\/icient for the discrete spectrum is explained in the appendix.

Presuming that the  variation of the potential is due to the time evolution of $L_1(x)$:
\[\delta L_1(x) = L_1(x,t+\delta t) - L_1(x,t)
\simeq \delta t \frac{\partial L_1 }{\partial t},
\]
one can get similar expansions for the time-derivative $\ad_{L_1}^{-1}  L_{1,t}(x)$:
\begin{gather*}
i\ad_{L_1}^{-1}  L_{1,t}(x)  = \frac{1}{2\pi} \int_{-\infty }^{\infty } \frac{ d \lambda}{\lambda}
\sum_{\alpha\in\Delta_+ } \left( \delta_t  \tau^{+}_{\alpha} (\lambda )
H_{-\alpha} (x, \lambda ) -  \delta_t  \tau_{\alpha}^{-} (\lambda )
H_{\alpha} (x, \lambda ) \right) \nonumber\\
\phantom{i\ad_{L_1}^{-1}  L_{1,t}(x)  =}{} - \sum_{k=1}^{N}
\sum_{\alpha\in \Delta_+ } \left\{ \frac{ 1}{\lambda_k^+} \left[ H_{\alpha; k}^+(x) \left( \delta_t \tau^{+}_{\alpha;k}
-\frac{ \tau^{+}_{\alpha;k} \lambda_{k,t}^+ }{\lambda_k^+} \right) + \dot{H}_{\alpha; k}^+(x) \tau^{+}_{\alpha;k}
 \lambda_{k,t}^+ \right] \right. \nonumber\\
\left.
\phantom{i\ad_{L_1}^{-1}  L_{1,t}(x)  =}{}
 +  \frac{ 1}{\lambda_k^-} \left[ H_{-\alpha; k}^-(x) \left( \delta_t \tau^{-}_{\alpha;k}
-\frac{ \tau^{-}_{\alpha;k}  \lambda_{k,t}^- }{\lambda_k^-} \right) + \dot{H}_{-\alpha; k}^-(x) \tau^{-}_{\alpha;k}
 \lambda_{k,t}^- \right] \right\}%\label{eq:exp-dtau'}
\end{gather*}
and
\begin{gather*}
i\ad_{L_1}^{-1}   L_{1,t}(x) = \frac{1}{2\pi} \int_{-\infty }^{\infty } \frac{ d \lambda}{\lambda}
\sum_{\alpha\in\Delta_+ } \left( \delta_t \rho^{+}_{\alpha}(\lambda )
H_{-\alpha}^+ (x, \lambda ) - \delta_t \rho_{\alpha}^{-}(\lambda )
H_{\alpha}^- (x, \lambda ) \right) \nonumber\\
\phantom{i\ad_{L_1}^{-1}   L_{1,t}(x) =}{}
+ \sum_{k=1}^{N}
\sum_{\alpha\in \Delta_+ } \left\{ \frac{ 1}{\lambda_k^+} \left[ H_{-\alpha; k}^+(x) \left( \delta_t \rho^{+}_{\alpha;k}
-\frac{ \rho^{+}_{\alpha;k} \delta \lambda_k^+ }{\lambda_k^+} \right) + \dot{H}_{-\alpha; k}^+(x) \rho^{+}_{\alpha;k}
 \lambda_{k,t}^+ \right] \right.\nonumber \\
  \left.
\phantom{i\ad_{L_1}^{-1}   L_{1,t}(x) =}{}
  +
\frac{ 1}{\lambda_k^-} \left[ H_{\alpha; k}^-(x) \left( \delta_t \rho^{-}_{\alpha;k}
-\frac{ \rho^{-}_{\alpha;k}  \lambda_{k,t}^- }{\lambda_k^-} \right) + \dot{H}_{\alpha; k}^-(x) \rho^{-}_{\alpha;k}
 \lambda_{k,t}^- \right] \right\},%\label{eq:exp-drho'}
\end{gather*}
where the expansion coef\/f\/icients $\delta_t \tau^{(1),\pm}_\alpha$ and $\delta_t \rho^{(1),\pm}_\alpha$ are given by:
\begin{gather*}
  \delta_t \tau_\alpha^{\pm}(\lambda)  =  \frac{d s^\pm_\alpha(\lambda) }{dt} \quad \mbox{for} \  \alpha \in \delta_1^+, \qquad
\delta_t \tau_{\alpha_3}^{\pm}(\lambda)  = \frac{d s^\pm_{\alpha_3}}{dt}  \pm \frac{1}{2}
\left( \frac{ ds^\pm_{\alpha_1}}{dt} s^\pm_{\alpha_2} -s^\pm_{\alpha_1} \frac{d s^\pm_{\alpha_2} }{dt} \right),   \nonumber\\
   \delta_t \rho_\alpha^{\pm}(\lambda) =  \frac{d r^\pm_\alpha(\lambda) }{dt} \quad \mbox{for} \ \alpha \in \delta_1^+, \qquad
\delta_t \rho_{\alpha_3}^{\pm}(\lambda)   = \frac{ d r^\pm_{\alpha_3}}{dt}  \pm \frac{1}{2}
\left( \frac{ d r^\pm_{\alpha_1}}{dt} r^\pm_{\alpha_2} -r^\pm_{\alpha_1} \frac{d r^\pm_{\alpha_2} }{dt} \right).  %\label{eq:dt-tau}
\end{gather*}
The details of calculating the expansion coef\/f\/icients relevant for the discrete spectrum of $L$ are given in~Appendix~\ref{appendixA}.

The completeness relation (\ref{eq:CR}) and the expansions that we constructed above can be viewed as spectral decompositions
for the recursion operators $\Lambda^\pm$ and $\tilde{ \Lambda}^\pm$ (\ref{Lambda_pm}). Indeed, one may check that
\begin{alignat}{3}
& \Lambda^{+} H_{\pm \alpha}^\pm (x, \lambda )  =\lambda^2 H_{\pm \alpha}^\pm (x, \lambda ), \!\!\qquad &&
\Lambda^{-} H_{\mp \alpha}^\pm (x, \lambda ) =\lambda^2 H_{\mp \alpha}^\pm (x, \lambda ), &\nonumber\\
& \Lambda^{+} H_{\pm \alpha;k}^\pm (x ) =(\lambda_k^\pm)^2 H_{\pm \alpha;k}^\pm (x),\!\! \qquad &&
\Lambda^{+} \dot{H}_{\pm \alpha;k}^\pm (x ) =(\lambda_k^\pm)^2 \dot{H}_{\pm \alpha;k}^\pm (x)
+2\lambda_k^\pm H_{\pm \alpha;k}^\pm (x), \!\!\!\!\!\!\! & \nonumber\\
& \Lambda^{-} H_{\mp \alpha;k}^\pm (x ) =(\lambda_k^\pm)^2 H_{\mp \alpha;k}^\pm (x), \!\! \qquad &&
\Lambda^{-} \dot{H}_{\mp \alpha;k}^\pm (x ) =(\lambda_k^\pm)^2 \dot{H}_{\mp \alpha;k}^\pm (x)
+2\lambda_k^\pm H_{\mp \alpha;k}^\pm (x) \!\!\!\!\!\!\! \label{eq:Lasq}&
\end{alignat}
and
\begin{alignat*}{3}
& \tilde{\Lambda}^{+} K_{\pm \alpha}^\pm (x, \lambda )  =\lambda^2 K_{\pm \alpha}^\pm (x, \lambda ),  \qquad &&
\tilde{\Lambda}^{-} K_{\mp \alpha}^\pm (x, \lambda ) =\lambda^2 K_{\mp \alpha}^\pm (x, \lambda ),&\nonumber \\
& \tilde{\Lambda}^{+} K_{\pm \alpha;k}^\pm (x ) =(\lambda_k^\pm)^2 K_{\pm \alpha;k}^\pm (x), \qquad&&
\tilde{\Lambda}^{+} \dot{K}_{\pm \alpha;k}^\pm (x ) =(\lambda_k^\pm)^2 \dot{K}_{\pm \alpha;k}^\pm (x)
+2\lambda_k^\pm K_{\pm \alpha;k}^\pm (x),&\nonumber \\
& \tilde{\Lambda}^{-} K_{\mp \alpha;k}^\pm (x ) =(\lambda_k^\pm)^2 K_{\mp \alpha;k}^\pm (x), \qquad&&
\tilde{\Lambda}^{-} \dot{K}_{\mp \alpha;k}^\pm (x ) =(\lambda_k^\pm)^2 \dot{K}_{\mp \alpha;k}^\pm (x)
+2\lambda_k^\pm K_{\mp \alpha;k}^\pm (x).&%\label{eq:Lasq2}
\end{alignat*}
In other words we have proved the completeness of the set of eigenfunctions of  $\Lambda_\pm$ and $\tilde{ \Lambda}_\pm$ (\ref{Lambda_pm}).

\subsection{Conjugation properties of the recursion operator}

Here we derive the  recursion operators, that are adjoint
with respect to the skew-scalar product, i.e.\ we def\/ine
$\Lambda_{1,2}^*$ by the relations:{\samepage
\begin{gather}\label{eq:La*1}
\biglb X_2, \Lambda_{1}^{\pm} X_1 \bigrb
= \biglb \left(\Lambda_{1}^{\pm}\right)^* X_2,  X_1 \bigrb   , \qquad
\biglb Y_2, \Lambda_{2}^{\pm} Y_1 \bigrb
= \biglb \left(\Lambda_{2}^{\pm}\right)^* Y_2,  Y_1 \bigrb,
\end{gather}
where $X_{1,2}=X_{1,2}^\perp\in \mathfrak{sl}^{1}(3)$
and $Y_{1,2}=Y_{1,2}^\perp\in\mathfrak{sl}^{0}(3)$.}

In doing this we use several times integration by parts and the
following properties of the operator $\ad_{L_1}^{-1}$
\begin{alignat*}{3}
& [L_1, \ad_{L_1}^{-1} X_{1,2}] = X_{1,2}  ,   \qquad &&   \langle X_1, \ad_{L_1}^{-1} X_2 \rangle
= - \langle \ad_{L_1}^{-1} X_1, X_2 \rangle, &\nonumber\\
 & [L_1, \ad_{L_1}^{-1} Y_{1,2}]= Y_{1,2}  ,   \qquad  && \langle Y_1, \ad_{L_1}^{-1} Y_2 \rangle
= - \langle \ad_{L_1}^{-1} Y_1, Y_2 \rangle .& %\label{eq:aa}
\end{alignat*}

The derivation of $\left(\Lambda_{1}^{\pm}\right)^* $ goes as follows:
 \begin{gather*}
 \biglb X_2, \Lambda_{1}^{\pm} X_1 \bigrb
= -\int_{-\infty}^{\infty}d x\,
\left\langle X_2, \left[L_1,  i \ad_{L_1}^{-1}
\frac{\partial X_1 }{\partial x} \right] \right\rangle \\
\phantom{\biglb X_2, \Lambda_{1}^{\pm} X_1 \bigrb=}{}  + \frac{i }{2} \int_{-\infty}^{\infty}  d x\,
\left\langle X_2, \left[L_1,  i \ad_{L_1}^{-1} L_{1,x}\right] \right\rangle
\int_{\pm\infty}^{x}  d y\,\left\langle L_1,\frac{\partial X_1 }{\partial y}
\right\rangle \\
\phantom{\biglb X_2, \Lambda_{1}^{\pm} X_1 \bigrb}{}  = i \int_{-\infty}^{\infty}  dx\,
\left\langle \frac{\partial X_2 }{\partial x}, X_1  \right\rangle
-\frac{i }{2} \int_{-\infty}^{\infty} dx\, \left\langle X_2,
L_{1,x} \right\rangle \int_{\pm\infty}^{x} dy\,
\left\langle X_1,L_{1,y}\right\rangle \\
\phantom{\biglb X_2, \Lambda_{1}^{\pm} X_1 \bigrb}{}
  = -i \int_{-\infty}^{\infty} {\rm d}x\,
\left\langle \ad_{L_1}^{-1}\frac{\partial X_2 }{\partial x},
[L_1, X_1]  \right\rangle \\
\phantom{\biglb X_2, \Lambda_{1}^{\pm} X_1 \bigrb=}{}
-\frac{i }{2} \int_{-\infty}^{\infty} dx\,
\left\langle \ad_{L_1}^{-1} L_{1,x}, X_1\right\rangle \left(
\int_{\pm\infty}^{x} dy\,  \left\langle L_{1,y},
\frac{\partial X_2 }{\partial y} \right\rangle \right) \\
\phantom{\biglb X_2, \Lambda_{1}^{\pm} X_1 \bigrb}{}
= \biglb \left(\Lambda_{1}^{\pm}\right)^* X_2, X_1 \bigrb .
\end{gather*}
The other recursion operators are treated analogously. Thus we obtain:
\begin{gather*}%\label{eq:Lam*3}
\left(\Lambda_{1}^{\pm}\right)^* = \Lambda_{1}^{\mp}  , \qquad   \left(\Lambda_{2}^{\pm}\right)^* = \Lambda_{2}^{\mp}.
\end{gather*}

\section{Fundamental properties of the NLEE}\label{section6}

\subsection{Integrals of motion}

In this subsection we are going to apply a method by Drinfel'd and
Sokolov \cite{DrSok*85} to derive the integrals of motion to system~(\ref{nee}). In order to do that it proves to be technically more
convenient to deal with the Lax pair (\ref{lax_1_g}), (\ref{lax_2_g}).
Then we map the operators $\tilde{L}$ and $\tilde{A}$ into
\begin{gather}
\mathcal{L}   =  \mathcal{P}^{-1}\tilde{L}\mathcal{P}
= i \partial_x + \lambda J +\mathcal{L}_0
+ \frac{\mathcal{L}_1}{\lambda} + \cdots,\label{Lambda}\\
\mathcal{A}  =  \mathcal{P}^{-1}\tilde{A}\mathcal{P}
= i \partial_t + \lambda^2 I + \lambda\mathcal{A}_{-1}
+ \mathcal{A}_0 + \frac{\mathcal{A}_1}{\lambda} + \cdots,\nonumber%\label{Mu}
\end{gather}
where all matrix coef\/f\/icients $\mathcal{L}_k$, $\mathcal{A}_{-1}$ and
$\mathcal{A}_k$, $k=0,1,\hdots$ are diagonal and make use of the following
asymptotic expansion for $\mathcal{P}(x,t,\lambda)$:
\begin{gather*}
\mathcal{P}(x,t,\lambda)= \openone +\frac{p_1(x,t)}{\lambda} +
\frac{p_2(x,t)}{\lambda^2} + \cdots. %\label{P}
\end{gather*}
In order to make all further considerations unambiguous
we assume that all coef\/f\/icients $p_l$ ($l=1,2,\ldots$) are
of\/f-diagonal matrices.

The zero curvature representation is gauge invariant, i.e.\
$[\mathcal{L}, \mathcal{A}] = 0$ is fulf\/illed. Since
$[\mathcal{L}_k, \mathcal{A}_l] =0$ the commutativity of
$\mathcal{L}$ and $\mathcal{A}$ is equivalent to the following
requirements
\begin{gather*}
\partial_x \mathcal{A}_{-1} =  0,\qquad
\partial_t\mathcal{L}_k - \partial_x \mathcal{A}_k =  0,\qquad
k=0,1,\dots.
\end{gather*}
Hence $\mathcal{L}_k$ represent densities of the integrals of motion
we are interested in.

Equality (\ref{Lambda}) rewritten as
\begin{gather*}
\tilde{L}\mathcal{P} = \mathcal{P}\mathcal{L}
%\label{split}
\end{gather*}
holds identically with respect to $\lambda$. This requirement leads
to the following set of recurrence relations:
\begin{gather}
U_0 + J p_1   = \mathcal{L}_0 + p_1 J,\nonumber\\ %\label{lambda_0}\\
i p_{1,x} + U_0 p_1  +  Jp_2 = \mathcal{L}_1 + p_1\mathcal{L}_0 + p_2 J ,
\label{lambda_m1}\\
\cdots \cdots\cdots\cdots\cdots\cdots\cdots\cdots\cdots\cdots\cdots\cdots\cdots\cdots
\nonumber\\
i p_{k,x} + U_0 p_k  +  J p_{k+1} = \mathcal{L}_k + p_{k+1}J +
\sum^{k-1}_{m=0}p_{k-m}\mathcal{L}_m ,\nonumber\\ %\label{lambda_mk}\\
\cdots \cdots\cdots\cdots\cdots\cdots\cdots\cdots\cdots\cdots\cdots\cdots\cdots\cdots
\nonumber
\end{gather}

\begin{figure}[t]
\centering
\includegraphics[width=0.48\textwidth]{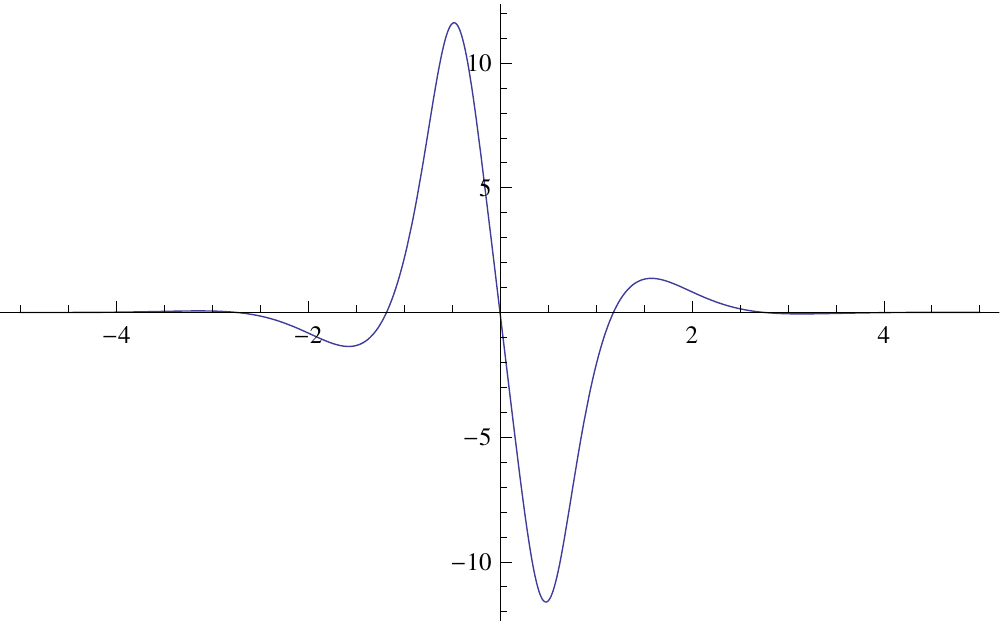} \includegraphics[width=0.48\textwidth]{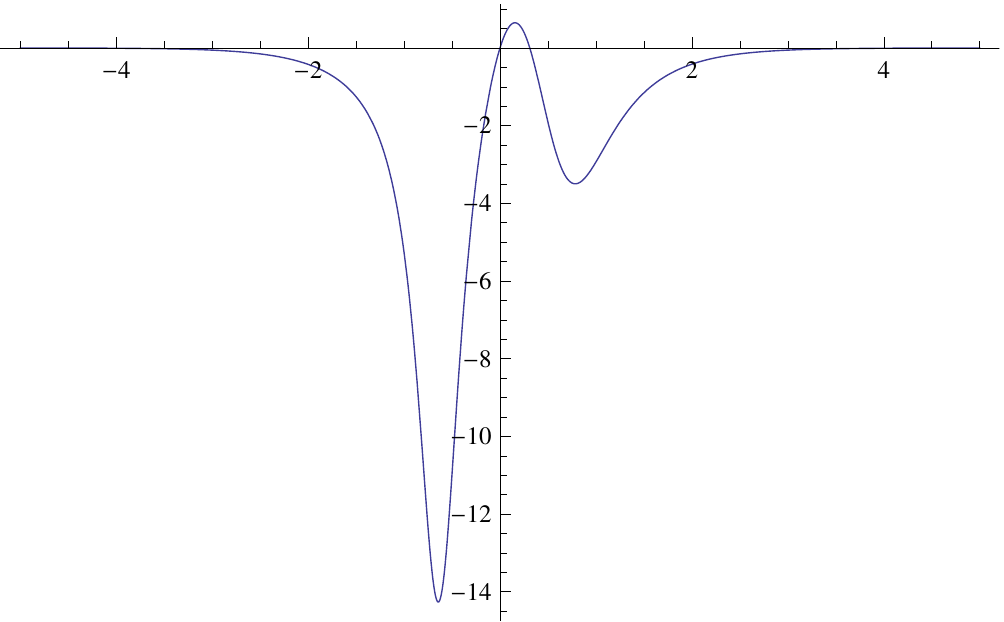}
\caption{Plots of the density of the f\/irst integral of motion as a
function of $x$  evaluated on the stationary quadruplet soliton~(\ref{solugamma0}) for $\alpha =\beta=\gamma=\omega=1$, $\delta=0$
(left panel) and for the stationary doublet soliton~(\ref{eq:kink}) for $\kappa=\sigma_0=1$, $\phi=0$ (right panel).}
\label{fig:integr}
\end{figure}

In order to solve it we apply the well-known procedure of splitting
each relation into a diagonal and of\/f-diagonal part. For example,
treating this way the f\/irst relation above one gets
\begin{gather}
\mathcal{L}_0 = U^d_0,\qquad U^{f}_0 = -[J,p_1],
\label{recursplit_1}\end{gather}
where the superscripts $d$ and $f$ above denote projection onto diagonal and
of\/f-diagonal part of a~matrix respectively. Taking into account the explicit
form of $U_0$ (see formula (\ref{u_0})) for $\mathcal{L}_0$ we have
\begin{gather*}
\mathcal{L}_0 = \frac{i}{2}(uu^*_x +vv^*_x)\left(\begin{array}{ccc}
1 & 0  & 0 \\
0 & -2 & 0 \\
0 & 0  & 1
\end{array}\right).%\label{Lambda_0}
\end{gather*}
Thus as a density of our f\/irst integral we can choose:
$\mathcal{I}_0 = u^*u_x + v^*v_x$. It represents momentum
density of our system.
For the stationary solutions~(\ref{solugamma0}) and~(\ref{eq:kink}) the momentum density is depicted on Fig.~\ref{fig:integr}. It is evidential that for both cases of solutions it is a well localised function of~$x$.
On the other hand
after inverting the commutator in the second equation
in~(\ref{recursplit_1}) one obtains
\begin{gather}
p_1 = -\ad^{-1}_J U^f_0 = -\frac{i}{2}\left(\!\begin{array}{ccc}
0  & \sqrt{2}(uv_x-vu_x) & (uu^*_x +vv^*_x)/2 \\
\sqrt{2}(u^*v^*_x-v^*u^*_x) & 0 & -\sqrt{2}(u^*v^*_x-v^*u^*_x)\\
(u^*u_x + v^*v_x)/2 & -\sqrt{2}(uv_x-vu_x) & 0
\end{array}\!\right).\!\!\!\!
\label{t_1}
\end{gather}

Similarly, for the second matrix $\mathcal{L}_1$ one needs to extract
the diagonal part of (\ref{lambda_m1}). The result reads
\begin{gather}
\big(U^f_0 p_1\big)^d = \mathcal{L}_1.
\label{recursplit_2d}
\end{gather}
After substituting the expression (\ref{t_1}) for $p_1$ into
(\ref{recursplit_2d}) one obtains{\samepage
\begin{gather*}
\mathcal{L}_1 = \frac{1}{8}\left[|uu^*_x+vv^*_x|^2+4|uv_x-vu_x|^2\right]
\left(\begin{array}{ccc}
1  & 0 & 0 \\
0 & 0 & 0 \\
0 & 0 & -1
\end{array}\right).%\label{Lambda_1}
\end{gather*}
Hence the second integral density is $\mathcal{I}_1=
|uu^*_x+vv^*_x|^2+4|uv_x-vu_x|^2$.}

In general, one is able to calculate the-$k$ integral of motion
through the formula
\begin{gather*}
\mathcal{L}_k = \big(U^f_0p_k\big)^d.
%\label{Lambda_k}
\end{gather*}
The matrix $p_k$ in its turn is obtained from the
following recursive formula
\begin{gather*}
p_k = -\ad^{-1}_J\left(i p_{k-1,x} + (U_0 p_{k-1})^f -
\sum^{k-1}_{m=0}p_{k-1-m}\mathcal{L}_m\right).
\end{gather*}

\subsection{The hierarchy of NLEE -- Lax pair approach}

Our goal here is to describe the hierarchy of integrable NLEEs
associated to the Lax opera\-tor~(\ref{lax_1}) in terms of the operators
$\Lambda_1$ and $\Lambda_2$. In order to achieve it we shall analyze
the recurrence relations obtained form the zero curvature representation
$[L,A] = 0$.

An arbitrary member of the integrable hierarchy under consideration
has a Lax pair in the form
\begin{gather*}
L(\lambda) =  i\partial_x +  \lambda L_1(x,t), \qquad
A(\lambda) =  i\partial_t   + \sum^N_{k=1}\lambda^k A_k(x,t).
\end{gather*}
As before the operators $L$ and $A$ are subject to the action of
the reductions (\ref{red1}), (\ref{red2}). Hence the coef\/f\/icients
of $A$ are hermitian matrices to fulf\/ill
\begin{gather*}%\label{red_gen}
\mathbf{C}A_{2q-1}\mathbf{C}  = -A_{2q-1}   \!\quad\Rightarrow\quad\!
A_{2q-1}  \in \mathfrak{sl}^1(3),\qquad
\mathbf{C}A_{2q}\mathbf{C} = A_{2q} \!\quad\Rightarrow\quad\!
A_{2q}  \in \mathfrak{sl}^0(3).
\end{gather*}
Thus the original Lax pair (\ref{lax_1}), (\ref{lax_2}) represents
the simplest nontrivial f\/low ($N=2$) of the above general f\/low
pair.

The compatibility condition $[L(\lambda),A(\lambda)]=0$ gives the following
set of recurrence relations:
\begin{gather}
 [L_1,A_N]=0, \label{lambda_Np1}\\
 i\partial_xA_N +   [L_1,A_{N-1}] = 0,\label{lambda_N}\\
\cdots \cdots \cdots \cdots\cdots \cdots \cdots \cdots  \nonumber\\
  i\partial_xA_k + [L_1,A_{k-1}] = 0,\qquad k=2,\ldots,N-1,\label{lambda_k}\\
\cdots \cdots \cdots \cdots\cdots \cdots \cdots \cdots \cdots \cdots\cdots \cdots \nonumber\\
  \partial_xA_1 - \partial_tL_1= 0.\label{lambda_1_gen}
\end{gather}
It directly follows from the f\/irst relation that the highest order
term is a polynomial of~$L_1$. Since $L_1\in \mathfrak{sl}^{1}(3)$
and $L_2\in \mathfrak{sl}^{0}(3)$ we have two options for $A_N$:
\begin{gather*}
{\rm a)}  \quad  A_N = f_{2p} L_2   \quad    \mbox{for} \quad N =2p, \qquad
{\rm b)}   \quad  A_N = f_{2p+1} L_1   \quad   \mbox{for} \quad N=2p+1.
\end{gather*}
It suf\/f\/ices to restrict ourselves with the case when $N=2p$ since the case $N=2p+1$
is completely analogous.

As we did many times in our exposition we shall split each element $A_k$
into two mutually orthogonal parts $A^{\bot}_{2q-1}$ and $f_{2q-1}L_{1}$
(resp. $A^{\bot}_{2q}$ and $f_{2q}L_{2}$ in the case even indices):
\begin{gather}
A_{2q-1}  = A^{\bot}_{2q-1} +  f_{2q-1} L_1,\qquad
A_{2q}  = A^{\bot}_{2q} + f_{2q}L_2.
\label{A_split}
\end{gather}
Substituting the splitting of $A_{N-1}$ into (\ref{lambda_N}) we have
\begin{gather*}
i f_{2p,x}L_{2} + i f_{2p}L_{2,x} + [L_1,A^{\bot}_{2p-1}] = 0.
\end{gather*}
After taking the Killing form $\langle \,\cdot\, ,L_2\rangle$ to separate
the $L_1$-commuting part and its orthogonal complement we deduce that
$f_N=c_N=\mbox{const}$ and
\begin{gather*}
A^{\bot}_{2p-1} = -i c_{2p} \ad^{-1}_{L_1} L_{2,x}.
\end{gather*}
Similarly, after inserting the splitting~(\ref{A_split}) into a generic recurrence
relation~(\ref{lambda_k}) we obtain
\begin{gather*}
i f_{2q-1,x} L_{1} + i  f_{2q-1} L_{1,x} +
i\big(A^{\bot}_{2q-1}\big)_x +  [L_1, A^{\bot}_{2q-2}] = 0,\\
i f_{2q,x} L_{2} +  i f_{2q}  L_{2,x} + i\big(A^{\bot}_{2q}\big)_x +  [L_1, A^{\bot}_{2q-1}] = 0.
\end{gather*}
After extracting the $L_1$-commuting part from the equations above one obtains
the coef\/f\/icient $f_{2q-1}$ (resp.~$f_{2q}$)
\begin{gather*}
f_{2q-1}  = c_{2q-1} - \frac{1}{2}\partial^{-1}_x
 \langle \big(A^{\bot}_{2q-1}\big)_x , L_{1} \rangle,\qquad
f_{2q}  = c_{2q} - \frac{3}{2} \partial^{-1}_x
 \langle \big(A^{\bot}_{2q}\big)_x , L_{2} \rangle,
%\label{f_k}
\end{gather*}
where $c_{2q-1}$ (resp.~$c_{2q}$) is a constant of integration. On the other hand for $A^{\bot}_{2q-1}$
(resp. $A^{\bot}_{2q}$) we have:
\begin{gather*}
A^{\bot}_{2q}  = \Lambda_{1} \big(A^{\bot}_{2q+1}\big)  - i  c_{2q+1} \ad^{-1}_{L_1} L_{1,x},\qquad
A^{\bot}_{2q-1}  = \Lambda_{2} \big(A^{\bot}_{2q}\big) - i  c_{2q}   \ad^{-1}_{L_1} L_{2,x},
\end{gather*}
where the integro-dif\/ferential operators $\Lambda_2$ and $\Lambda_1$ are given by (\ref{b1bot})
and (\ref{l1t_tau}) respectively.

The last recurrence relation (\ref{lambda_1_gen}) yields to
\begin{gather*}%\label{eq:f1}
f_1= c_1 - \frac{1}{2}\partial^{-1}_x \langle (A^{\bot}_{1})_x, L_1\rangle,\qquad
i\ad_{L_1}^{-1}\partial_t L_{1} + \Lambda_1 A^{\bot}_1
- i c_1 \ad^{-1}_{L_1}{L}_{1,x} = 0.
\end{gather*}
What remains is to substitute consequently the expressions for $A^{\bot}_k$, $k=1,\ldots,2p-1$
above in order to write the NLEE in terms of the operators~$\Lambda_1$ and~$\Lambda_2$. Here
we give the results for both cases~a) and~b):
\begin{gather}
{\rm a)}\quad  \ad^{-1}_{L_1}\partial_t L_1 - \sum_{q=1}^{p} c_{2q} (\Lambda_1\Lambda_2)^{q-1} \Lambda_1 \ad_{L_1}^{-1} L_{2,x}
- \sum_{q=0}^{p-1} c_{2q +1}(\Lambda_1\Lambda_2)^{q} \ad_{L_1}^{-1} L_{1,x}  =0, \nonumber\\
 {\rm b)} \quad  \ad^{-1}_{L_1}\partial_t L_1  - \sum_{q=1}^{p} c_{2q} (\Lambda_1\Lambda_2)^{q-1} \Lambda_1 \ad_{L_1}^{-1}L_{2,x}
-  \sum_{q=0}^{p} c_{2q+1} (\Lambda_1\Lambda_2)^{q}\ad_{L_1}^{-1} L_{1,x}    =0.\label{eq:nlee}
\end{gather}
The coef\/f\/icients $c_k$ are involved in dispersion laws of NLEEs. By analogy with
(\ref{f_lambda}) the dispersion law of a general NLEE is def\/ined through
\begin{gather*}%\label{f_lambda_gen}
f(\lambda)=\lim_{x\to\pm \infty} g_{\pm}^{-1}
\sum^{N}_{k=1} \lambda^k A_k(x,t) g_{\pm}.
\end{gather*}

The dispersion law governs the time evolution
of scattering matrix (\ref{tmatrix}) through the linear equation
\begin{gather}
i\partial_tT(t,\lambda)  +  [f(\lambda), T(t,\lambda)] = 0
\quad\Rightarrow\quad
T(t,\lambda)=e^{i f(\lambda)t}T(0,\lambda)e^{-i f(\lambda)t}.
\label{eq:dTt}
\end{gather}

It is not hard to check that the equalities below are valid
\begin{gather}
{\rm a)} \quad f(\lambda)   = \sum_{q=0}^{p-1} c_{2q+1} \lambda^{2q+1}J   +
\sum_{q=1}^{p} c_{2q} \lambda^{2q}I, \nonumber\\
 {\rm b)} \quad f(\lambda)   = \sum_{q=0}^{p} c_{2q+1} \lambda^{2q+1}J
 + \sum_{q=1}^{p} c_{2q} \lambda^{2q}I.
\label{eq:flam}
\end{gather}

We remind that the constant diagonal matrices $J$ and $I$ represent the diagonal forms
of the asymptotic of~$L_1$ and~$L_2$ respectively, see~(\ref{lax_1_g}),~(\ref{lax_2_g}).
As one can easily convince himself the initial NLEE~(\ref{nee}) can be derived from the above
formulae in the simplest case $N=2$ after plugging $c_2=- 1$ and $c_1= 0$.

\subsection{The hierarchy of NLEE and generalized Fourier transforms}

In the previous subsection we described the class of NLEE related to the Lax operator~$L$.
In particular we showed that if~$L_1(x,t)$ is a solution to one of the NLEE~(\ref{eq:nlee}) then
the corresponding scattering matrix $T(\lambda,t)$ satisf\/ies the linear evolution equation~(\ref{eq:dTt})
with dispersion law $f(\lambda)$ given by~(\ref{eq:flam}).

Here we will prove a more general theorem.
\begin{theorem}\label{thm:1}
Let the Lax operator $L$ be such that its potential satisfies the conditions {\bf C1)}--{\bf C3)}.
Then the NLEE \eqref{eq:nlee} are equivalent to each of the following set of linear evolution equations:
\begin{alignat}{4}
& i \frac{\partial S^+}{ \partial t } + [f(\lambda), S^+(\lambda) ]  =0,  \qquad &&
i \frac{\partial D^+ }{ \partial t}  =0,  \qquad && i \frac{\partial S^-}{ \partial t } + [f(\lambda), S^-(\lambda) ]  =0,& \nonumber\\
& i \frac{\partial s^+_k}{ \partial t } + [f(\lambda_k^+), s^+_k  ]  =0,  \qquad && i \frac{\partial \lambda_k^\pm }{ \partial t } =0,
 \qquad && i \frac{\partial s^-_k}{ \partial t } + [f(\lambda_k^-), s^-_k  ]  =0;& \label{eq:dTdt1}
\\
& i \frac{\partial T^-}{ \partial t } + [f(\lambda), T^-(\lambda) ] =0, \qquad &&
i \frac{\partial D^- }{ \partial t} =0, \qquad && i \frac{\partial T^+}{ \partial t } + [f(\lambda), T^+(\lambda) ] =0,& \nonumber\\
& i \frac{\partial r^-_k}{ \partial t } + [f(\lambda_k^+), r^-_k  ] =0, \qquad && i \frac{\partial \lambda_k^\pm }{ \partial t }=0,
\qquad  && i \frac{\partial r^+_k}{ \partial t } + [f(\lambda_k^-), r^+_k  ] =0. & \label{eq:dTdt2}
\end{alignat}
\end{theorem}

\begin{proof}
Let us f\/irst consider the NLEE (\ref{eq:nlee}a) and let us denote its left hand side by~$N_{\rm a)}(x,t)$.
Next let us expand it over the complete set of eigenfunctions~$H^\pm_{\pm \alpha}(x,\lambda)$. If we make
use of the expansion~(\ref{eq:compl}) then we need to calculate the expansion coef\/f\/icients~(\ref{eq:gammas}).
Using the properties of the recursion operators~(\ref{eq:Lasq}) and~(\ref{eq:La*1}) we have:
\begin{gather}
\gamma^\pm_{N_{\rm a) }} (\lambda)  = \biglb e^\pm_{\mp\alpha}(y,\lambda), \ad_{L_1}^{-1} N_{\rm a)}(y)\bigrb \nonumber\\
\phantom{\gamma^\pm_{N_{\rm a) }} (\lambda)}{}
= \biglb H^\pm_{\mp\alpha}(y,\lambda),  L_{1,t}\bigrb
- \sum_{q=1}^{p} c_{2q} \biglb K^\pm_{\mp\alpha}(y,\lambda), (\Lambda^{-,\ast}) ^{q-1} \Lambda_{1}^- \ad_{L_1}^{-1} L_{2,x} \bigrb \nonumber\\
\phantom{\gamma^\pm_{N_{\rm a) }} (\lambda)=}{}
- \sum_{q=0}^{p-1} c_{2q+1} \biglb K^\pm_{\mp\alpha}(y,\lambda), (\Lambda^{-,\ast}) ^{q}  \ad_{L_1}^{-1} L_{1,x} \bigrb
\nonumber\\
\phantom{\gamma^\pm_{N_{\rm a) }} (\lambda)}{}
= \frac{ i}{\lambda } \delta_t \tau_\alpha^\pm(\lambda) +
\sum_{q=1}^{p} c_{2q} \biglb \Lambda_{1}^+(\Lambda ^+) ^{q-1} K^\pm_{\mp\alpha}(y,\lambda),   \ad_{L_1}^{-1} L_{2,x} \bigrb
\nonumber\\
\phantom{\gamma^\pm_{N_{\rm a) }} (\lambda)=}{}
+ \sum_{q=0}^{p-1} c_{2q+1} \biglb (\Lambda ^+) ^{q} K^\pm_{\mp\alpha}(y,\lambda),   \ad_{L_1}^{-1} L_{1,x} \bigrb \nonumber\\
\phantom{\gamma^\pm_{N_{\rm a) }} (\lambda)}{}
= \frac{ i}{\lambda } \delta_t \tau_\alpha^\pm(\lambda) +
\sum_{q=1}^{p} c_{2q}\lambda^{2q-1} \biglb H^\pm_{\mp\alpha}(y,\lambda),   \ad_{L_1}^{-1} L_{2,x} \bigrb \nonumber\\
\phantom{\gamma^\pm_{N_{\rm a) }} (\lambda)=}{}
+ \sum_{q=0}^{p-1} c_{2q+1} \lambda^{2q+1} \biglb  K^\pm_{\mp\alpha}(y,\lambda),   \ad_{L_1}^{-1} L_{1,x} \bigrb \nonumber\\
\phantom{\gamma^\pm_{N_{\rm a) }} (\lambda)}{}
= \frac{1}{\lambda} \left( i\left\langle \hat{S}^\pm \frac{\partial S^\pm}{ \partial t }, E_{\mp\alpha}\right\rangle
+ \langle \hat{S}^\pm f(\lambda) S^\pm ( \lambda), E_{\mp\alpha}\rangle \right)
=0 \quad \mbox{for all} \quad \alpha \in\Delta_+ .\label{eq:prf1}
\end{gather}
From equation (\ref{eq:prf1}) we f\/ind that
\begin{gather}\label{eq:prf2}
i \hat{S}^\pm \frac{\partial S^\pm}{ \partial t } + \hat{S}^+ f(\lambda) S^\pm ( \lambda) = \mathcal{H}(\lambda),
 \end{gather}
where $\mathcal{H}(\lambda) \in \mathfrak{h}$. It remains to take the Killing form of~(\ref{eq:prf2}) with the Cartan
elements~$J$ and~$I$ of $\mathfrak{sl}$ to determine that~$\mathcal{H}(\lambda) = f(\lambda)$. Thus we have
proved that from the NLEE one gets the f\/irst of the equations in (\ref{eq:dTdt1}):
\begin{gather*}%\label{eq:dSpmt}
i  \frac{\partial S^\pm}{ \partial t } + [f(\lambda), S^\pm( \lambda)] =0 .
 \end{gather*}

Next we rewrite equation~(\ref{eq:T}) in the form:
\begin{gather*}%\label{eq:Dpm}
 \hat{ D}^+ \hat{T}^- T^+ D^-(\lambda) = \hat{S}^+S^-(\lambda),
 \end{gather*}
and equate the principle upper- and lower-minors of both sides. As a result we get:
\begin{gather}
\frac{1}{m_1^+ m_2^-}  = 1 - s^+_{\alpha_1}s^-_{\alpha_1} -s^+_{\alpha_3}s^-_{\alpha_3} +
\frac{1}{2} (s^+_{\alpha_1}s^+_{\alpha_2}s^-_{\alpha_3} -s^-_{\alpha_1}s^-_{\alpha_2}s^+_{\alpha_3})
+ \frac{1}{4} s^+_{\alpha_1}s^-_{\alpha_1}s^+_{\alpha_2}s^-_{\alpha_2}, \nonumber\\
\frac{1}{m_2^+ m_1^-}  = 1 - s^+_{\alpha_2}s^-_{\alpha_2} -s^+_{\alpha_3}s^-_{\alpha_3} -
\frac{1}{2} (s^+_{\alpha_1}s^+_{\alpha_2}s^-_{\alpha_3} -s^-_{\alpha_1}s^-_{\alpha_2}s^+_{\alpha_3})
+ \frac{1}{4} s^+_{\alpha_1}s^-_{\alpha_1}s^+_{\alpha_2}s^-_{\alpha_2} \label{eq:disp1}
\end{gather}
and
\begin{gather}
\frac{1}{m_2^+ m_1^-}  = 1 - r^+_{\alpha_2}r^-_{\alpha_2} -r^+_{\alpha_3}r^-_{\alpha_3} -
\frac{1}{2} (r^+_{\alpha_1}r^+_{\alpha_2}r^-_{\alpha_3} -r^-_{\alpha_1}r^-_{\alpha_2}r^+_{\alpha_3})
+ \frac{1}{4} r^+_{\alpha_1}r^-_{\alpha_1}r^+_{\alpha_2}r^-_{\alpha_2},\nonumber\\
\frac{1}{m_1^+ m_2^-}  = 1 - r^+_{\alpha_1}r^-_{\alpha_1} -r^+_{\alpha_3}r^-_{\alpha_3} +
\frac{1}{2} (r^+_{\alpha_1}r^+_{\alpha_2}r^-_{\alpha_3} -r^-_{\alpha_1}r^-_{\alpha_2}r^+_{\alpha_3})
+ \frac{1}{4} r^+_{\alpha_1}r^-_{\alpha_1}r^+_{\alpha_2}r^-_{\alpha_2} \label{eq:disp2}
\end{gather}
for $\lambda\in\bbbr$.
We remind that the scalar functions $m_1^+(\lambda)$ and $m_2^+(\lambda)$ (resp.~$m_1^-(\lambda)$ and $m_2^-(\lambda)$)
are analytic functions for $\lambda \in \bbbc_+$ (resp.~$\lambda \in \bbbc_-$). Then the relations~(\ref{eq:disp1})
can be viewed as a~RHP which can be solved by the Plemelj--Sokhotsky formulae. Thus equations~(\ref{eq:disp1})
allow us to recover $m_1^\pm(\lambda)$ and $m_2^\pm(\lambda)$, and ef\/fectively $D^\pm(\lambda)$ in their whole
regions of analyticity from~$S^\pm(\lambda)$. Likewise, equations~(\ref{eq:disp2})
allow us to recover  $D^\pm(\lambda)$ from $T^\pm(\lambda)$.

From equations~(\ref{eq:disp1}) there also follows that
\begin{gather*}%\label{eq:dDpm}
\frac{ dm_k^\pm }{dt} =0, \qquad k=1,2, \qquad \mbox{i.e.} \qquad \frac{ dD^\pm}{dt} =0.
 \end{gather*}

We can also reconstruct $T^\pm(\lambda)$ as the Gauss factors of  $ D^+ \hat{S}^+ S^- \hat{ D}^-(\lambda) $
and check that
\begin{gather*}%\label{eq:t_pm}
i\partial_t T^\pm(t,\lambda) + [f(\lambda), T^\pm(t,\lambda)]=0.
 \end{gather*}
Thus we have proved the statement of the theorem for the scattering data on the continuous spectrum of $L$.
Similar procedure allows us to recover the equations for the data on the discrete spectrum of $L$.

In order to complete the proof of the equivalence between the NLEE (\ref{eq:nlee}) and (\ref{eq:dTdt1}) it
remains to apply Corollary~\ref{cor:1}.

The equivalence of the NLEE (\ref{eq:nlee}) and (\ref{eq:dTdt2}) is proved along the same lines
analyzing the expansion coef\/f\/icients $\theta^\pm_{N_{\rm a) }} (\lambda)$.
\end{proof}

As an immediate consequence of the theorem we obtain:
\begin{corollary}\label{cor:2}
Each of the minimal sets of scattering data $\mathcal{S}_k$, $k=1,2$:
\begin{gather*}
\mathcal{S}_1  = \big\{ s_\alpha^+(\lambda), s_\alpha^-(\lambda), \  \lambda\in\bbbr, \
s_{\alpha;k}^+, s_{\alpha;k}^-, \lambda_k^\pm, \  \alpha \in \Delta_+,\  k=1,\dots , 2N_1+N_2 \big\}, \nonumber\\
\mathcal{S}_2  = \big\{ r_\alpha^+(\lambda), r_\alpha^-(\lambda), \  \lambda\in\bbbr, \
r_{\alpha;k}^+, r_{\alpha;k}^-, \lambda_k^\pm, \  \alpha \in \Delta_+, \  k=1,\dots , 2N_1+N_2 \big\},%\label{minsets}
\end{gather*}
determines uniquely both the scattering matrix $T(\lambda)$ and the potential $L_1(x)$ of the Lax
opera\-tor~$L$.
\end{corollary}

\begin{proof}
The fact that each of the sets $\mathcal{S}_k$ determine the scattering matrix $T(\lambda)$
follows easily from the Theorem. Indeed, we showed how, starting from  $\mathcal{S}_k$ one can construct
each of the Gauss factors of~$T(\lambda)$; so it remains just to take the corresponding products.

In order to reconstruct the corresponding potential we have to solve the RHP with canonical normalization
for $\lambda =0$.  Considering the Taylor expansion of $\chi^\pm (x,\lambda)$:
\begin{gather*}%\label{chi_as}
\chi^\pm(x,\lambda) = \openone + \sum_{n=1}^{\infty} \lambda^s \chi^\pm_{(s)}(x),
\end{gather*}
we get
\begin{gather*}%\label{eq:L1}
L_1(x) = -i \frac{d\chi^\pm_{(1)}(x) }{dx}.\tag*{\qed}
\end{gather*}
\renewcommand{\qed}{}
\end{proof}

\subsection{Hierarchy of Hamiltonian formulations}

Our analysis here is again based on the Wronskian relations, extending the
ones in Section~\ref{section4}. We f\/irst use slightly modif\/ied equation~(\ref{eq:Wrd2}):
\begin{gather*}
 \left\langle \hat{\chi}^\pm \delta \chi^\pm (x,\lambda),J \right\rangle \big|_{-\infty}^{\infty}
= i \lambda  \int_{-\infty}^{\infty}  d x\, \left\langle  \delta L_1 , h_J^\pm (x,\lambda)\right\rangle , \nonumber\\
  \left\langle \hat{\chi}^\pm \delta \chi^\pm (x,\lambda),I \right\rangle \big|_{-\infty}^{\infty}
= i \lambda  \int_{-\infty}^{\infty}  d x\, \left\langle  \delta L_1 , h_I^\pm (x,\lambda)\right\rangle ,%\label{eq:Wrd3}
\end{gather*}
where
\begin{gather*}%\label{eq:h_J}
  h_J^\pm (x,\lambda)=  \chi^\pm (x,\lambda) J\hat{\chi}^\pm (x,\lambda) , \qquad
 h_I^\pm (x,\lambda)=  \chi^\pm (x,\lambda) I\hat{\chi}^\pm (x,\lambda).
\end{gather*}

A third class of Wronskian relations connects the variation of the derivative
$\dot{\chi}(x,\lambda)\equiv \partial_{\lambda}\chi(x,\lambda)$ with~$L_1$, i.e.
\begin{gather*}
  \left\langle \hat{\chi}^\pm \dot{\chi}^\pm (x,\lambda),J \right\rangle \big|_{-\infty}^{\infty}
= -  \int_{-\infty}^{\infty}  d x\, \left\langle L_1 , h_J^\pm (x,\lambda)\right\rangle , \nonumber\\
  \left\langle \hat{\chi}^\pm \dot{\chi}^\pm (x,\lambda),I \right\rangle \big|_{-\infty}^{\infty}
= -  \int_{-\infty}^{\infty}  d x\, \left\langle L_1 , h_I^\pm (x,\lambda)\right\rangle .%\label{eq:Wrd4}
\end{gather*}

Thus we derive the interrelations between the generating functionals of integrals of motion and
the corresponding potential:
\begin{gather*}
\langle \hat{D}^\pm \delta D^\pm , J\rangle  \equiv  \pm \delta \ln (m_1^\pm m_2^\pm) =
-i\lambda \biglb \ad_{L_1}^{-1}\delta L_1, h_J^\pm(x,\lambda)\bigrb , \nonumber\\
\langle \hat{D}^\pm \delta D^\pm , I\rangle  \equiv  \delta \ln \frac{m_1^\pm}{m_2^\pm} =
-i\lambda \biglb \ad_{L_1}^{-1} \delta L_1, h_I^\pm(x,\lambda)\bigrb .%\label{eq:dD0}
\end{gather*}

Note that the second reduction in equation~(\ref{jostred2}) applied to $h_J^\pm(x,\lambda)$ and $h_I^\pm(x,\lambda)$
gives
\begin{gather*}%\label{eq:h-red}
 \mathbf{C}h_J^\pm(x,-\lambda)\mathbf{C} = -h_J^\pm(x,\lambda), \qquad  \mathbf{C}h_I^\pm(x,-\lambda)\mathbf{C} = h_I^\pm(x,\lambda),
 \end{gather*}
which means that
\begin{gather*}%\label{eq:IMC}
\pm \ln (m_1^\pm m_2^\pm)  = \sum_{s=1}^{\infty} \lambda^{-2s-1} C_{J;2s+1}, \qquad
  \ln \frac{m_1^\pm}{m_2^\pm}  = \sum_{s=1}^{\infty} \lambda^{-2s} C_{I;2s},
 \end{gather*}
where $C_{J;2s+1}$ and $C_{I;2s}$ are integrals of motion.

Let us note that neither $h_J^\pm(x,\lambda)$ nor $h_I^\pm(x,\lambda)$ are eigenfunctions of the recursion operators.
Indeed, if we separate the `orthogonal' and `parallel' to~$L_1$ and $L_2$ parts and split them into:
\begin{gather*}%\label{eq:hH}
h_J^{\pm,\perp}(x,\lambda) = H_J^\pm(x,\lambda) + K_J^\pm(x,\lambda), \qquad
h_I^{\pm,\perp}(x,\lambda) = H_I^\pm(x,\lambda) + K_I^\pm(x,\lambda),
 \end{gather*}
we get:
\begin{alignat*}{3}
& \Lambda_1^\pm K_J^{\pm} (x,\lambda) = \lambda H_J^{\pm} (x,\lambda) + i\ad_{L_1}^{-1} L_{1,x}, \qquad&&
\Lambda_2^\pm H_J^{\pm} (x,\lambda)  = \lambda K_J^{\pm} (x,\lambda),& \\ %\label{eq:LamH}
& \Lambda_1^\pm K_I^{\pm} (x,\lambda) = \lambda H_I^{\pm} (x,\lambda),  \qquad&&
\Lambda_2^\pm H_I^{\pm} (x,\lambda) = \lambda K_I^{\pm} (x,\lambda)+ i\ad_{L_1}^{-1} L_{2,x}.&\nonumber
\end{alignat*}
Therefore
\begin{alignat*}{3}
& (\Lambda_2^\pm \Lambda_1^\pm -\lambda^2 )K_J^{\pm} (x,\lambda)   = i \Lambda_2^\pm \ad_{L_1}^{-1} L_{1,x}, \qquad&&
(\Lambda_1^\pm \Lambda_2^\pm -\lambda^2 ) H_J^{\pm} (x,\lambda)  = i \lambda  \ad_{L_1}^{-1} L_{1,x}, & \nonumber\\
& (\Lambda_2^\pm \Lambda_1^\pm -\lambda^2) K_I^{\pm} (x,\lambda)  = i \lambda \ad_{L_1}^{-1} L_{2,x}, \qquad &&
(\Lambda_1^\pm \Lambda_2^\pm -\lambda^2) H_I^{\pm} (x,\lambda)  = i \Lambda_1^\pm \ad_{L_1}^{-1} L_{1,x} & %\label{eq:lamH3}
\end{alignat*}
and
\begin{gather*}
\pm \delta \ln (m_1^\pm m_2^\pm)  =\sum_{s=0}^{\infty} \lambda^{-2s-1} \delta C_{J,2s+1} =
\lambda \biglb \ad_{L_1}^{-1}\delta L_1,  (\Lambda_2\Lambda_1 -\lambda^2)^{-1} \Lambda_2  \ad_{L_1}^{-1}  L_{1,x}\bigrb,\nonumber \\
\delta \ln \frac{m_1^\pm  }{m_2^\pm}  =\sum_{s=0}^{\infty} \lambda^{-2s} \delta C_{I,2s} =
\lambda^2 \biglb \ad_{L_1}^{-1}\delta L_1,  (\Lambda_2\Lambda_1 -\lambda^2)^{-1}  \ad_{L_1}^{-1}  L_{2,x} \bigrb ,%\label{m_12_exp}
\end{gather*}
or in other words
\begin{gather}
\delta C_{J,2s+1}  = -\biglb \ad_{L_1}^{-1}\delta L_1,  (\Lambda_2\Lambda_1)^{s} \Lambda_2  \ad_{L_1}^{-1}\delta L_{1,x} \bigrb, \nonumber\\
\delta C_{I,2s}  = -\biglb \ad_{L_1}^{-1}\delta L_1,  (\Lambda_2\Lambda_1)^{s}   \ad_{L_1}^{-1}\delta L_{2,x}\bigrb.\label{eq:dCs}
\end{gather}
We can rewrite equations~(\ref{eq:dCs}) in the form:
\begin{gather*}%\label{eq:nabla}
 \frac{ \delta C_{J,2s+1} }{\delta L_1(x)} = (\Lambda_2\Lambda_1)^{s} \Lambda_2  \ad_{L_1}^{-1} L_{1,x}, \qquad
  \frac{ \delta C_{I,2s} }{\delta L_1(x)} = (\Lambda_2\Lambda_1)^{s}   \ad_{L_1}^{-1}  L_{1,x}.
 \end{gather*}
Thus in fact we have proved that the conserved quantities of these NLEE satisfy the well known Lenart relation
\begin{gather}\label{eq:Len}
 \frac{ \delta C_{J,2s+1} }{\delta L_1(x)} = \Lambda_2\Lambda_1  \frac{ \delta C_{J,2s-1} }{\delta L_1(x)}, \qquad
 \frac{ \delta C_{I,2s} }{\delta L_1(x)} = \Lambda_2\Lambda_1  \frac{ \delta C_{I,2s-2} }{\delta L_1(x)}.
 \end{gather}

We f\/inish by formulating the Hamiltonian properties of these NLEE.
It is only natural that the Hamiltonians must be linear combinations of  integrals of motion:
\begin{gather*}
H_{\rm a)}  = \sum_{q=0}^{p-1} c_{2q+1} C_{J,2q+1} + \sum_{q=0}^{p} c_{2q} C_{I,2q} , \qquad
H_{\rm b)}  = \sum_{q=0}^{p} c_{2q+1} C_{J,2q+1} + \sum_{q=0}^{p} c_{2q} C_{I,2q} .%\label{eq:Hab0}
\end{gather*}

Next we introduce a symplectic structure using the  symplectic form:
\begin{gather*}%\label{eq:Ome1}
\Omega_{1} = \biglb \ad_{L_1}^{-1}\delta L_1 \wedgecomma  \Lambda_2  \ad_{L_1}^{-1}\delta L_1 \bigrb.
\end{gather*}
Using (\ref{eq:Len}) we f\/ind that the equation of motion (\ref{eq:nlee}) can be written down as
\begin{gather}\label{eq:OmeV}
\Omega_{1}^\vee \equiv \biglb \ad_{L_1}^{-1}\delta L_1 \cdot \Lambda_2  \ad_{L_1}^{-1} L_{1,t} \bigrb = \delta H,
 \end{gather}
with $H= H_{\rm a)}$ or  $H= H_{\rm b)}$ respectively.

In particular, choosing $H=C_{I,2}$ we f\/ind that  equation (\ref{eq:OmeV}) becomes
\begin{gather*}%\label{eq:Ham}
\Omega_{1} \big(\,\cdot\,  ,   \ad_{L_1}^{-1}\delta L_{1,t}\big) = \delta C_{I,2}
 \end{gather*}
and coincides with the reduced HF equation (\ref{nee}) we started with.

Along with $\Omega_{1}$ we can also introduce the following hierarchy of symplectic forms:
\begin{gather}\label{eq:Ome0}
\Omega_{2k-1} = \biglb \ad_{L_1}^{-1}\delta L_1 \wedgecomma (\Lambda_2 \Lambda_1)^{k-1} \Lambda_2  \ad_{L_1}^{-1}\delta L_1 \bigrb .
 \end{gather}
Then we can write the equation of motion (\ref{eq:nlee}) in Hamiltonian form
using each one of the above 2-forms as follows:
\begin{gather*}%\label{eq:Heq}
\Omega_{2k-1}^\vee  \equiv \biglb \ad_{L_1}^{-1}\delta L_1 (\Lambda_2 \Lambda_1)^{k-1} \Lambda_2  \ad_{L_1}^{-1} L_{1,t} \bigrb
 = \delta H_{2k}
 \end{gather*}
with $H= H_{2k; {\rm a)}}$ or  $H= H_{2k; {\rm b)}}$ respectively, where
\begin{gather*}
H_{2k; {\rm a)}}  = \sum_{q=0}^{p-1} c_{2q+2k+1} C_{J,2q+1} + \sum_{q=0}^{p} c_{2q} C_{I,2q+2k} ,\nonumber \\
H_{2k; {\rm b)}}  = \sum_{q=0}^{p} c_{2q+1} C_{J,2q+2k+1} + \sum_{q=0}^{p} c_{2q} C_{I,2q+2k} .%\label{eq:Hab}
\end{gather*}

Using the expansions of $\ad_{L_1}^{-1}\delta L_1$ over the `squared solutions'
\begin{gather*}
\Omega_{2k-1} = \biglb \ad_{L_1}^{-1}\delta L_1 \wedgecomma (\Lambda_2 \Lambda_1)^{k-1} \Lambda_2  \ad_{L_1}^{-1}\delta L_1 \bigrb
= \frac{1}{2\pi} \int_{ -\infty}^{\infty} d\lambda  \lambda^{2k-1} (\Omega_0^+(\lambda) -\Omega_0^-(\lambda) ) \nonumber\\
\phantom{\Omega_{2k-1} =}{}
+ \sum_{k=1}^{2N_1+N_2} \left( \Res_{\lambda =\lambda_k^+} \lambda^{2k-1}\Omega_0^+(\lambda) +
\Res_{\lambda =\lambda_k^-} \lambda^{2k} \Omega_0^-(\lambda) \right),%\label{eq:Omek}
\end{gather*}
where
\begin{gather*}%\label{eq:Ome3}
\Omega_0^+(\lambda) = \sum_{ \alpha\in \Delta_+}^{} \delta' \rho_\alpha^\pm (\lambda) \wedge \delta' \tau_\alpha^\pm (\lambda) .
 \end{gather*}

Thus we conclude that the family of  symplectic forms (\ref{eq:Ome0}) are dynamically compatible.

\section{Discussion and conclusions}\label{section7}

A system of coupled equations, which generalize Heisenberg
ferromagnet equations has been studied. The system
is associated with a polynomial bundle Lax operator $L$
related to the symmetric space $SU(3)/S(U(1)\times U(2))$.

The spectral properties of the operator $L$ in the case of
the simplest constant boundary condition (\ref{const_bc})
have been described. The continuous spectrum of $L$ f\/ills up the
real axis in the complex $\lambda$-plane and divides it
 into two regions: the upper half plane $\bbbc_{+}$
and the lower half plane $\bbbc_{-}$. Each region is an analyticity domain
of a fundamental analytic solution to the auxiliary linear problem.

 By using the  dressing method \cite{brown-bible,zakharovshabat,zm1,zm2} we have constructed 1-soliton solutions of  the ($\bbbz_2$-HF) models.
Due to the additional symmetries the Lax operator $L$ may possess two types of discrete eigenvalues:
generic ones and purely imaginary ones. Therefore we will have two types of soliton solutions -- quadruplet and doublet solitons respectively. We outlined the purely algebraic construction for deriving the $N$-soliton solutions for both types of solitons.

Using the Wronskian relations one is able to construct `squared
solutions' and an integro-dif\/ferential operator called recursion
operator whose eigenfunctions they are. There exists another
viewpoint on recursion operator~-- they generate hierarchy of
symmetries of NLEEs. Thus one can derive the recursion operator
of a NLEE from purely symmetry considerations.

Using the interpretation of the ISM as a
generalized Fourier transform and the expansions over the `squared solutions'  we studied the fundamental properties of the class of NLEE admitting a~Lax operator of the form (\ref{lax_1}), including   the description of the whole class of NLEE, the inf\/inite set of integrals of motion and the Hamiltonian formulation of the corresponding hierarchy.

The results of the present article can be extended in several directions.
Firstly, one can develop the theory in the case of a rational bundle $L$:
\[
L= i \partial_x + \lambda L_1 + \frac{1}{\lambda} L_{-1}.
\]
The simplest NLEEs related to Lax pair of this type take the form
\begin{gather}
i u_t =  u_{xx}-(u(u^*u_x+v^*v_x))_x+8 vv^*u,\nonumber\\
i v_t = v_{xx}-(v(u^*u_x+v^*v_x))_x-8uu^*v  ,\label{nee2}
\end{gather}
where $u$ and $v$ are functions of $x$ and $t$ subject to the same condition as
in the polynomial case, i.e.~$|u|^2 + |v|^2=1$. The system~(\ref{nee2})  can also be seen as an anisotropic
deformation of (\ref{nee}) with $k=1$, $N=3$ and is a special case of the models proposed in~\cite{golsok}.

That modif\/ication is required when one imposes an additional
$\bbbz_2$ reduction of the form $\lambda\to 1/\lambda$, see~\cite{ours}.
This case is more complicated and much richer than the one we have
studied. It requires the construction of automorphic Lie algebras
and studying their properties following the ideas of~\cite{lomsan,miklom,wang}.

The second direction of generalization concerns considering Lax operator $L$ related to a~ge\-ne\-ric symmetric space
of the type $\mathbf{A.III}\cong SU(n+k)/S(U(n)\times U(k))$
or, more generally, related to other types of symmetric spaces.
This will allow one to treat various multi-component generalizations of our NLEEs.
Geometric properties of similar models related to other symmetric
spaces are studied in~\cite{wu}.

\appendix

\section{Parametrization of Gauss factors}\label{appendixA}

Here we will list useful formulae allowing one, given the scattering matrix
$T(\lambda)$ to evaluate its Gauss factors $S^\pm(\lambda)$, $T^\pm(\lambda)$  or rather
 $s^\pm(\lambda)$, $r^\pm(\lambda)$ %in equation~(\ref{gauss_par_a}).
\begin{alignat*}{3}
& S^\pm (\lambda) =\exp s^\pm(\lambda) , \qquad && T^\pm (\lambda) =\exp r^\pm(\lambda) ,&\nonumber\\
& s^+ (\lambda) = \left(\begin{array}{ccc} 0 & s^+_{\alpha_1} &  s^+_{\alpha_3} \\ 0 & 0 &  s^+_{\alpha_2} \\
0 & 0 & 0  \end{array}\right), \qquad  &&r^+ (\lambda) = \left(\begin{array}{ccc} 0 & r^+_{\alpha_1} &  r^+_{\alpha_3} \\ 0 & 0 &
r^+_{\alpha_2} \\ 0 & 0 & 0  \end{array}\right),& \nonumber\\
& s^- (\lambda) = \left(\begin{array}{ccc} 0 & 0 & 0 \\ s^-_{\alpha_1} & 0 &  0 \\ s^-_{\alpha_3} & s^-_{\alpha_2} & 0  \end{array}\right),
\qquad && r^- (\lambda) = \left(\begin{array}{ccc} 0 & 0 &  0 \\ r^-_{\alpha_1} & 0 &  \\ r^-_{\alpha_3} & r^-_{\alpha_2} & 0
\end{array}\right), &\nonumber \\
& D^+(\lambda) = \diag (m_{1}^+, m_{2}^+/m_{1}^+, 1/m_{2}^+), \qquad  && D^-(\lambda) = \diag (1/m_{2}^-, m_{2}^-/m_{1}^-, m_{1}^-).& %\label{gauss_par_a}
\end{alignat*}

Next we insert these formulae in the left hand sides of the Wronskian relations (\ref{eq:rho}), (\ref{eq:tau}) and (\ref{eq:ssp-3}).
Skipping the details we f\/ind
\begin{gather}
s^+_{\alpha_1}   = - \frac{T_{12}}{m_1^+},   \qquad s^+_{\alpha_2}   = - \frac{T_{13}T_{21}-T_{23}T_{11} }{m_2^+},  \qquad
s^+_{\alpha_3}   = \frac{T_{12}T_{23}-T_{22}T_{13} }{2m_2^+} - \frac{ T_{13}}{2m_1^+} ,\nonumber\\
r^-_{\alpha_1}   =  \frac{T_{21}}{m_1^+},   \qquad r^-_{\alpha_2}   =  \frac{T_{11}T_{32}-T_{12}T_{31} }{m_2^+},  \qquad
r^-_{\alpha_3}   = -\frac{T_{21}T_{32}-T_{22}T_{31} }{2m_2^+} + \frac{ T_{31}}{2m_1^+} ,\nonumber\\
s^-_{\alpha_2}   =  -\frac{T_{32}}{m_1^-},   \qquad s^-_{\alpha_1}   =  \frac{T_{23}T_{31}-T_{21}T_{33} }{m_2^-},  \qquad
s^-_{\alpha_3}   = \frac{T_{33}T_{21}-T_{22}T_{31} }{2m_2^-} - \frac{ T_{31}}{2m_1^-} ,\nonumber\\
r^+_{\alpha_2}   =  \frac{T_{23} }{m_1^-},  \qquad r^+_{\alpha_1}   = - \frac{T_{12} T_{33} -T_{13} T_{32}}{m_2^-},   \qquad
r^+_{\alpha_3}   = \frac{T_{13}T_{22}-T_{23}T_{11} }{2m_2^-} + \frac{ T_{13}}{2m_1^-} ;\nonumber %\label{eq:Sp}
\\
\langle \hat{S}^+ J S^+(\lambda) ,E_{\alpha}\rangle = (\alpha, J) s^+_\alpha(\lambda), \nonumber\\
\langle \hat{S}^+ I S^+(\lambda) ,E_{\alpha}\rangle = (\alpha, I) s^+_\alpha(\lambda) \quad \mbox{for} \quad \alpha = e_1-e_2, e_2-e_3, \nonumber\\
\langle \hat{S}^+ I S^+(\lambda) ,E_{e_1-e_3}\rangle  = s^+_{12}s^+_{23};\label{eq:SpJ}
\\
\langle \hat{S}^+ \delta S^+(\lambda) ,E_{\alpha}\rangle  = \delta s^+_\alpha(\lambda) \quad \mbox{for} \quad \alpha = e_1-e_2, e_2-e_3, \nonumber\\
\langle \hat{S}^+ \delta S^+(\lambda) ,E_{e_1-e_3}\rangle  = \delta S^+_{13} +\frac{1}{2} \delta ( s^+_{12}s^+_{23}) .\nonumber %\label{eq:dSp}
\end{gather}
Thus we are able to express all expansions coef\/f\/icients in terms of $s^\pm_\alpha$. The results are:
\begin{gather*}
\tau_\alpha^{(1),\pm}(\lambda)  \equiv \langle \hat{S}^\pm J S^\pm(\lambda),E_{\mp\alpha} \rangle = \pm (\alpha, J) s^\pm_\alpha(\lambda),
\nonumber\\
\tau_\alpha^{(2),\pm}(\lambda)  \equiv \langle \hat{S}^\pm I S^\pm(\lambda),E_{\mp\alpha} \rangle = \left\{ \left(
\pm (\alpha, J) s^\pm_\alpha(\lambda) \ \mbox{for}\  \alpha \in \delta_1^+
 \right) \right\}
\pm (\alpha, J_2) s^\pm_\alpha(\lambda).%\label{eq:}
\end{gather*}

\begin{table}[t]
  \centering
    \caption{The expansion coef\/f\/icients of $\ad_{L_1}^{-1}L_{j,x} $, $\ad_{L_1}^{-1}\delta L_1 $
  and $\ad_{L_1}^{-1}L_{1,t} $.}\label{tab:1}
  \vspace{1mm}

  \begin{tabular}{|l|c|c|c|}
    \hline
Roots    & $\alpha_1=e_1-e_2$ & $\alpha_1=e_1-e_2$ & $\alpha_1=e_1-e_2$ \tsep{1pt} \bsep{1pt}\\
\hline \hline
$(\alpha,J)$ & 1 & 1 &  2 \tsep{1pt}\\
$(\alpha,I)$ & 1 & $-1$ &  0 \bsep{1pt} \\
\hline
  $\tau_\alpha^{(1),+}(\lambda)$ \tsep{4pt} & $s^+_{\alpha_1}(\lambda)$  & $s^+_{\alpha_2}(\lambda)$ & $2s^+_{\alpha_3}(\lambda)$ \\
  $\tau_\alpha^{(1),-}(\lambda)$ \bsep{2pt}   & $-s^-_{\alpha_1}(\lambda)$  & $-s^-_{\alpha_2}(\lambda)$ & $-2s^-_{\alpha_3}(\lambda)$ \\
\hline
  $\tau_\alpha^{(2),+}(\lambda)$\tsep{4pt}  & $s^+_{\alpha_1}(\lambda)$  & $-s^+_{\alpha_2}(\lambda)$ & $s^+_{\alpha_1}s^+_{\alpha_2}(\lambda)$ \\
  $\tau_\alpha^{(2),-}(\lambda)$\bsep{2pt}  & $-s^-_{\alpha_1}(\lambda)$  & $s^-_{\alpha_2}(\lambda)$ & $s^-_{\alpha_1}s^-_{\alpha_2}(\lambda)$ \\
\hline
  $\rho_\alpha^{(1),+}(\lambda)$  & $ -\frac{(m_1^+)^2 }{m_2^+}r^+_{\alpha_1}(\lambda)$\tsep{6pt}\bsep{3pt}   & $ -\frac{(m_2^+)^2 }{m_1^+}r^+_{\alpha_2}(\lambda)$ &
  $-2 m_1^+ m_2^+ r^+_{\alpha_3}(\lambda)$ \\[2mm]
  $\rho_\alpha^{(1),-}(\lambda)$ & $\frac{(m_2^-)^2 }{m_1^-}r^-_{\alpha_1}(\lambda)$ \bsep{5pt}  & $ \frac{(m_1^-)^2 }{m_2^-}r^-_{\alpha_2}(\lambda)$ &
  $2  m_1^- m_2^- r^-_{\alpha_3}(\lambda)$ \\
\hline
  $\rho_\alpha^{(2),+}(\lambda)$  & $ -\frac{(m_1^+)^2 }{m_2^+}r^+_{\alpha_1}(\lambda)$\tsep{6pt}\bsep{3pt}  & $-\frac{(m_2^+)^2 }{m_1^+}r^+_{\alpha_2}(\lambda)$ &
  $ m_1^+ m_2^+ r^+_{\alpha_1}s^+_{\alpha_2}(\lambda)$ \\[2mm]
  $\rho_\alpha^{(2),-}(\lambda)$  & $\frac{(m_2^-)^2 }{m_1^-} r^-_{\alpha_1}(\lambda)$\bsep{5pt}  & $-\frac{(m_1^-)^2 }{m_2^-}s^-_{\alpha_2}(\lambda)$ &
  $ m_2^-m_1^-r^-_{\alpha_1}s^-_{\alpha_2}(\lambda)$ \\
\hline
  $\delta' \tau_\alpha^{+}(\lambda)$  & $\delta s^+_{\alpha_1}(\lambda)$  & $\delta s^+_{\alpha_2}(\lambda)$ &
  $\delta s^+_{\alpha_3}(\lambda) + \frac{1}{2} \delta s^+_{\alpha_1}s^+_{\alpha_2}(\lambda)
  - \frac{1}{2} \delta s^+_{\alpha_2} s^+_{\alpha_1}(\lambda)$\tsep{3pt}\bsep{3pt}  \\[1mm]
  $\delta ' \tau_\alpha^{-}(\lambda)$  & $\delta s^-_{\alpha_1}(\lambda)$  & $\delta s^-_{\alpha_2}(\lambda)$ & $
  \delta s^-_{\alpha_3}(\lambda) - \frac{1}{2} \delta s^-_{\alpha_1}s^+_{\alpha_2}(\lambda)
+ \frac{1}{2} \delta s^-_{\alpha_2} s^-_{\alpha_1}(\lambda)$\bsep{2pt} \\
\hline
  $\delta'_t \tau_\alpha^{+}(\lambda)$  & $ \frac{d}{dt}s^+_{\alpha_1}(\lambda)$  & $ \frac{d}{dt} s^+_{\alpha_2}(\lambda)$ &
  $\frac{d}{dt} s^+_{\alpha_3}(\lambda) + \frac{1}{2} s^+_{\alpha_2}(\lambda)\frac{d}{dt} s^+_{\alpha_1}
  - \frac{1}{2} s^+_{\alpha_1}(\lambda) \frac{d}{dt} s^+_{\alpha_2} $\tsep{3pt}\bsep{3pt} \\[1mm]
  $\delta '_t \tau_\alpha^{-}(\lambda)$  & $ \frac{d}{dt} s^-_{\alpha_1}(\lambda)$  & $ \frac{d}{dt} s^-_{\alpha_2}(\lambda)$ & $
  \frac{d}{dt} s^-_{\alpha_3}(\lambda) - \frac{1}{2} s^+_{\alpha_2}(\lambda) \frac{d}{dt} s^-_{\alpha_1}
+ \frac{1}{2} s^-_{\alpha_1}(\lambda) \frac{d}{dt} s^-_{\alpha_2} $\bsep{2pt} \\
\hline
    \hline
  \end{tabular}
\end{table}

Note that varying $L_1$ may lead to varying the positions of the discrete eigenvalues.
In order to evaluate the corresponding expansion coef\/f\/icients, now we make use of the Wronskian relations~(\ref{eq:ssp-3}).
We use also the condition {\bf C2} and the explicit formulae that allow one to express all the Gauss
factors $S^\pm(\lambda)$, $T^\pm (\lambda)$ and $D^\pm (\lambda)$ as rational expressions of the matrix elements of~$T(\lambda)$.
Condition {\bf C2} ensures that the Gauss factors have at most f\/irst order poles in the vicinity of~$\lambda_k^\pm$. Therefore the
variations of $S^\pm(\lambda)$ and $T^\pm(\lambda)$ in the vicinity of $\lambda_k^\pm $ take the form:
\begin{gather*}
\delta s_\alpha^\pm(\lambda)   \simeq (\lambda -\lambda_k^\pm)^{\mp p_\alpha +1} \left(
\frac{s_{\alpha; k}^\pm}{(\lambda - \lambda_k^\pm)^2} \delta\lambda_k^\pm +
\frac{\delta s_{\alpha; k}^\pm}{\lambda - \lambda_k^\pm} + \mathcal{O}(1)\right), \nonumber\\
\delta r_\alpha ^\mp(\lambda)   \simeq (\lambda -\lambda_k^\pm)^{\mp p_\alpha +1} \left(
\frac{r_{\alpha; k}^\mp}{(\lambda - \lambda_k^\pm)^2} \delta\lambda_k^\pm +
\frac{\delta r_{\alpha; k}^\mp}{\lambda - \lambda_k^\pm} + \mathcal{O}(1) \right),%\label{eq:dSdT}
\end{gather*}
where $s_{\alpha; k}^\pm$ and $r_{\alpha; k}^\pm$ determine the values of the corresponding Gauss factors for $\lambda =\lambda_k^\pm$.
Thus, along with the (\ref{eq:ssp-3}) we  obtain:
\begin{alignat*}{3}
& \rho^\pm_{\alpha;k} \delta\lambda_k^\pm  = i\lambda_k^\pm \biglb e^\pm_{\pm\alpha;k} , \ad_{L_1}^{-1} \delta L_1\bigrb, \qquad &&
\delta'\rho^\pm_{\alpha;k}  = i\biglb e^\pm_{\pm\alpha;k} +\lambda_k^\pm \dot{e}^\pm_{\pm\alpha;k} , \ad_{L_1}^{-1} \delta L_1\bigrb,&\nonumber \\
& \tau^\pm_{\alpha;k} \delta\lambda_k^\pm = -i\lambda_k^\pm \biglb e^\pm_{\mp\alpha;k} , \ad_{L_1}^{-1} \delta L_1\bigrb, \qquad&&
\delta'\tau^\pm_{\alpha;k}  = -i \biglb e^\pm_{\mp\alpha;k} +\lambda_k^\pm \dot{e}^\mp_{\pm\alpha;k}
 , \ad_{L_1}^{-1} \delta L_1\bigrb .& %\label{eq:drho}
\end{alignat*}

\begin{remark}\label{rem:5}
More precise treatment of the contribution from the discrete spectrum starts by considering the potentials
for which $L_1(x) - L_{1,{\rm as}}$ is on f\/inite support. Then the Jost solutions of $L$, as well as the scattering
matrix~$T(\lambda)$ become meromorphic functions of $\lambda$ which ensures the validity of all considerations
above. The next step would be taking the limit to potentials of Schwartz-type. These considerations come out of
the scope of the present paper.
\end{remark}

When we analyze the discrete spectrum we make use of the conditions {\bf C2} and {\bf C3}. Since
the discrete eigenvalues are zeroes of the principal minors, and assuming that~$\lambda_k^+$ (resp.~$\lambda_k^-$)
is a zero of, say~$m_1^+(\lambda)$, then we have:
\begin{gather*}%\label{eq:m1pm}
m_1^\pm(\lambda)  = (\lambda -\lambda_k^\pm) \dot{m}_{1;k}^\pm + \mathcal{O}\big((\lambda -\lambda_k^\pm)^2\big),
 \end{gather*}
and similar expansions for $m_2^\pm(\lambda)$. Using remark \ref{rem:3} on formulae (\ref{eq:SpJ})  we easily f\/ind, that
$\tau_\alpha^{(j),\pm}(\lambda)$ have simple zeroes in the vicinity of $\lambda_k^\pm$.

We end by reminding a few useful formulae of frequently appearing expressions in the main text;
they are derived in~\cite{ours,ours2}.

Since $L_1$ satisf\/ies the characteristic equation $L_1^3 =L_1$
then $\ad_{L_1}$ has eigenvalues $\pm 2$, $\pm 1$, $0$
and satisf\/ies the characteristic equation:
\begin{gather*}%\label{eq:chareq1}
\big(\ad_{L_1}^2 - 4\big) \big(\ad_{L_1}^2 - 1\big)  \ad_{L_1} =0
\end{gather*}
and therefore
\begin{gather*}%\label{eq:chareq2}
\ad_{L_1}^{-1} = \frac{1}{4} \big( 5\ad_{L_1} - \ad_{L_1}^3 \big).
\end{gather*}
If we choose $X\in \mathfrak{sl}^{1}$ we have
\begin{gather*}%\label{eq:adL1-0}
\ad_{L_1}^{-1} X = \left(\begin{array}{ccc}  w-w^* & 0 & 0 \\ 0 & u^*a - a^* u  & u^*b - va^*  \\
0 &  v^*a - ub^*   & v^*b - b^* v \end{array}\right) +
\frac{3}{4} (w - w^*) \left(\begin{array}{ccc} -1 & 0 & 0 \\
0 & |u|^2 & u^*v \\ 0 & v^*u & |v|^2 \end{array}\right) ,
\end{gather*}
where $w=ua^* + vb^*$. If in particular, $X=L_{1,x}$ we f\/ind:
\begin{gather*}%\label{eq:adL1x}
\mathcal{L}_1 \equiv \ad_{L_1}^{-1} L_{1,x}  = \left(\!\begin{array}{ccc} \frac{1}{2}w_0 & 0 & 0\\
0 & u^*u_x -u^*_x u +\frac{3}{2} w_0|u|^2 &  u^*v_x -v u^*_x  +\frac{3}{2} w_0u^* v \\
0 & v^*u_x - uv^*_x  +\frac{3}{2} w_0 uv^* & v^*v_x -v^*_x v +\frac{3}{2} w_0|v|^2
\end{array}\!\right),
\end{gather*}
where $w_0 = uu^*_x + vv^*_x$.
Similarly for  $Y\in \mathfrak{sl}^{0}$ and $\ad_{L_1}^{-1} Y$ we have
\begin{gather*}
Y  = \left(\!\begin{array}{c@{\,\,\,}c@{\,\,\,}c} -k-n & 0 & 0 \\ 0 & k & m \\
0 & m^* & n \end{array}\!\right) , \qquad
\ad_{L_1}^{-1} Y =\frac{1}{4}\left(\!\begin{array}{c@{\,\,\,}c@{\,\,\,}c} 0 & u\alpha_0 +vm^* & v\alpha_1+um \\
-u^*\alpha_0 -mv^*  & 0  & 0  \\ -v^* \alpha_1- u^*m^* &  0   & 0 \end{array}\!\right),\nonumber  \\
\alpha_0  = 5k +n-3W  , \qquad   \alpha_1 =k +5n-3W.%\label{eq:Y1}
\end{gather*}
Choosing $Y=L_{2,x}$ we get:
\begin{gather*}
\mathcal{L}_2 \equiv \ad_{L_1}^{-1} L_{2,x}
 = \left(\!\begin{array}{c@{\,\,\,}c@{\,\,\,}c} 0 & - u|u|^2_x - v(uv^*)_x & - v|v|^2_x - u(v^*u)_x \\
u^* |u|^2_x + v^*(u^*v)_x & 0 & 0 \\ v|v|^2_x + u(u^*v)_x & 0 & 0 \end{array}\!\right)  \\
\phantom{\mathcal{L}_2 \equiv \ad_{L_1}^{-1} L_{2,x}}{}
= \left(\!\begin{array}{c@{\,\,\,}c@{\,\,\,}c} 0 & - u_x  - u(uu^*_x +vv^*_x) & - v_x  - v(uu^*_x +vv^*_x) \\
u^*_x  + u^*(u^*u_x +v^*v_x) & 0 & 0 \\ v^*_x  + v^*(u^*u_x +v^*v_x)  & 0 & 0 \end{array}\!\right)  .%\label{eq:aL2}
\end{gather*}

\subsection*{Acknowledgements}

The authors have the pleasure to thank Professor
Allan Fordy  for numerous useful discussions. The authors acknowledge support from the Royal Society and
the Bulgarian academy of sciences via joint research project
``Reductions of Nonlinear Evolution Equations and analytic
spectral theory''.  The work of G.G.G.\ is supported by the
Science Foundation of Ireland (SFI), under Grant no.~09/RFP/MTH2144. Finally we would like to
thank one of the referees for useful suggestions.

%\pdfbookmark[1]{References}{ref}
\LastPageEnding

\end{document}